\RequirePackage{fix-cm}
\documentclass[twocolumn,epjc3,runningheads,smartrunhead]{svjour3}  
\smartqed  
\RequirePackage[colorlinks,breaklinks]{hyperref}

\hypersetup{linkcolor=blue,citecolor=blue,filecolor=black,urlcolor=blue} 

\RequirePackage{graphicx}
\usepackage{atlasphysics}
\usepackage{mathrsfs}
\usepackage{subfig}

\usepackage{amssymb}

\newcommand{\antiKT}{anti-$k_t$~}

\journalname{Eur.\ Phys.\ J.\ C}

\usepackage{preprintcover} 
\PreprintCoverPaperTitle{Single hadron response measurement and calorimeter jet energy scale uncertainty with the ATLAS detector at the LHC} 
\PreprintIdNumber{CERN-PH-EP-2012-005}
\PreprintCoverAbstract{The uncertainty on the calorimeter energy response to jets of particles is derived for the ATLAS experiment at the Large Hadron Collider (LHC). First, the  calorimeter response to single isolated charged hadrons is measured and compared to the Monte Carlo simulation using proton-proton collisions at centre-of-mass energies of $\sqrt{s} = 900$ GeV and 7 TeV collected during 2009 and 2010. Then, using the decay of $K_s$ and $\Lambda$ particles, the calorimeter response to specific types of particles (positively and negatively charged pions, protons, and anti-protons) is measured and compared to the Monte Carlo predictions. Finally, the jet energy scale uncertainty is determined by propagating the response uncertainty for single charged and neutral particles to jets. 
The response uncertainty is 2--5\% for central isolated hadrons and 1--3\% for the final calorimeter jet energy scale.} 
\PreprintJournalName{Eur.\ Phys.\ J.\ C} 

\begin{document}

\title{Single hadron response measurement and calorimeter jet energy scale uncertainty with the ATLAS detector at the LHC}
\titlerunning{Single hadron response measurement and calorimeter jet energy scale uncertainty}

\author{The ATLAS Collaboration}
\institute{CERN, Geneva, Switzerland}

\date{Received: date / Accepted: date}

\maketitle

\begin{abstract}
The uncertainty on the calorimeter energy response to jets of particles is derived for the ATLAS experiment at the Large Hadron Collider (LHC). First, the  calorimeter response to single isolated charged hadrons is measured and compared to the Monte Carlo simulation using proton-proton collisions at centre-of-mass energies of $\sqrt{s} = 900$ GeV and 7 TeV collected during 2009 and 2010. Then, using the decay of $K_s$ and $\Lambda$ particles, the calorimeter response to specific types of particles (positively and negatively charged pions, protons, and anti-protons) is measured and compared to the Monte Carlo predictions. Finally, the jet energy scale uncertainty is determined by propagating the response uncertainty for single charged and neutral particles to jets. 
The response uncertainty is 2--5\% for central isolated hadrons and 1--3\% for the final calorimeter jet energy scale.


\end{abstract}

\section{Introduction}

Partons scattered in proton-proton interactions are measured with the ATLAS detector as collimated jets of hadrons.
The uncertainty on the jet energy scale is the largest source of detector-related systematic uncertainty for many physics analyses carried out by the ATLAS Collaboration, from the di-jet cross-section and top mass measurements, to searches for new physics with jets in the final state. It is thus the subject of an extensive and detailed study~\cite{JES_Summary_r16}.

The jet energy measured by the calorimeter is corrected for calorimeter non-compensation and energy loss in dead material. The corresponding jet energy scale correction factor is referred to as the JES. The JES is derived from Monte Carlo (MC) simulations by comparing the  
calorimeter energy of an isolated reconstructed jet to that of the particle jet\footnote{Particle jets (in MC simulated events) are defined as the jets obtained by running the jet finding algorithm on the stable particles from the event generator, including those with lifetimes longer than 10 ps and excluding neutrinos and muons.} that points to it~\cite{JES_Summary_r16}.
 
The uncertainty on the calorimeter energy response is a significant component of the total uncertainty on the JES. It is derived in this paper by convolving the measured uncertainty on the single charged hadron energy response and the estimated uncertainty on the neutral particle energy response with the expected particle spectrum within a jet.

The calorimeter response to single isolated charged hadrons, and the accuracy of its Monte Carlo simulation description, can be evaluated from the ratio of the calorimeter energy $E$ to the associated isolated track momentum $p$. The aim of the measurement is to estimate the systematic uncertainty on jet calorimeter response and therefore the focus is on data-to-MC comparison. The ratio $E/p$ is measured using proton-proton collisions at centre-of-mass energies of $\sqrt{s} = 900$~GeV and 7 TeV over a wide range of track momenta in the central region of the calorimeter. Possible additional uncertainties introduced by certain particle species are addressed by measuring the response to hadrons identified through the reconstruction of known short-lived particles.

This paper is organised as follows.
The relevant features of the ATLAS detector are summarised in Section~\ref{sec:AtlasDetector}.
Then the measurement of the single charged hadron response, at centre-of-mass energies of $\sqrt{s} = 900$~GeV and 7~TeV, is discussed in Section~\ref{sec:E/P}. 
Section~\ref{sec:resonances} describes the measurement of $E/p$ for charged particles identified using $K_s$ and $\Lambda$ particle decays.
In Section~\ref{sec:method}, the determination of the JES uncertainty is discussed. This includes the propagation of the charged particle response uncertainty to jets and additional systematic uncertainties related to the calorimeter response of neutral particles. The total uncertainty on the calorimeter energy response to jets and its correlations are shown for jets with transverse momenta between 15~GeV and 2.5~TeV and in the pseudorapidity\footnote{ATLAS uses a right-handed coordinate system with its origin at the nominal interaction point (IP) in the centre of the detector and the $z$-axis along the beam pipe. The $x$-axis points from the IP to the centre of the LHC ring, and the $y$-axis points upward. Cylindrical coordinates $(r,\phi)$ are used in the transverse plane, $\phi$ being the azimuthal angle around the beam pipe. The pseudorapidity is defined in terms of the polar angle $\theta$ as $\eta=-\ln\tan(\theta/2)$.} range $|\eta|<0.8$.
The total JES uncertainty, including all calorimeter regions and effects not related to the calorimeter response, is discussed in Ref.~\cite{JES_Summary_r16}.
Throughout the paper, all uncertainties are combined in quadrature, unless stated otherwise.

\section{The ATLAS detector}
\label{sec:AtlasDetector}

The ATLAS detector covers almost the whole solid angle around the collision point with layers of tracking
detectors, calorimeters and muon chambers and is described in detail in Ref.~\cite{atlasDetectorPaper}. Here, the features relevant for this analysis are summarised.

The inner detector (ID) is immersed in a 2~T axial magnetic field and provides tracking for charged particles with $|\eta|<2.5$. 
The ID consists of a silicon pixel tracker and silicon microstrip tracker (SCT) covering $|\eta|<2.5$ and a transition radiation tracker (TRT) covering $|\eta|<2.0$.

The calorimeter system covers $|\eta|<4.9$, using a variety of technologies. High granularity liquid-argon (LAr)
electromagnetic (EM) sampling calorimeters, with excellent performance in terms of energy and position
resolution, cover $|\eta|<3.2$. They use accordion-shaped electrodes and lead absorbers and consist of a barrel (EMB, $|\eta|<1.475$) 
and an end-cap (EMEC, $1.375<|\eta|<3.2$). They are longitudinally segmented in depth into three layers, with a pre-sampler behind the solenoid.
For $|\eta|<1.7$ hadronic calorimetry is provided by a sampling calorimeter made of iron and scintillating tiles (TileCal). TileCal comprises a large barrel ($|\eta|<0.8$) and two smaller extended barrel cylinders ($0.8<|\eta|<1.7$).  It is segmented longitudinally into three layers, with a total thickness of about eight interaction lengths at $\eta = 0$. The hadronic end-cap calorimeters
(HEC, covering $1.5<|\eta|<3.2$) are LAr sampling calorimeters with copper absorbers. The copper/tungsten-LAr forward calorimeters (FCal) provide
both electromagnetic and hadronic energy measurements and extend the coverage to $|\eta|<4.9$.

The data used for this analysis were triggered using the minimum bias trigger scintillators (MBTS) \cite{minbias}. The MBTS are mounted at each end of the detector in front of the LAr end-cap calorimeter cryostats at $z = \pm 3.56$~m and are segmented into eight sectors in azimuth and two rings in pseudorapidity ($2.09 < |\eta| < 2.82$ and $2.82 < |\eta| < 3.84$).

\section{Single particle response for charged hadrons}
\label{sec:E/P}

In this section, the response of the calorimeters to isolated charged hadrons is compared to the predictions from MC simulation. The ratio of the energy, $E$, deposited by an isolated charged particle in the calorimeter to the track momentum, $p$, is used to evaluate the uncertainty on the calorimeter response modelling in the MC simulation.

\subsection{Event selection}
\label{sec:event_selection}
Events are required to have at least one reconstructed vertex with at least four associated tracks. The total number of events satisfying this selection is about 25~million collected in 2009 (approximately one million events at $\sqrt{s}=900$ GeV) and 2010 (approximately 24 million events at $\sqrt{s}=7$ TeV). 

For every selected event, each track candidate is extrapolated to the second longitudinal layer of the EM calorimeter. A track is defined as isolated if its impact point has a distance $\Delta R = \sqrt{\left (\Delta \eta \right)^2 + \left(\Delta \phi \right) ^2} > 0.4$ from all other track candidate impact points. The isolation conditions are studied in detail in Section~\ref{sec:nearbyParticles}.

The isolated tracks must also have:

\begin{itemize}
\item a transverse momentum of $\pt > 500$~MeV\footnote{Below $500$~MeV a charged particle loops in the ID and does not reach the barrel calorimeter.}, 
\item a minimum of one hit in the pixel detector and six hits in the SCT, and
\item small transverse and longitudinal impact parameters (defined in Ref.~\cite{ATLAS:2010ir}) computed with respect to
 the primary vertex\footnote{In case of multiple vertices, the primary vertex is taken to be the one for which the sum of the square of momenta of the attached tracks $\pt^2$ is the largest.}, $|d_0| < 1.5$ mm and $|z_0| \sin \theta < 1.5$ mm. 
\end{itemize}

The above requirements ensure a good quality of the track and reduce contributions from fake tracks to a negligible level.

\subsection{Monte Carlo simulation}
\label{sec:mcsim}
A sample of about 10 (20) million non-diffractive proton-proton collision events at $\sqrt{s} = 900$ GeV ($\sqrt{s} = 7$ TeV) are generated using {\sc Pythia}~6.421~\cite{pythia} with the ATLAS minimum bias tune 1 (AMBT1)~\cite{ATLAS:2010ir}.
For isolated, high-momentum ($\pt > 15$ GeV) tracks this corresponds to $\sim$~60\% of the available events in data. 
All the events are run through a full detector simulation \cite{2010arXiv1005.4568T} based on {\sc Geant4} \cite{geant4}. 
The set of {\sc Geant4} physics models used is QGSP\_BERT~\cite{geant4Physics}. 
The reconstruction and analysis software used for the MC simulation is the same as for the data.

\subsection{Defining the $E/p$~observable}
\label{sec:definitions}
\begin{figure*}[tb]
\centering
\subfloat[]{\label{fig:EoverPcentral} \includegraphics[width=.48\textwidth]{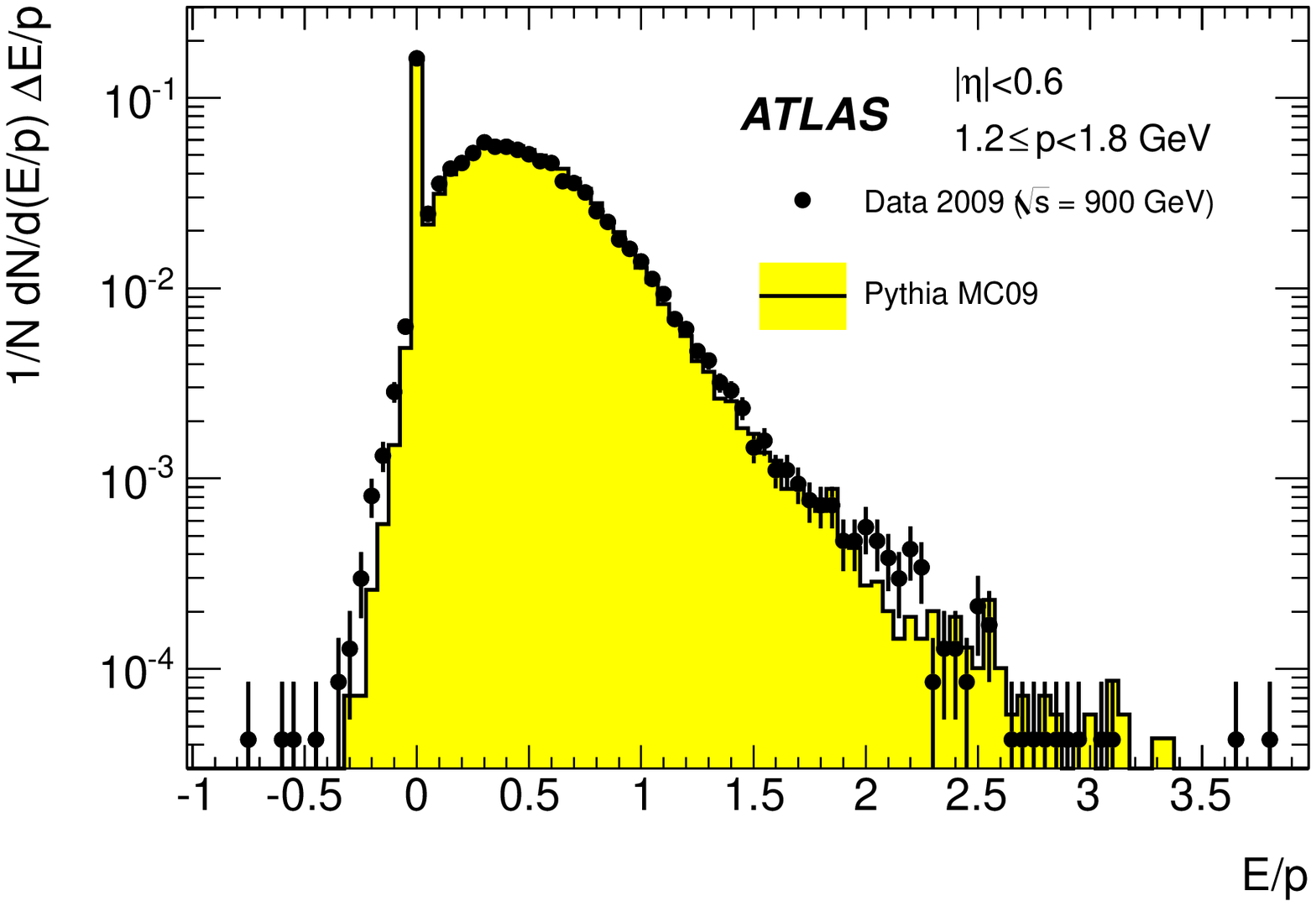}}
\subfloat[]{\label{fig:EoverPforward} \includegraphics[width=.48\textwidth]{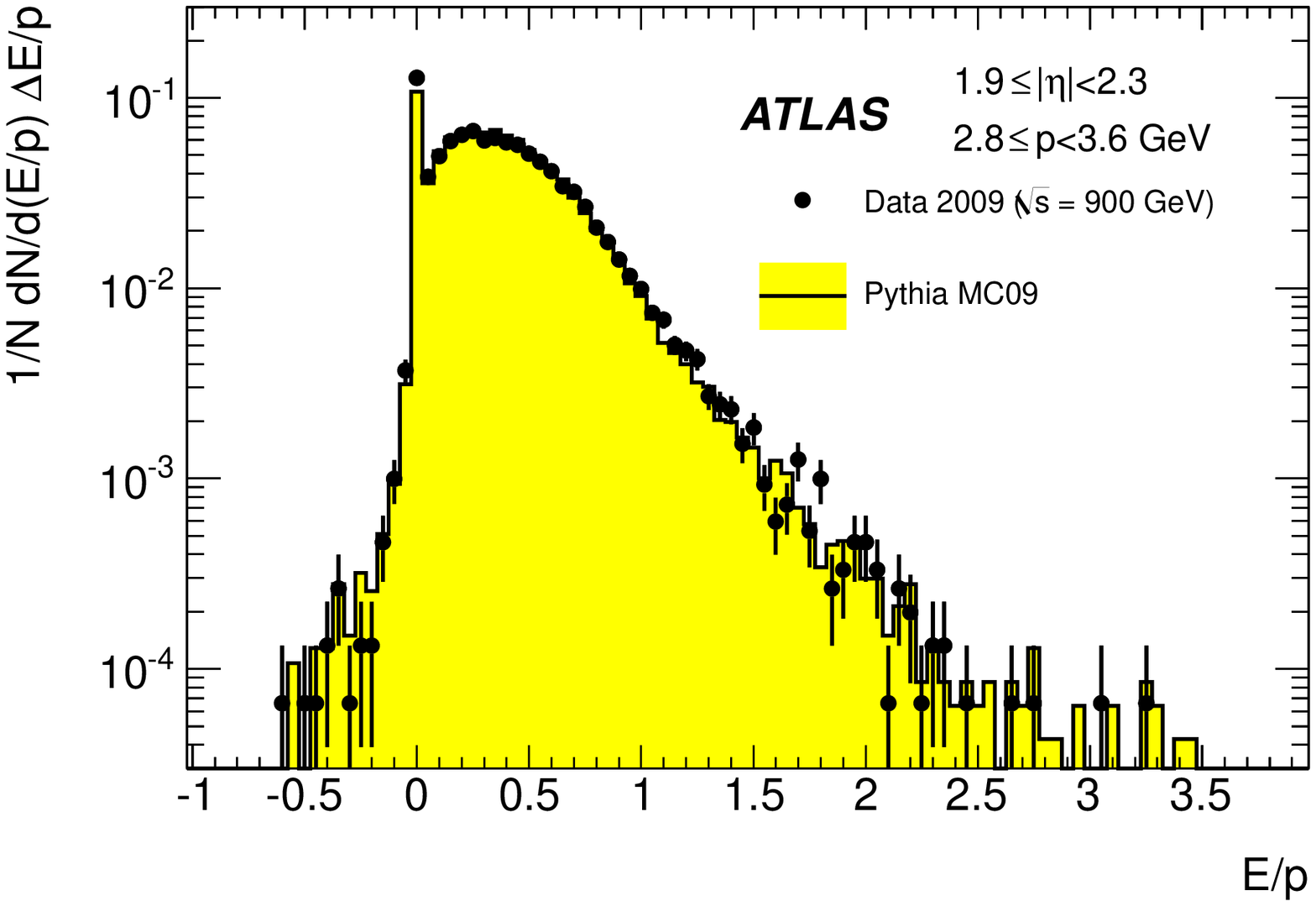}}
\caption{\protect\subref{fig:EoverPcentral} $E/p$ distribution for isolated tracks with an impact point in the region $|\eta| < 0.6$ and with a momentum in the range $1.2 \leq p < 1.8$ GeV. \protect\subref{fig:EoverPforward} $E/p$ distribution for tracks with impact points in the region $1.9 \leq |\eta| < 2.3$ and with momenta in the range $2.8 \leq p < 3.6$ GeV.}
\label{fig:E_P}
\end{figure*}

\begin{figure*}[htp]
\centering
\subfloat[]{\label{fig:correlationPositive} \includegraphics[width=.48\textwidth]{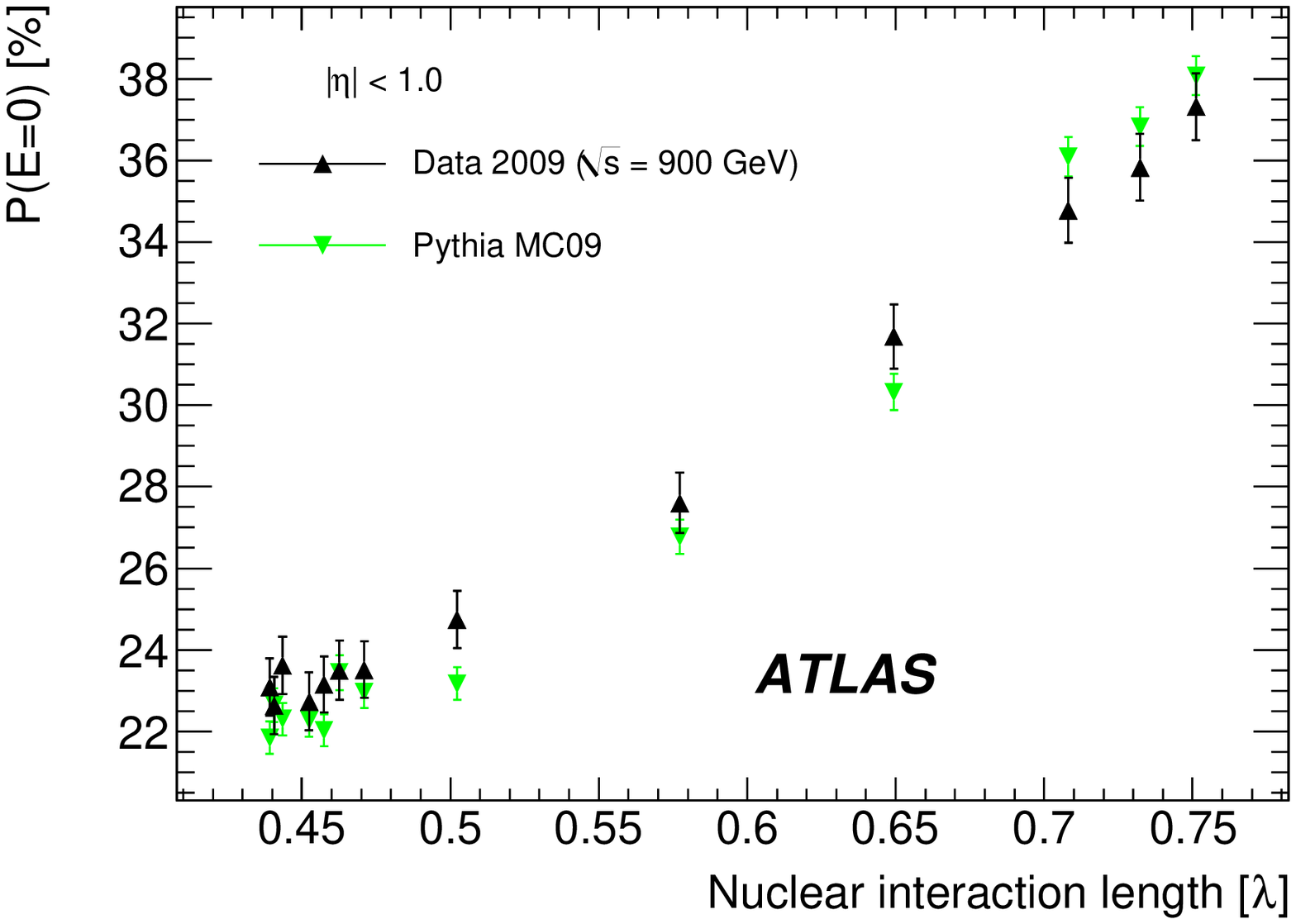}}
\subfloat[]{\label{fig:zp_vs_momentum} \includegraphics[width=.48\textwidth]{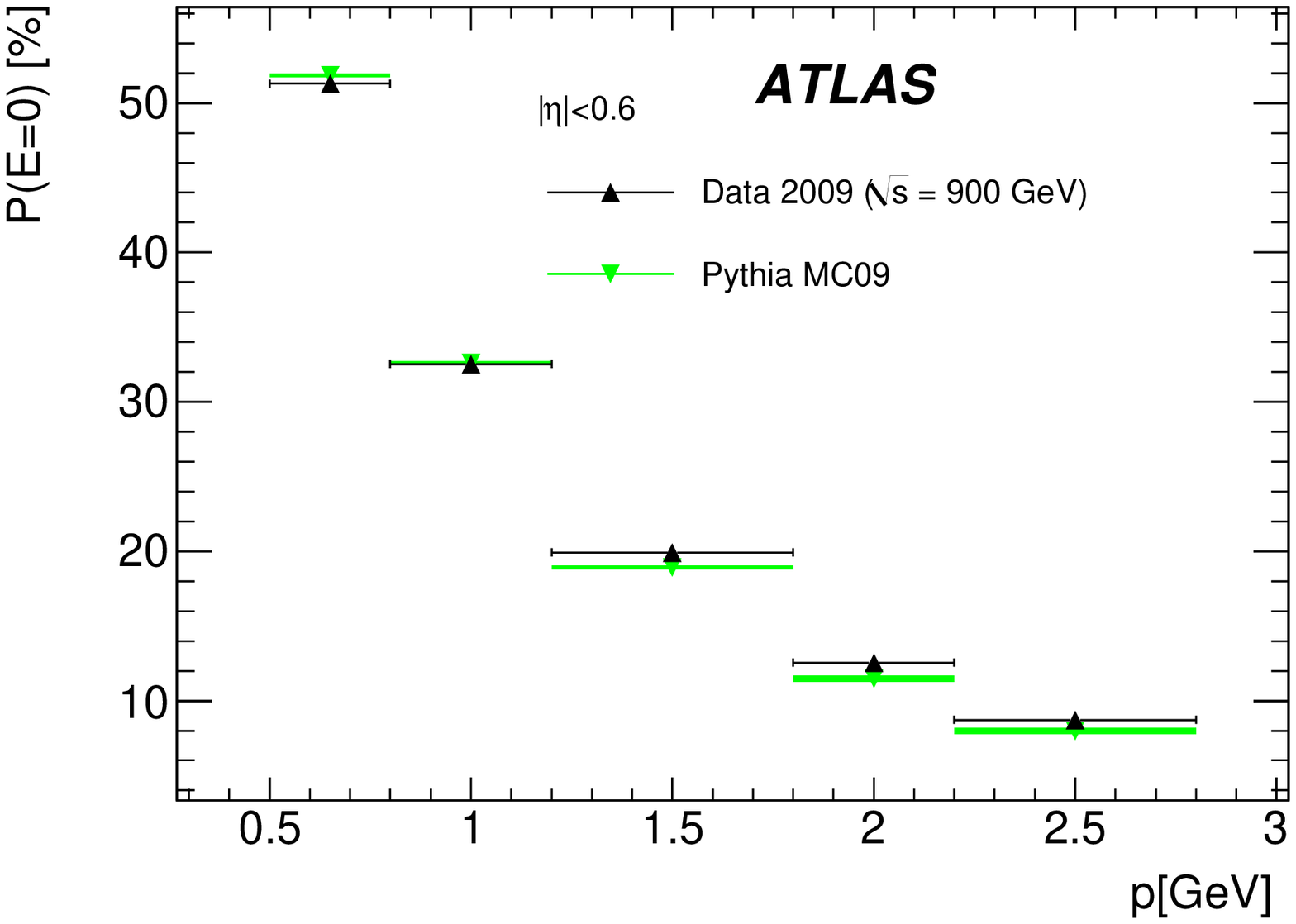}}
\caption{\protect\subref{fig:correlationPositive} Probability to measure a calorimeter response consistent with zero
$P(E=0)$ as a function of the amount of material in nuclear interaction lengths in front of active
volume of the calorimeter for $|\eta| < 1.0$. \protect\subref{fig:zp_vs_momentum} $P(E=0)$ as a function of track
momentum for $|\eta|<0.6$.}
\label{fig:zpCorrelation}
\end{figure*}

The sum of the energy deposits in layers of calorimeter cells associated to a selected track is computed using topological clusters \cite{topoclusters} at the electromagnetic scale, i.e., without applying any correction for the calorimeter non-compensation or for energy loss in dead material.  The purpose of the topological clustering algorithm is to identify areas of connected
energy deposits in the calorimeter, based on the significance of the energy deposits in cells with
respect to the expected noise level. Topological clusters are formed around cells with energy $|E_{\mathrm{cell}}| > 4\sigma_{\mathrm{noise}}$ (``seeds''), where $\sigma_{\mathrm{noise}}$ is the RMS of that cell noise. Then, iteratively, the cluster is expanded by adding all neighbouring cells with $|E_{\mathrm{cell}}| >
2\sigma_{\mathrm{noise}}$. Finally, the cells surrounding the resulting cluster are added, regardless of their energy.
The $\eta-\phi$ position of a cluster $i$ in a given calorimeter layer $j$,
$(\eta_{cl}^{ij}, \phi_{cl}^{ij})$ is computed as the energy--weighted position of the cells in layer $j$
belonging to the cluster. 

The position of the track $k$ extrapolated to the layer $j$ is $(\eta_{tr}^{kj},\phi_{tr}^{kj})$. The energy of a cluster in the layer $j$ $(E_j)$ is associated to the track if: 

\begin{equation}
\sqrt{(\eta_{tr}^{kj} - \eta_{cl}^{ij})^2 + (\phi_{tr}^{kj} - \phi_{cl}^{ij})^2} < R_{\mathrm{coll}} .
\end{equation}

The parameter $R_{\mathrm{coll}}$ is set to $0.2$ based on a trade-off between maximising the particle shower
containment and minimising the background contribution coming from neutral particles produced close
to the track. Roughly 90\% of the shower energy is collected in a cone of such size.

The energy $E$ associated to a track is computed as the sum of the associated energy deposits in all layers, $E = \sum_j E_j$, and the ratio $E/p$ is formed with the reconstructed track momentum. Note that because of calorimeter noise fluctuations, $E$ (and therefore $E/p$) can assume negative values.

\subsection{$E/p$ distributions} 
\label{sec:distributionsEoP}

The $E/p$ distributions in two representative regions of $\eta$ and track momentum are shown in \figurename~\ref{fig:E_P}. 
The large number of entries with $E/p = 0$ corresponds to isolated tracks that have no associated cluster in the calorimeter. 
Several effects may be responsible for this:
\begin{itemize}
\item Particles can interact hadronically before reaching the calorimeter (in the ID, cryostat or solenoid magnet). Such particles can change their direction, or produce a large number of low momentum secondary particles. 
\item A cluster is created only if a seed is found. Hadrons with low momentum and an extended shower topology sometimes do not have a single cell energy deposit large enough to seed a topological cluster.   
\end{itemize}

The cases where the calorimeter response is compatible with zero have been further studied.
The probability that the calorimeter response is suppressed by noise threshold requirements, $P(E=0)$, is shown in \figurename~\ref{fig:correlationPositive} as a
function of the amount of material (in nuclear interaction lengths) in front of the active
volume in the central $\left(|\eta| < 1.0\right)$ calorimeter region. When the hadron passes through more material, the probability that no energy is associated to the track increases.

Figure~\ref{fig:zp_vs_momentum} shows $P(E=0)$ as a function of the
track momentum in the central region of the calorimeter ($|\eta| < 0.6$). In this region, the dead material in front of the calorimeter is approximately constant. The  probability
decreases with increasing track momentum. In general, $P(E=0)$ is well predicted by the MC simulation.

\subsection{Background subtraction}
\label{sec:background}
The energy measured inside the cone of $\Delta R < R_{\mathrm{coll}} = 0.2$ centered around the track impact point may be contaminated by energy deposits from the showers of close-by particles produced in the proton-proton collision. The track isolation requirement suppresses possible shower contamination from charged particles. There is no obvious way to suppress shower contamination from photons, mostly produced in  $\pi^0\rightarrow \gamma \gamma$ decays, and neutral hadrons. The neutral particle background contribution to the $E/p$ measurement in the MC simulation depends on the event generator settings of the parameters governing non-perturbative QCD processes and on the modelling of the calorimeter response to low momentum neutral particles, and it is therefore difficult to model correctly. 
The neutral background is thus subtracted from the measured response using an in situ background estimate. In the following, the expression ``EM (HAD) energy'' will refer to the energy deposited in the electromagnetic (hadronic) calorimeter. 

\begin{figure}[tb]
\centering
\includegraphics[width=.2\textwidth]{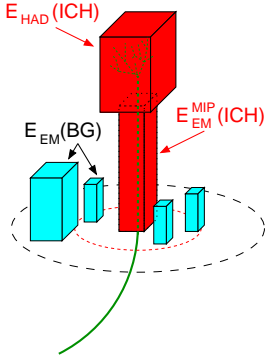}
\caption{Sketch of the background subtraction method, selecting isolated charged hadrons (ICH) that pass the EM calorimeter as minimum ionising particles (MIP) and shower in the hadronic calorimeter.}
\label{fig:sketch_bg_sub}
\end{figure}

The background subtraction relies on the assumption that the EM energy from photons and neutral hadrons is independent of the energy deposited by the selected track.
Charged hadrons are selected that behave like minimum ionising particles in the EM calorimeter and start their shower in the hadronic calorimeter (late-showering hadrons). Excluding a narrow region around the late-showering hadron track, the remaining EM energy is mainly due to showers from neutral particles. The strategy is sketched in \figurename~\ref{fig:sketch_bg_sub}. 

Late-showering hadrons are selected by requiring a small amount of EM energy  in a cone of $\Delta R < 0.1$, $E_{\mathrm{EM}}^{0.1} < 1.1$~GeV, and a large HAD energy fraction, $E_{\mathrm{HAD}}^{0.1}/p > 0.4$. The background is measured in the EM calorimeter in an annulus around the late-showering charged hadrons. The mean of the background distribution over many events in a given momentum and pseudorapidity bin
estimates the energy deposition of photons and neutral hadrons showering in the EM calorimeter:

\begin{equation}
\label{eop:background}
\langle E/p\rangle_{\mathrm{BG}}~=~  \left\langle \frac{E_{\mathrm{EM}}^{0.2}~-~E_{\mathrm{EM}}^{0.1}}{p}\right\rangle .
\end{equation}

\noindent where $E_{\mathrm{EM}}^{0.2}$ is the EM energy in a cone of  $\Delta R < 0.2$.

The background contribution from neutral hadrons depositing their energy in the hadronic calorimeter was estimated with a similar technique applied to different hadronic calorimeter layers and found to be negligibly small. 

\begin{figure*}[tbp]
\centering
\subfloat[]{\label{fig:eopBG900_etaBin7}\includegraphics[width=.45\textwidth]{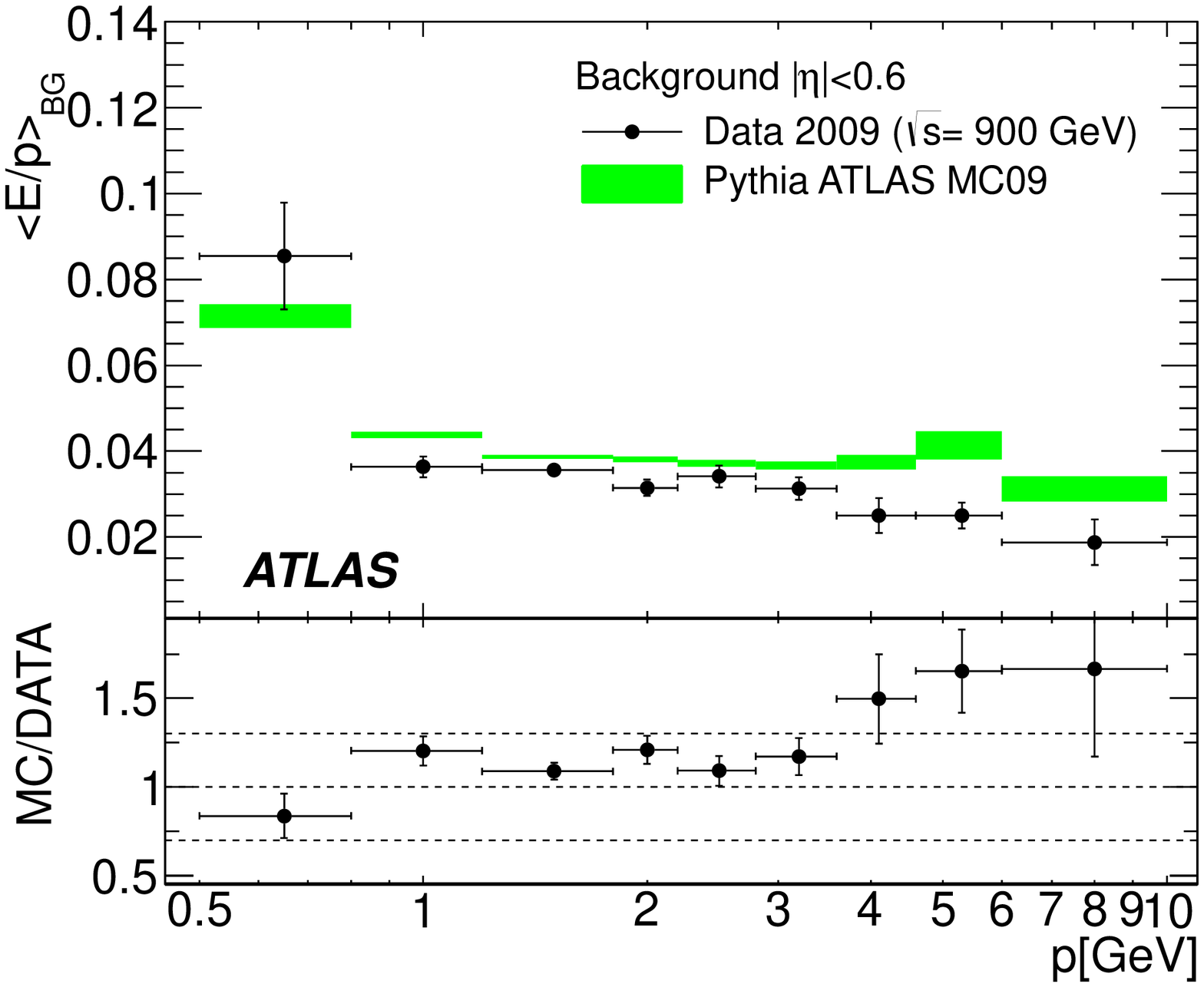}}
\subfloat[]{\label{fig:eopBG900_etaBin6}\includegraphics[width=.45\textwidth]{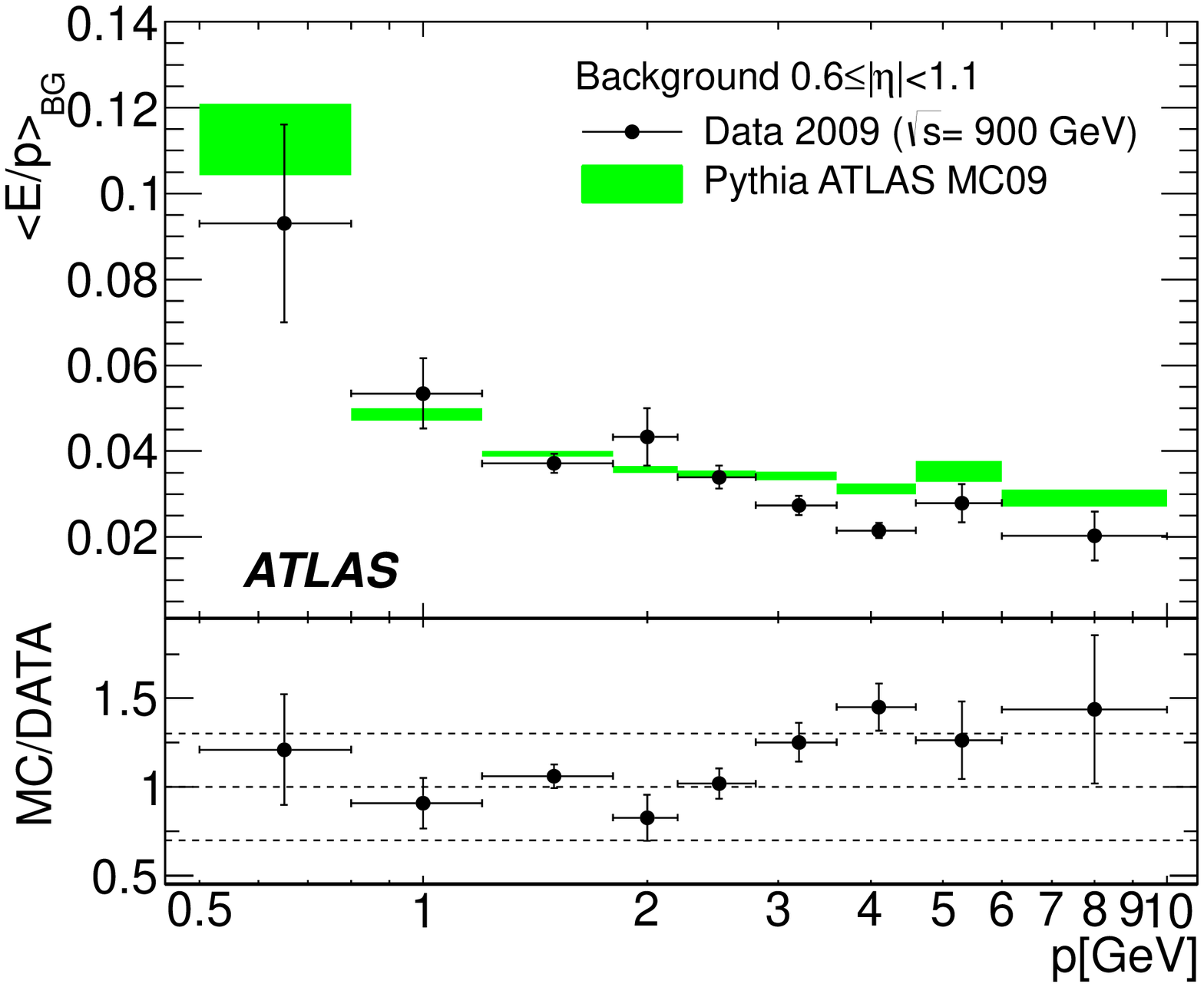}}
\caption{$\langle E/p \rangle_{\mathrm{BG}}$ as a function of the track momentum at $\sqrt{s} = 900$~GeV for \protect\subref{fig:eopBG900_etaBin7} $|\eta| < 0.6$ and \protect\subref{fig:eopBG900_etaBin6} $0.6 \leq |\eta| < 1.1$. The black markers represent the background estimated from collision data, while the green rectangles represent the MC prediction, with the vertical width showing its statistical uncertainty. The lower panes show the ratio of the MC prediction to collision data. 
 The MC prediction of the background energy deposit is obtained using the same procedure as applied to the data.
}
\label{fig:back900}
\end{figure*}

\begin{figure*}[tbp]
\centering
\subfloat[]{\label{fig:eopBG_etaBin7}\includegraphics[width=.45\textwidth]{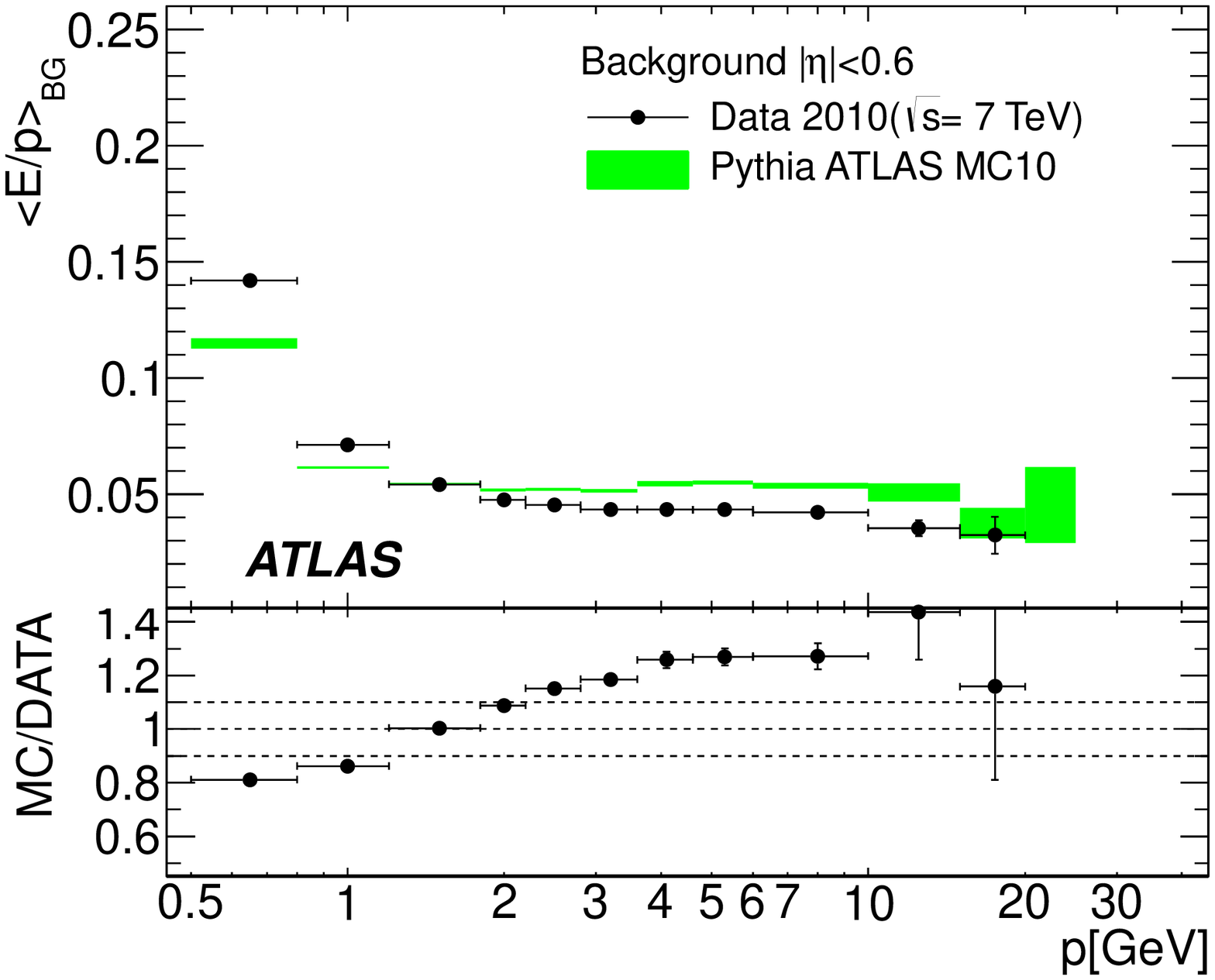}}
\subfloat[]{\label{fig:eopBG_etaBin6}\includegraphics[width=.45\textwidth]{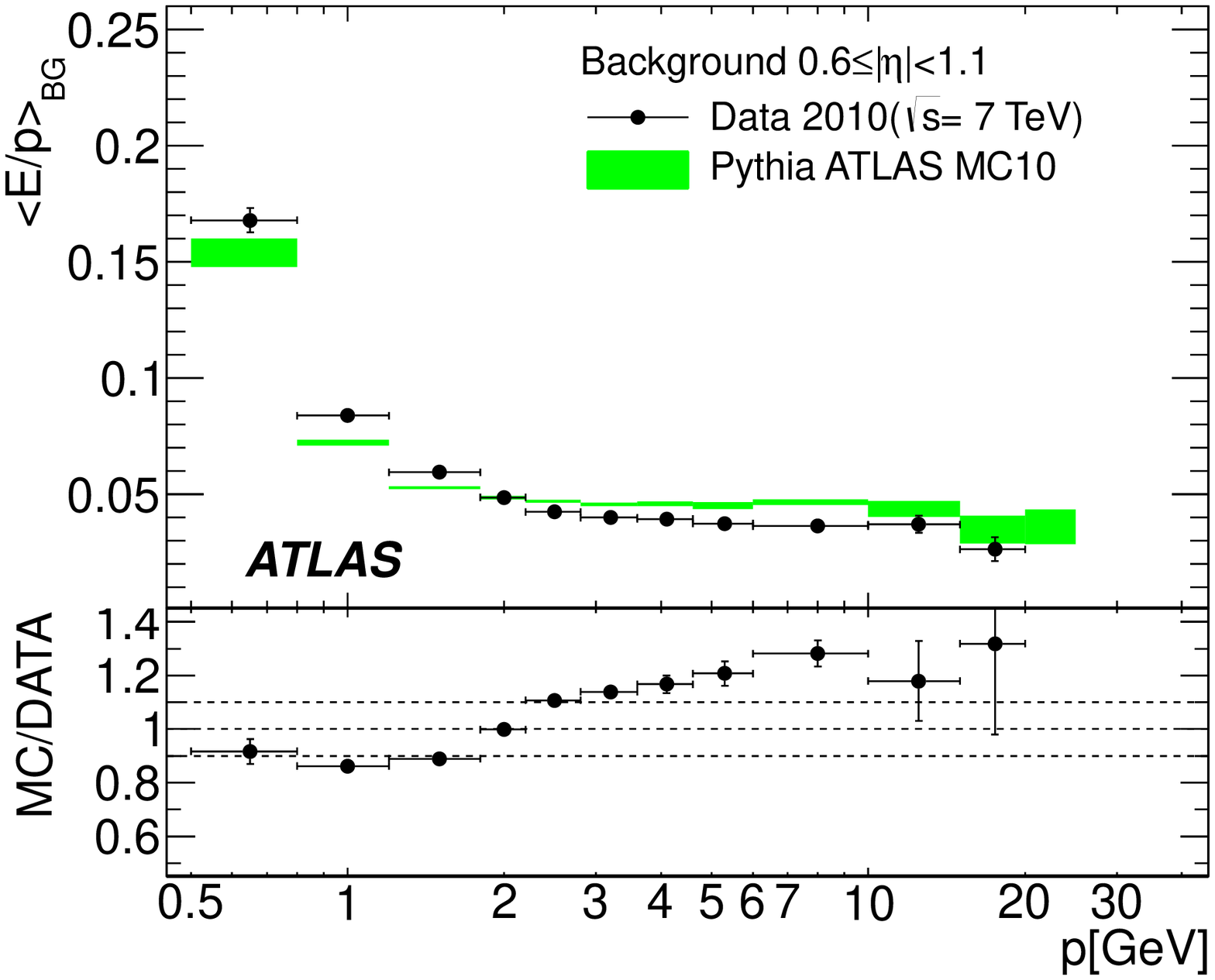}}
\caption{$\langle E/p \rangle_{\mathrm{BG}}$ as a function of the track momentum at $\sqrt{s} = 7$~TeV for \protect\subref{fig:eopBG_etaBin7} $|\eta| < 0.6$ and \protect\subref{fig:eopBG_etaBin6} $0.6 \leq |\eta| < 1.1$. The black markers represent the background estimated from collision data, while the green rectangles represent the MC prediction, with the vertical width showing its statistical uncertainty. The lower panes show the ratio of the MC prediction to collision data. The MC prediction of the background energy deposit is obtained using the same procedure as applied to the data.
}
\label{fig:back}
\end{figure*}

Figures~\ref{fig:back900} and~\ref{fig:back} show $\langle E/p\rangle_{\mathrm{BG}}$ as a function of $p$ in two bins of pseudorapidity at $\sqrt{s} = 900$~GeV and~$\sqrt{s} = 7$~TeV. 
The background $\langle E/p\rangle_{\mathrm{BG}}$ at $\sqrt{s} = 7$~TeV in both bins is about $0.04-0.08$ for $p<$~10~GeV ($0.02-0.08$ at $\sqrt{s} = 900$~GeV) and decreases to $\sim 0.04$ for track momenta of $\sim$20~GeV. 
The general trend is confirmed by the MC simulation, although the two show some significant differences. The discrepancy is attributed to an imperfect modelling of non-perturbative QCD processes by the {\sc Pythia} settings used for the MC event simulation.

The quantity that is studied in detail in the next section as a function of the track momentum and calorimeter impact point pseudorapidity is:

\begin{equation}
\langle E/p \rangle = \langle E/p \rangle_{\mathrm{raw}} - \frac{4}{3} \langle E/p \rangle_{\mathrm{BG}}
\end{equation}

\noindent where $\langle E/p \rangle_{\mathrm{raw}}$ is the mean value of the $E/p$ distribution before background subtraction (cf. \figurename~\ref{fig:E_P}). 

The background  (Eq.~\ref{eop:background}) is rescaled by the factor of 4/3, which is the ratio of the area of the full $\Delta R<0.2$~cone to that of an annulus with $0.1 \leq \Delta R<0.2$, to take into account the background contribution in $E_{\mathrm{EM}}^{0.1}$, since the background is only measured in this annulus.  A uniform energy density of the background in the $\Delta R<0.2$~cone is assumed.  The assumption has been validated by comparing its prediction with that of a more complex procedure, which is described in  Appendix A. 

The background subtraction procedure is applied in the following to all $E/p$ measurements, both at $\sqrt{s} = 900$ GeV and $\sqrt{s} = 7$ TeV. Because the background contribution increases with $\sqrt{s}$, the JES uncertainty estimation relies on the 900 GeV data for the measurement of calorimeter response to tracks with momenta below 2.2~GeV.

\subsection{Results} 
\label{sec:eoverp_results}
Several systematic uncertainties on the measurement are estimated. Each is taken to be completely correlated between all pseudorapidity and momentum bins in the $\langle E/p \rangle$ measurement:
 
\begin{itemize}
  \item Track selection: The dependence of $\langle E/p \rangle$ on the track selection requirements in Section~\ref{sec:event_selection} has been estimated by varying the number of silicon tracker hits required and the impact parameter selection with respect to the primary vertex within reasonable ranges. The MC-to-data ratio of $\langle E/p \rangle$ was found to be almost unaffected by variations in the track selection. The maximum variation found in the ratio (0.5\%) is taken as a systematic uncertainty.
  \item Track momentum scale: The uncertainty on the momentum scale $p$ as measured by the inner detector is negligibly small for $p < 5$ GeV~\cite{Aad:2011hd}. For $p > 5$~GeV, a conservative 1\% uncertainty has been assumed on the momentum scale.
  \item Background subtraction: The difference between the background estimate obtained with the method described in Section~\ref{sec:background} and with the validation method described in Appendix A is taken as a systematic uncertainty. This results in a 1\% uncertainty on the $E/p$ measurements. 
\end{itemize}

\begin{figure*}[htp]
\centering
\subfloat[]{\label{fig:eopSUBBG_etaBin7}\includegraphics[width=.48\textwidth]{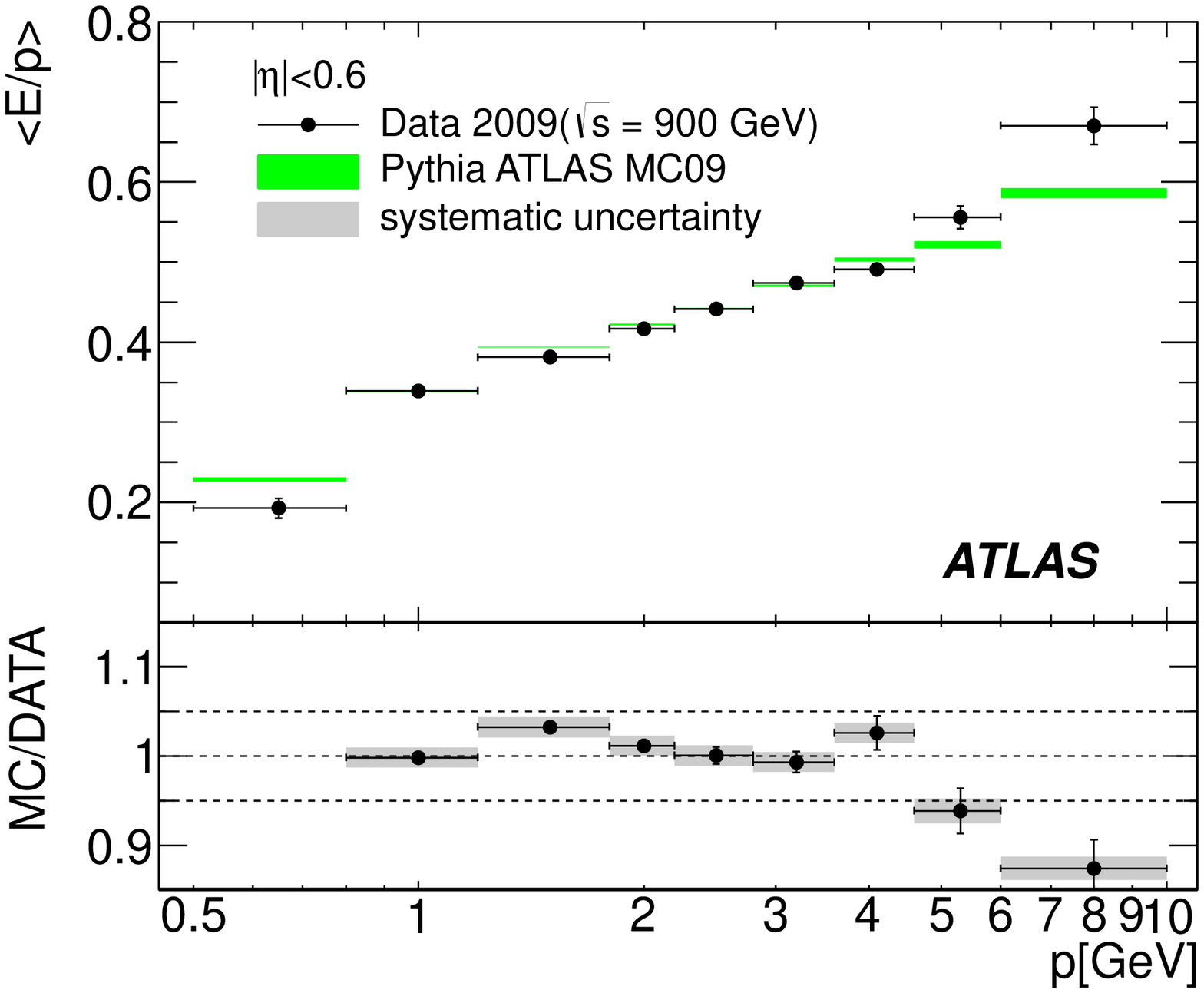}}
\subfloat[]{\label{fig:eopSUBBG_etaBin6}\includegraphics[width=.48\textwidth]{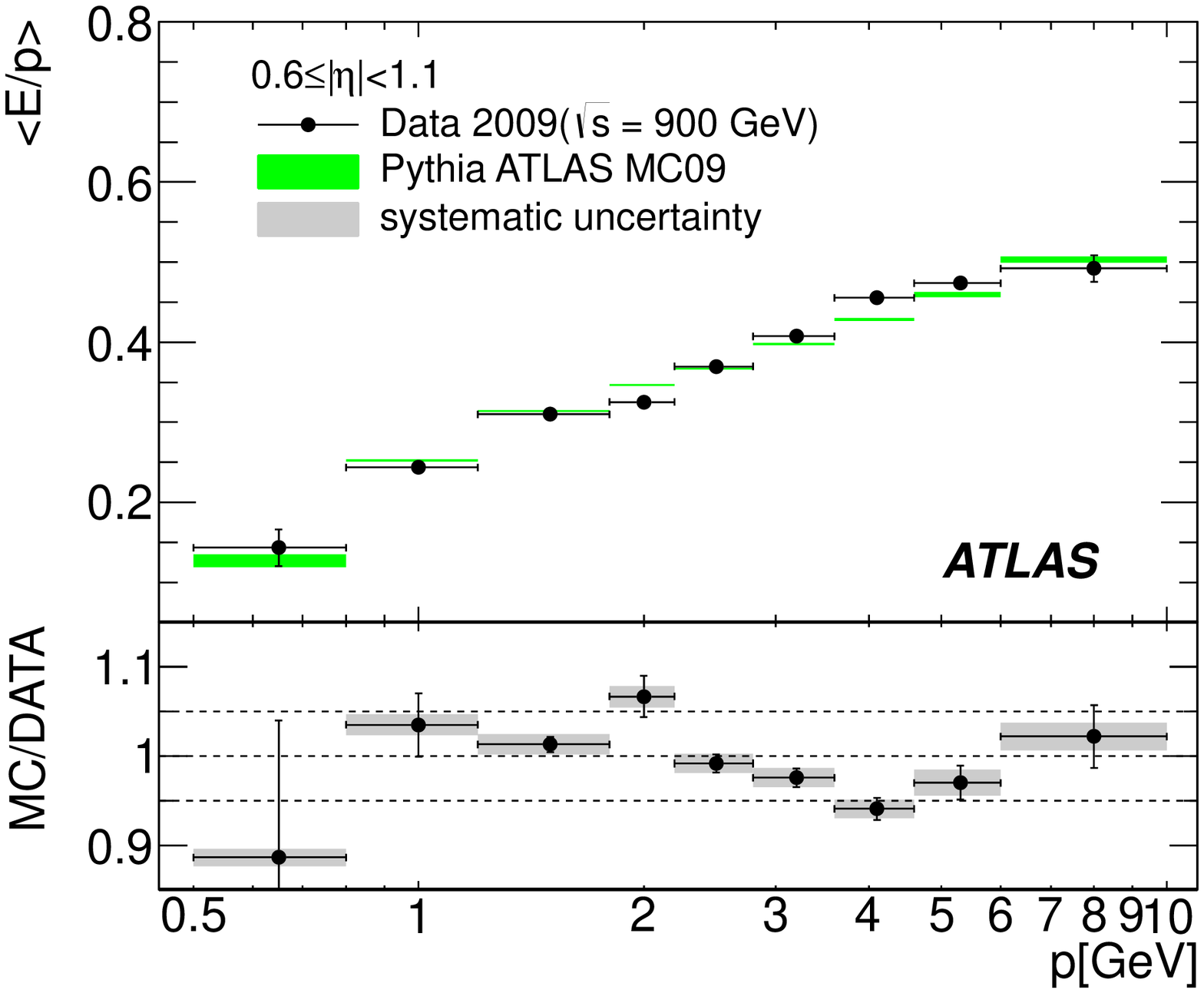}}
\caption{$\langle E/p \rangle$ at $\sqrt{s} = 900$ GeV as a function of the track momentum for \protect\subref{fig:eopSUBBG_etaBin7} $|\eta| < 0.6$ and \protect\subref{fig:eopSUBBG_etaBin6} $0.6 \leq |\eta| < 1.1$. The black markers represent the collision data, while the green rectangles represent the MC prediction, with the vertical width showing its statistical uncertainty. The lower panes show the ratio of the MC simulation prediction to collision data. The grey band indicates the size of the systematic uncertainty on the measurement. The dotted lines are placed at $\pm 5\%$ of unity and at unity.}
\label{fig:result_P_900GeV}
\end{figure*}

\begin{figure*}[htp]
\centering
\subfloat[]{\label{fig:eopSUBBG_etaBin7_7TeV}\includegraphics[width=.48\textwidth]{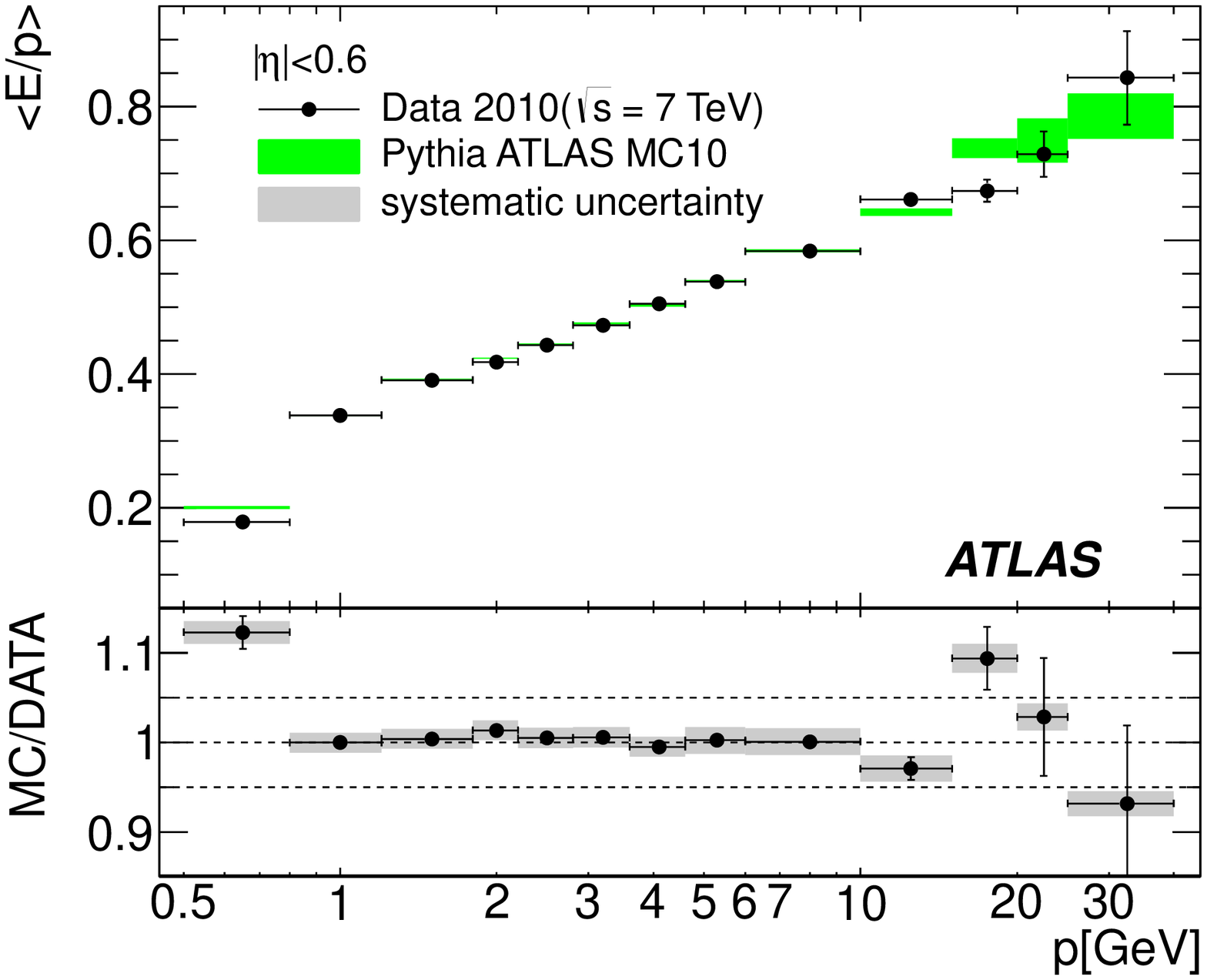}}
\subfloat[]{\label{fig:eopSUBBG_etaBin6_7TeV}\includegraphics[width=.48\textwidth]{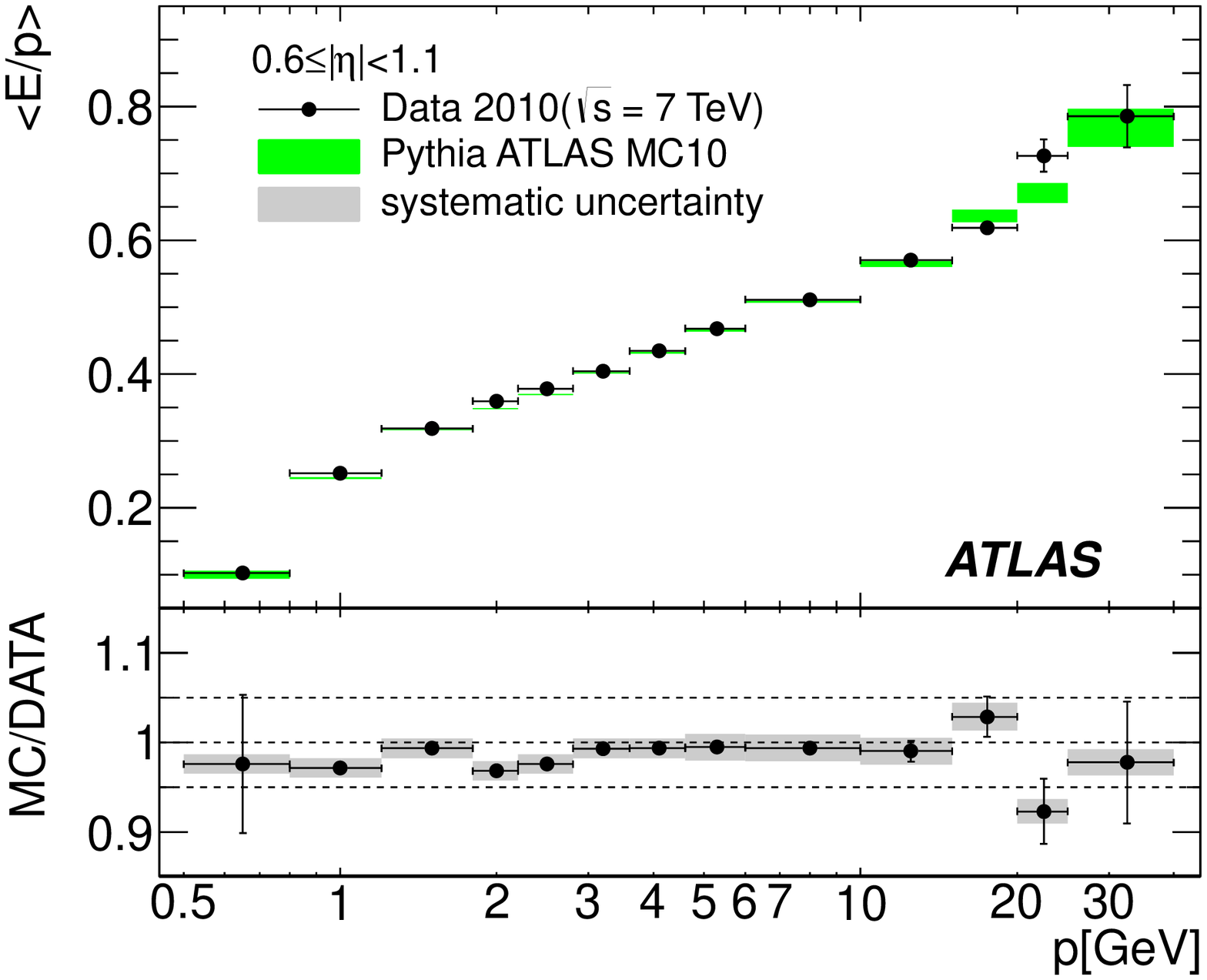}}
\caption{$\langle E/p \rangle$ at $\sqrt{s} = 7$ TeV as a function of the track momentum for \protect\subref{fig:eopSUBBG_etaBin7_7TeV} $|\eta| < 0.6$ and \protect\subref{fig:eopSUBBG_etaBin6_7TeV} $0.6 \leq |\eta| < 1.1$. The black markers represent the collision data, while the green rectangles represent the MC prediction, with the vertical width showing its statistical uncertainty. The lower panes show the ratio of the MC simulation prediction to collision data. The grey band indicates the size of the systematic uncertainty on the measurement. The dotted lines are placed at $\pm 5\%$ of unity and at unity.}
\label{fig:result_P_7TeV}
\end{figure*}

The mean $E/p$ value after background subtraction is evaluated in bins of momentum and pseudorapidity. 
Figures~\ref{fig:result_P_900GeV} and~\ref{fig:result_P_7TeV} show $\langle E/p \rangle$ as a function of the track momentum, 
in two different $|\eta|$ bins up to $|\eta| = 1.1$, at $\sqrt{s} = 900$ GeV and $\sqrt{s} = 7$ TeV.  
The lower parts of the figures present the ratio of MC simulation to data. The data with $\sqrt{s} = 900$ GeV and $\sqrt{s} = 7$ TeV agree within the statistical uncertainty. The maximum momentum that can be probed with the data considered is approximately $30\ \textrm{GeV}$.  The agreement between data and MC simulation is within $\sim$2\% for particles with momenta in the 1--10~GeV range, and it is around 5\% for momenta in the 10--30~GeV range. Below 1~GeV, where tracks are just at the kinematic threshold of entering the calorimeter volume, large differences of $\sim$10\% or more between data and MC simulation are visible. However, due to the low absolute calorimeter response to very low momentum particles, these differences are not critical for the JES determination.

\section{Calorimeter response to identified hadrons}
\label{sec:resonances}

The extrapolation of the previous single particle response studies into the environment of a jet requires understanding of two additional effects.
A jet includes a variety of hadrons that may differ from the inclusive sample of isolated hadrons.
Therefore, measuring the average response to different species of particles is valuable to ensure that the Monte Carlo correctly models all aspects of the jet shower.
Additionally, the hadrons in a jet are not isolated.
Threshold effects and hadronic shower widths affect the calorimeter response to the multi-hadron system.
In order to address these two points, this study adds two additional features to help complete the understanding of calorimeter response.
To minimise the impact of the background and to simplify its estimation, the measurements are presented as a difference between the ratio $E/p$ for two different particle types or as a ratio to the inclusive measurement.

First, single hadrons are identified using decays of $K_S$ (for positive and negative pions), $\Lambda$ (for protons), and $\overline{\Lambda}$ (for anti-protons) particles.
These single hadrons are required to be isolated from all other charged particles in the event.
Single pions will have manifestly lower energy response distributions from those of anti-protons, because of the eventual annihilation of the anti-proton.
By identifying and isolating single pions, single protons, and single anti-protons, the effects of hadronic interactions and annihilation can be separated at low to moderate energies, where they are most important.

Second, when the mother particle is highly boosted, the decay products are more collimated.
For track momenta of $\sim$2--6~GeV from $K_S$ decays, the range of opening angles of the decay products allows a measurement of calorimeter response as more tightly collimated pairs of pions are selected.
There are two related effects probed by such a measurement.
Energy deposited in the calorimeter may fall below the thresholds for reconstruction and be neglected as consistent with noise.
As two showers overlap, the addition of energies that might have individually been below threshold can produce a signal above these noise thresholds.
The increase of the signal is related to both the thresholds themselves and the width of the hadronic shower.
For this study, the $\pi^+\pi^-$ system is required to be isolated from other charged particles in the calorimeter.

\subsection{Event selection and observable definition}

The event selection for this measurement follows closely the event selection of the inclusive measurement already presented, except that a secondary vertex is required in the event.
The events are collected using random triggers at $\sqrt{s} = 7$ TeV and correspond to an integrated luminosity of  about 800~$\mu$b$^{-1}$. 
The same primary vertex requirements are applied, and the same requirements are placed on tracks entering the measurement of $E/p$, except the requirements on the tracks' impact parameters.
The Monte Carlo used is described in Section~\ref{sec:mcsim}.
The definition of the ratio $E/p$ and the isolation requirement for tracks is described in Section~\ref{sec:definitions}.

The ratio $E/p$ is presented as a function of \emph{available energy}, $E_a$.
For pions, this is simply the particle's total energy: $E_a = \sqrt{p^2+m^2}$.
For protons, only the kinetic energy is included: $E_a = \sqrt{p^2+m^2} - m$.
For anti-protons, the kinetic energy plus double the rest-mass is included, in order to take into account annihilation, $E_a = \sqrt{p^2+m^2}+m$.
The available energy is therefore calculated using the information from the  tracker.

\subsection{Reconstruction of short-lived particle candidates}

The reconstruction and selection of long-lived particles is based on previous ATLAS results~\cite{Aad:2011hd}.
The decay $K_S\rightarrow\pi^+\pi^-$, which dominates the $K_S$ decays to other charged particles by several orders of magnitude, is used to identify pions.
The decay $\Lambda\rightarrow\pi^-p$ ($\overline{\Lambda}\rightarrow\pi^+\overline{p}$), which dominates the $\Lambda$ ($\overline{\Lambda}$) decays to other charged particles by several orders of magnitude, is used to identify protons (anti-protons).
Both decay product tracks are required to have $\pt >100$~MeV, and the tracks used for the $E/p$ measurement must have $\pt >500$~MeV.
In the case of $\Lambda$ ($\overline{\Lambda}$) candidate decays, the charge of the higher momentum track is used to label the candidate as a $\Lambda$ (positively-charged higher momentum track) or $\overline{\Lambda}$ (negatively-charged higher momentum track), as is kinematically favored. 
The tracks entering the $E/p$ distributions are additionally required to have a pseudorapidity-dependent number of hits in the TRT, the outermost tracking system.
This requirement suppresses the effect of nuclear interactions in the material of the inner detector, particularly in the outer layers of the silicon tracker.

\begin{figure}[tb]
\begin{center}
\subfloat[]{\label{fig:LambdaMass}\includegraphics[width=0.48\textwidth]{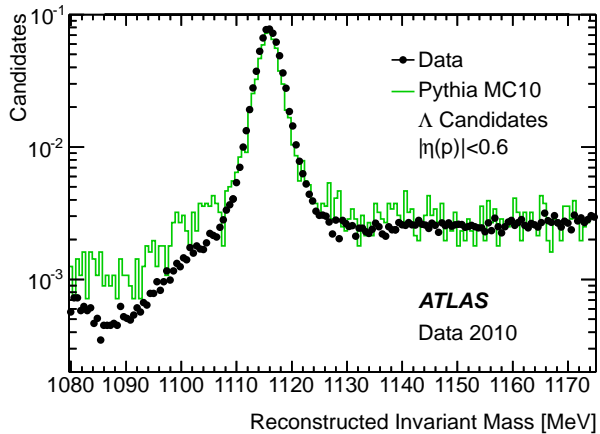}}\\
\subfloat[]{\label{fig:LambdaMassFit}\includegraphics[width=0.48\textwidth]{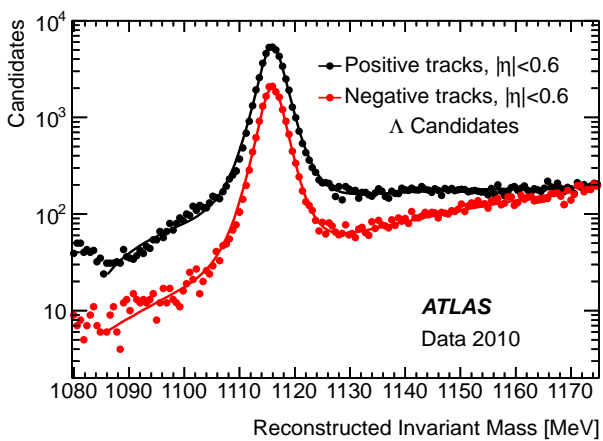}}
\end{center}
\caption{
\protect\subref{fig:LambdaMass} The reconstructed mass peak of $\Lambda$ candidates with central ($|\eta|<0.6$) proton-candidate tracks in data (black points) and MC simulation (green histogram).
\protect\subref{fig:LambdaMassFit} The reconstructed mass peak of $\Lambda$ candidates in data.
The distribution is shown separately for positive (i.e. proton candiate) and negative (i.e. pion candidate) tracks in the central region ($|\eta|<0.6$).
\label{fig:recoMass}}
\end{figure}

The individual tracks entering the response distributions are divided into two bins of pseudorapidity, $|\eta|<0.6$ and $0.6\leq|\eta|<1.1$.
Several higher pseudorapidity bins show consistent results, albeit with significantly lower statistics. 
Example distributions of reconstructed mass for candidates with at least one central ($|\eta|<0.6$) track are shown in \figurename~\ref{fig:recoMass}.
The mass peaks stand out clearly over the background.
The composition of the background will be discussed further in Section~\ref{sec:resonancebg}.

The signal purity for a given pseudorapidity interval, where the binning is always in terms of the kinematic properties of the track of interest, is estimated using a fit to the signal over a cubic polynomial fit to the background, following Ref.~\cite{Aad:2011hd}.
For the $K_S$, a double Gaussian signal function is used, constrained to have identical central values for both Gaussians, and for the $\Lambda$ and $\overline{\Lambda}$, a modified Gaussian is used of the form

\begin{equation}
y = a \times \exp\left[ -\frac{1}{2} \times x^{1+\frac{1}{1+x/2} } \right] ,
\end{equation}

\noindent where $x \equiv \left| \frac{ m - b } {c} \right|$, $m$ is the reconstructed mass, and $a$, $b$ and $c$ are free parameters of the fit.
The peak mean, $b$, is stable to within a few hundred keV over all pseudorapidity bins.
The fitted width of the mass peak, $c$, increases with track pseudorapidity, due to the track resolution.
Example fit results for the $\Lambda$ are shown in \figurename~\ref{fig:recoMass} for positively charged tracks (proton candidates) and negatively charged tracks (pion candidates) in the central pseudorapidity bin for the $E/p$ measurement.
The negatively charged tracks are softer, which is reflected in the smaller statistics for the negatively charged track mass distribution.

For each mass peak and each bin of track pseudorapidity, an acceptance window is constructed in order to optimise both signal purity and statistics.
The width of the window is set at three times the width of the narrower Gaussian for the $K_S$ and three times the width of the modified Gaussian for the $\Lambda$ ($\overline{\Lambda}$).
The purities are over 97\% for the $K_S$ candidates and above 92\% for the central $\Lambda$ candidates.
The purities in the MC simulation are within 2\% of those in the data.
The number of tracks from signal candidates, extracted from the fits, in the data and 20 million MC simulation events are shown in Table~\ref{table:stats}.
Roughly twice as many $K_S$ candidates and three to four times more $\Lambda$ and $\overline{\Lambda}$ candidates are found in the data than in the MC simulation, because of the limited MC simulation statistics.
The difference in $K_S$ and $\Lambda$ yields between data and MC simulation are discussed further in Ref.~\cite{Aad:2011hd}, and the response measurements here are insensitive to these differences.
Only the positively and negatively charged tracks that form short-lived particle candidates within the defined acceptance window are considered in the remainder of this section.

\begin{table}[htb]
\begin{center}
\begin{tabular}{l|cc|cc}
\hline\noalign{\smallskip}
                                         & \multicolumn{2}{c|}{$|\eta|<0.6$} & \multicolumn{2}{c}{$0.6\leq|\eta|<1.1$} \\
Particle                                 & Data                & MC         & Data          &      MC              \\
\noalign{\smallskip}\hline\noalign{\smallskip}
$\pi^+$ from $K_S$                       & 4.04$\times10^5$  & 2.04$\times10^5$  & 2.52$\times10^5$   & 1.28$\times10^5$  \\
$\pi^-$ from $K_S$                       & 3.94$\times10^5$  & 1.98$\times10^5$  & 2.49$\times10^5$   & 1.26$\times10^5$  \\
$p$ from $\Lambda$                       & 2.63$\times10^4$  & 8.03$\times10^3$  & 1.81$\times10^4$   & 5.62$\times10^3$  \\
$\overline{p}$ from $\overline{\Lambda}$ & 2.64$\times10^4$  & 7.21$\times10^3$  & 1.53$\times10^4$   & 4.10$\times10^3$  \\
\noalign{\smallskip}\hline
\end{tabular}
\caption{
The number of signal candidate tracks, extracted from the fits described in the text, in data and MC simulation.
\label{table:stats}}
\end{center}
\end{table}

\subsection{Background subtraction}
\label{sec:resonancebg}

There are three sources of charged backgrounds entering the $E/p$ distributions.
Nuclear interactions in the material of the inner detector that fake short-lived particle candidates are suppressed by the narrow mass acceptance window.
The smoothness of the distribution of secondary vertex positions confirms that these are at the level of a few percent.
The tracks may also undergo nuclear interactions prior to entering the calorimeter.
These interactions are suppressed by requiring that the track have hits in the TRT.
An additional charged background comes from combinatorics, particularly in high track-multiplicity events.
Because the purity of $K_S$ candidates is typically 5--10\% higher than that for $\Lambda$ or $\overline{\Lambda}$ candidates, and because the majority of the background is from charged pions, charged background corrections are applied only for calculating proton and anti-proton response.
The response to pions is taken from the $K_S$ candidates and is used to correct the response to protons and anti-protons.

When charged pions enter the proton energy response distribution, they are given an incorrect track mass hypothesis.
The track extrapolation to the calorimeter does not calculate energy loss differently for pions and (anti-)protons, so that the measured associated energy in the calorimeter remains the same.
However, the available energy changes significantly when the different mass hypothesis is used.
Thus, in order to subtract the charged pion background from the (anti-)proton distribution, the pions from $K_S$ candidates are given the (anti-)proton mass hypothesis and their response is re-calculated.
The (anti-)proton response in a bin $i$ of available energy and $\eta$, $\langle E/p \rangle_i$, is then given by

\begin{equation}
\langle E/p \rangle_i = \frac{1}{\epsilon_i} \left( \langle E/p \rangle_i^\mathrm{raw} - \left(1-\epsilon_i\right) \langle E/p\rangle_i^{\pi} \right) ,
\end{equation}

\noindent where $\epsilon_i$ is the purity of the sample in that pseudorapidity bin, $\langle E/p\rangle_i^\mathrm{raw}$ is the measured response to the (anti-)protons, and $ \langle E/p \rangle_i^{\pi}$ is the measured response to the charged pions with a proton mass hypothesis.

The corrections are small in the region $|\eta|<1.1$, falling from $\sim$5\% at low available energy to $\sim$1\% at high available energy for both protons and anti-protons.
The corrections are derived independently for data and MC simulation to ensure that the differences in purity and any differences in response are taken into account.

As the momenta of the daughter tracks increase, the position resolution of the secondary vertex broadens.
This introduces a larger combinatorial background in the high-momentum bins.
The purities in MC simulation follow, to $\sim$5\%, those of the data with increasing momentum.
The systematic uncertainty on the corrected value of $E/p$ introduced by the $\pt$-dependence of the purity is determined to be well below 1\% and is therefore neglected.

There is an additional contribution to the (anti-) proton response measurement for $\Lambda$ ($\overline{\Lambda}$) decays faking $\overline{\Lambda}$ ($\Lambda$) decays.
Particularly when the $\Lambda$ or $\overline{\Lambda}$ has low energy, the pion may be the higher-momentum track.
These ``fakes'' are suppressed by the kinematic cuts applied in the candidate selection, but are still present at some level, particularly in the lowest bin of available energy.
This background is dominated by well-understood two-body decay kinematics.
It should be well-described by the MC simulation and is therefore taken into account using MC predictions.

As discussed in Section~\ref{sec:background}, there is an additional contribution to $E/p$ from neutral particles in the event, since isolation is only defined relative to charged particles.
Only differences in response between particle species are reported here, because the neutral background should cancel in the difference (except insofar as the $K_S$ or $\Lambda$ production occurs near additional activity).
The charged particle isolation criterion should be sufficient to ensure that the neutral background is uncorrelated with any jet-like activity in the event.
The cancellation of the neutral background is tested using a simulation sample of single particles and is found to be valid up to the available statistical accuracy.
Therefore, no additional correction or uncertainty is added for this background.
Thus, background systematic uncertainties are found to be negligible with respect to the statistical errors of the MC simulation sample.

\subsection{Isolated identified single particle response}
\label{sec:ResonantSinglePart}

\begin{figure*}[tb]
\begin{center}
\subfloat[]{\label{fig:RawEoverP}\includegraphics[width=0.48\textwidth]{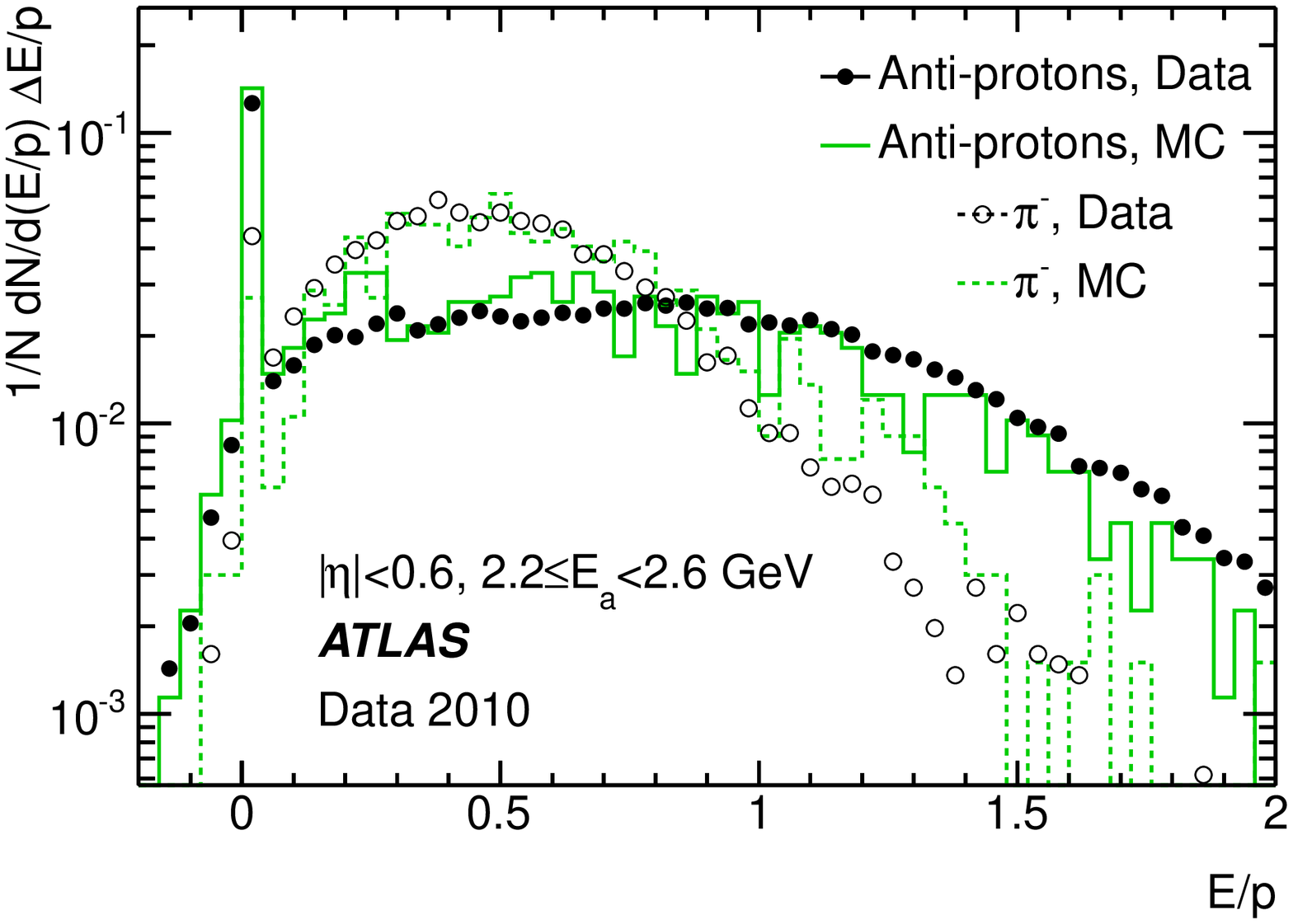}}
\subfloat[]{\label{fig:RawEoverP_PPi}\includegraphics[width=0.48\textwidth]{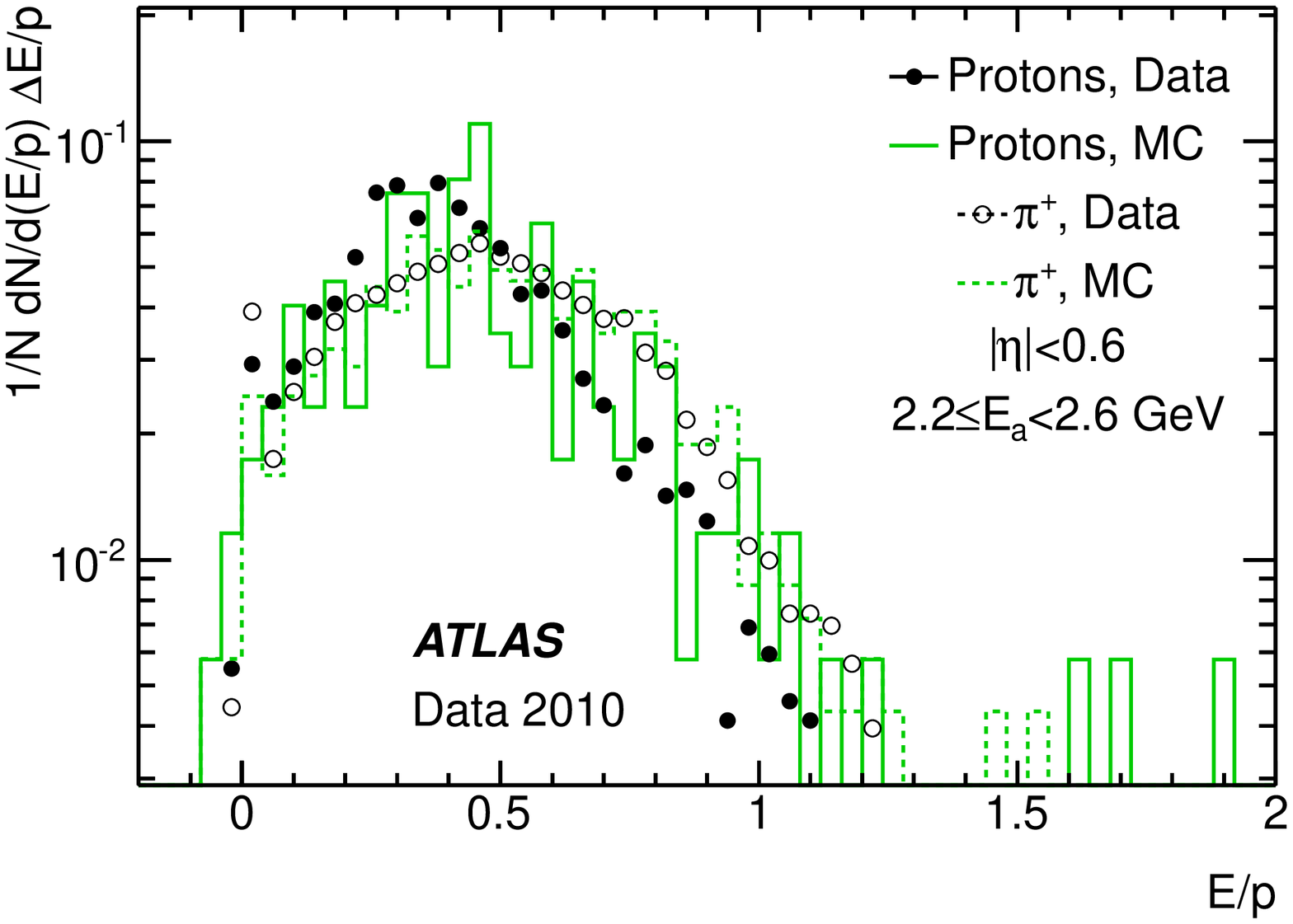}}
\end{center}
\caption{
The uncorrected $E/p$ distribution for  \protect\subref{fig:RawEoverP} single $\pi^{-}$ and $\overline{p}$ and \protect\subref{fig:RawEoverP_PPi} $\pi^{+}$ and $p$ in a single bin of available energy and pseudorapidity, $2.2\leq E_a<2.8$~GeV and $|\eta|<0.6$.
\label{fig:EopRaw}}
\end{figure*}

The uncorrected distribution of $E/p$ for $\pi^{-}$, $\overline{p}$, $\pi^{+}$ and $p$ in a single bin of available energy and pseudorapidity, $2.2
\leq E_a<2.8$~GeV and $|\eta|<0.6$, is shown in \figurename~\ref{fig:EopRaw} (cf. \figurename~\ref{fig:E_P} for inclusive hadrons in a different $p$ range).
All the distributions have a small negative tail from noise in the calorimeter and a long positive tail from the neutral background.
There is a much more prominent positive tail in the $\overline{p}$ response distribution, where annihilation plays a significant role.
A significant fraction of tracks have $E=0$, for which no cluster of energy was found near the track in the calorimeter, as discussed in Section~\ref{sec:distributionsEoP}.
The average response, $\langle E/p \rangle$, is defined as the arithmetic mean of the full distribution.
The response of protons and positively charged pions is well modeled in the MC simulation.
The response to pions is significantly lower than that to anti-protons in this available energy range, although fewer pions have $E=0$.
The agreement in the fraction of pions with $E=0$ builds confidence in the modelling of the material in front of the calorimeter.

\begin{figure*}[tb]
\begin{center}
\subfloat[]{\label{fig:PiPPiMDiff}\includegraphics[width=0.48\textwidth]{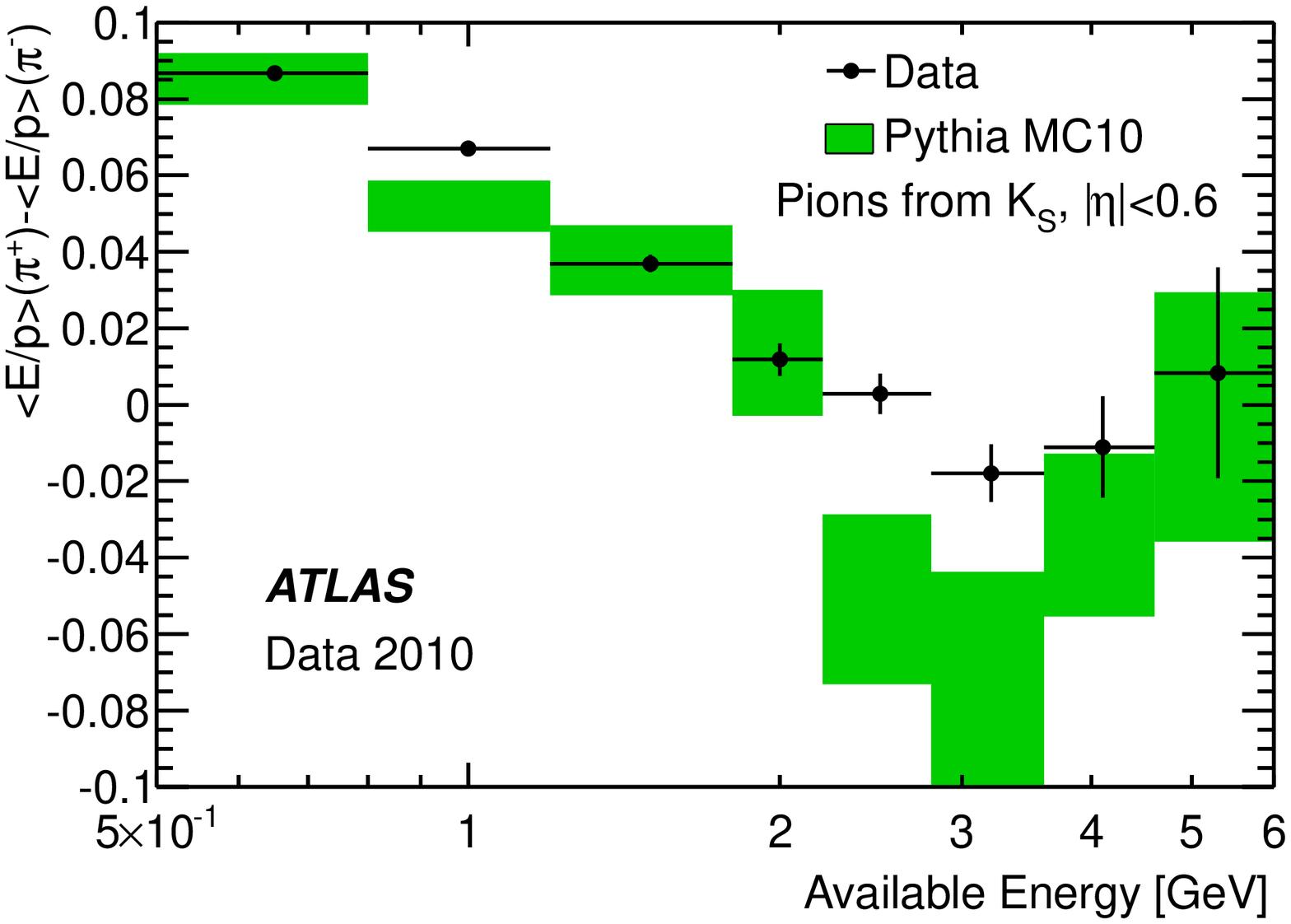}}
\subfloat[]{\label{fig:PiPPiMDiffFwd}\includegraphics[width=0.48\textwidth]{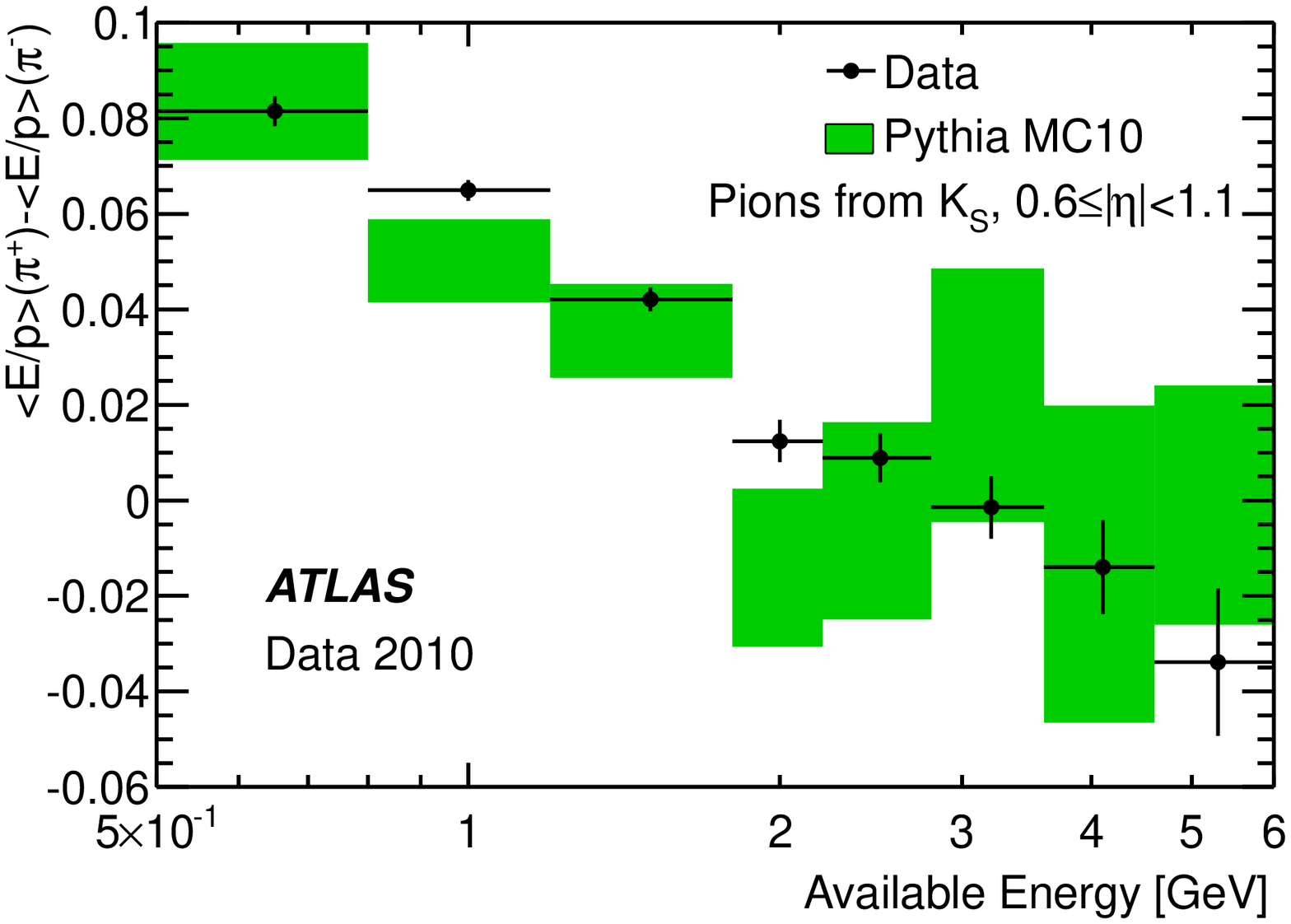}}
\end{center}
\caption{
The difference in $\langle E/p \rangle$ between $\pi^+$ and $\pi^-$ from $K_S$ candidates in tracks with \protect\subref{fig:PiPPiMDiff} $|\eta|<0.6$ and \protect\subref{fig:PiPPiMDiffFwd} $0.6\leq|\eta|<1.1$.
\label{fig:RespDiff}}
\end{figure*}

In order to reduce the effects of neutral background subtraction and focus on the differences in response amongst particle species, the difference in response between several pairs of particles is measured.
Figure~\ref{fig:RespDiff} shows the difference in response between $\pi^+$ and $\pi^-$ in the central ($|\eta|<0.6$) and forward ($0.6\leq|\eta|<1.1$) pseudorapidity bins.
The response to $\pi^+$ is higher than that to $\pi^-$ at low available energy, in agreement with Ref.~\cite{CMSEoP}, wherein this difference is attributed to a material dependent charge-exchange effect.
The difference may also be related to the ``Barkas correction'' described in Ref.~\cite{PDG}.
The MC simulation is, however, reasonably consistent with the data.

\begin{figure*}[tb]
\begin{center}
\subfloat[]{\label{fig:PiPProtonDiff}\includegraphics[width=0.48\textwidth]{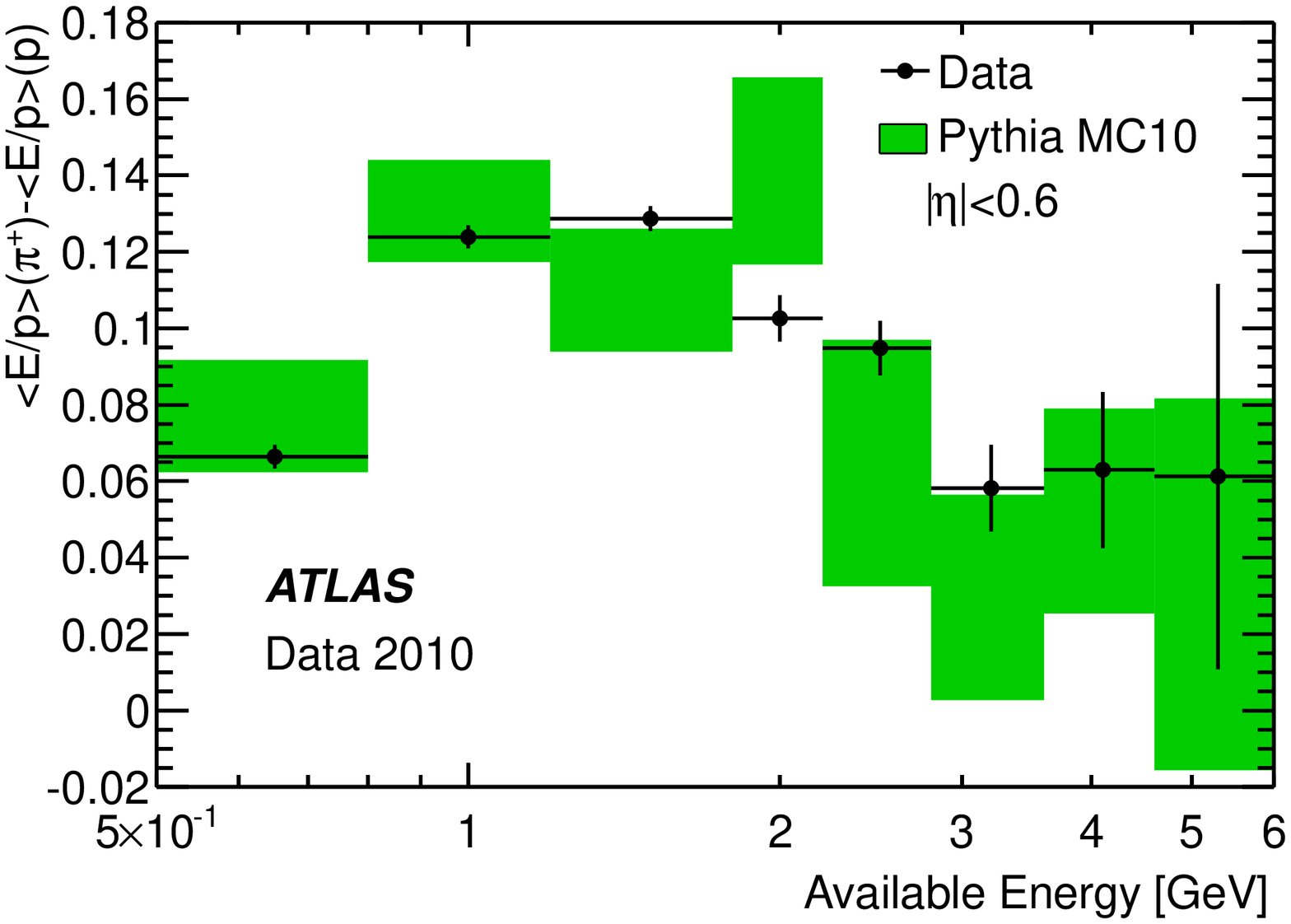}}
\subfloat[]{\label{fig:PiPProtonDiffFwd}\includegraphics[width=0.48\textwidth]{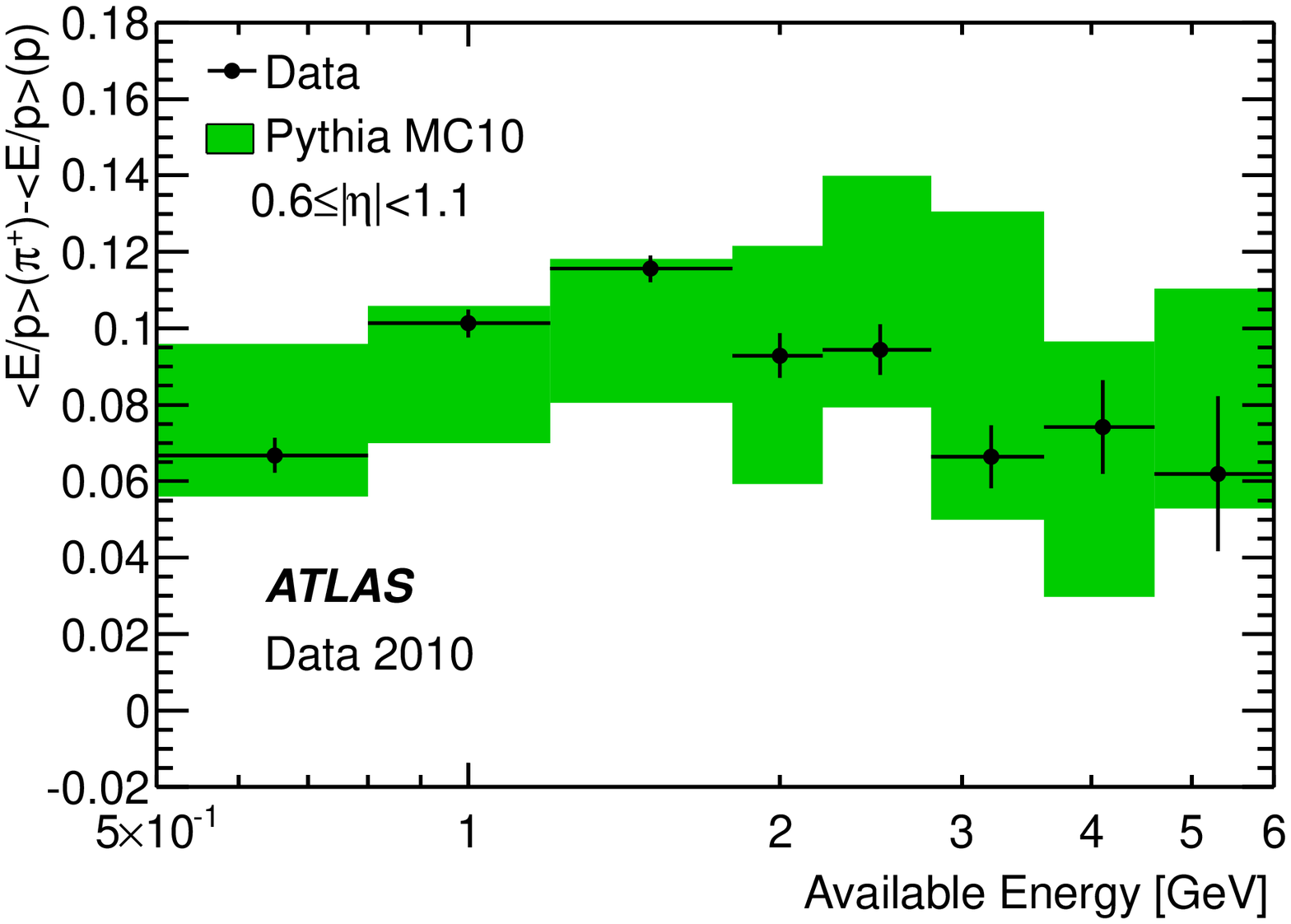}}
\end{center}
\caption{
The difference in $\langle E/p \rangle$ between $\pi^+$ from $K_S$ candidates and $p$ from $\Lambda$ candidates in tracks with \protect\subref{fig:PiPProtonDiff} $|\eta|<0.6$  and \protect\subref{fig:PiPProtonDiffFwd} $0.6\leq|\eta|<1.1$.
The responses to the protons are corrected for charged particle backgrounds (see text).
\label{fig:RespDiff2}}
\end{figure*}

Figure~\ref{fig:RespDiff2} shows the difference in response between $\pi^+$ and $p$ in the central ($|\eta|<0.6$) and forward ($0.6\leq|\eta|<1.1$) pseudorapidity bins.
Again, the difference between $\pi^+$ and proton response in MC simulation is consistent with the data over the entire range of available energies.
At low available energy, a larger average fraction of the initial hadron energy is converted into an electromagnetic shower for pions than for protons~\cite{PDG}.
This leads to a lower response for protons, particularly at low available energy, due to the non-compensation of the calorimeter.

\begin{figure*}[tb]
\begin{center}
\subfloat[]{\label{fig:PiMAProtonDiff}\includegraphics[width=0.48\textwidth]{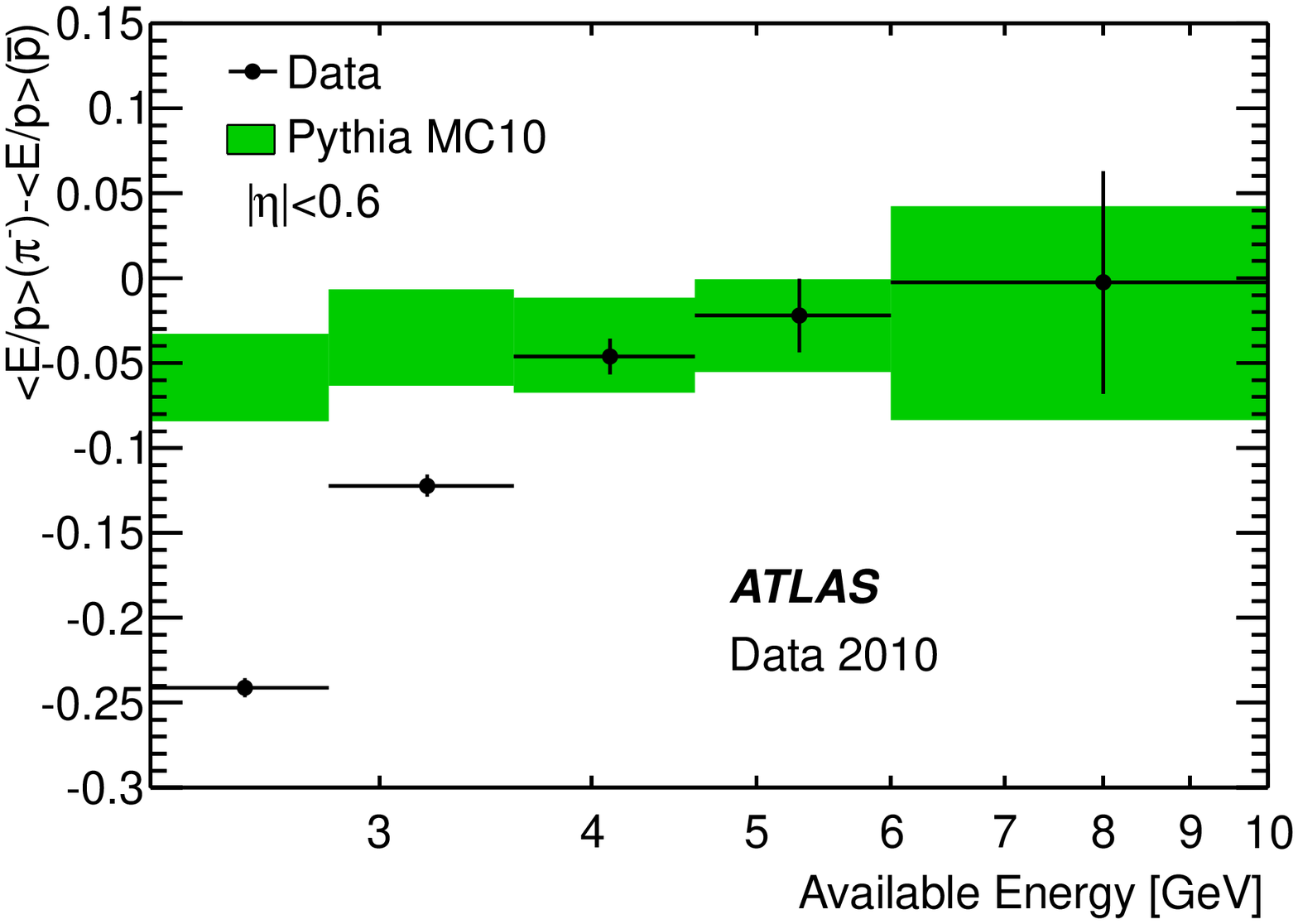}}
\subfloat[]{\label{fig:PiMAProtonDiffFwd}\includegraphics[width=0.48\textwidth]{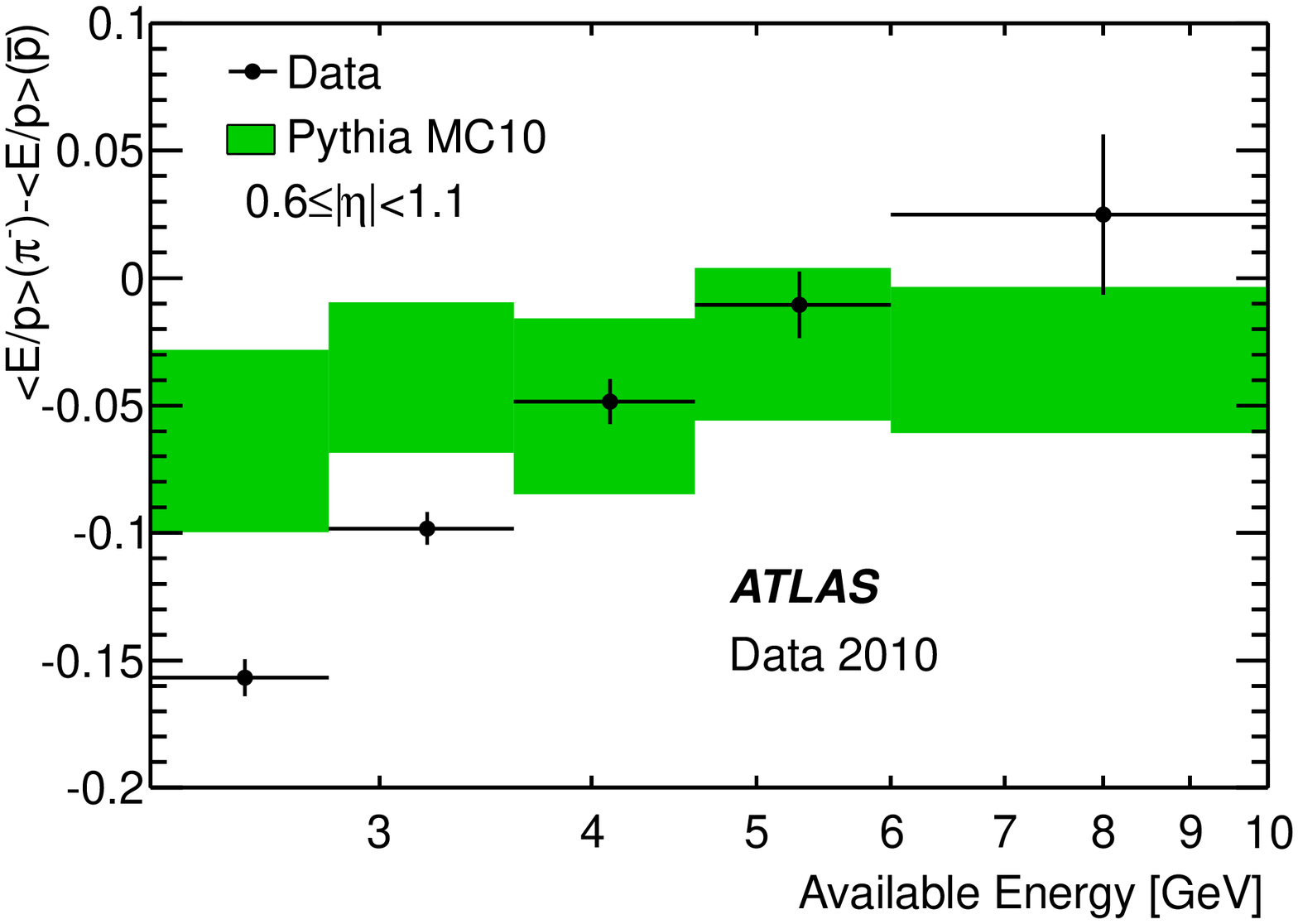}}
\end{center}
\caption{
The difference in $ \langle E/p \rangle$ between $\pi^-$ from $K_S$ candidates and $\overline{p}$ from $\overline{\Lambda}$ candidates in tracks with \protect\subref{fig:PiMAProtonDiff} $|\eta|<0.6$ and \protect\subref{fig:PiMAProtonDiffFwd} $0.6\leq|\eta|<1.1$.
The responses to the anti-protons are corrected for charged particle backgrounds (see text).
\label{fig:RespDiff3}}
\end{figure*}

Figure~\ref{fig:RespDiff3} shows the difference in response between $\pi^-$ and $\overline{p}$ in the central ($|\eta|<0.6$) and forward ($0.6\leq|\eta|<1.1$) pseudorapidity bins.
The difference shows a $\sim$30\%$\times \langle E/p \rangle$ disagreement between data and MC simulation for $E_a<3$~GeV, though they are consistent for $E_a>4$~GeV.
At these low available energies, the anti-proton response should be dominated by the annihilation and the subsequent shower.
The difference indicates large contributions from processes not well-modelled by {\sc Geant4}.
To test the contribution from the calorimeter acceptance to this effect, the response is constructed excluding the tracks with $E=0$.
The pions and protons show the same level of agreement between data and MC simulation, and the disagreement in the response difference between pions and anti-protons remains near 10\%.

\subsection{Response to nearby particles}
\label{sec:nearbyParticles}

To study the effects of calorimeter thresholds and noise suppression as more tightly collimated pairs of particles are selected, isolation is not required with respect to the other daughter of the particle candidate.
That is, when constructing the energy response for the $\pi^+$ from a $K_S$ decay, the $\pi^+$ is required to be isolated from all tracks in the event other than the $\pi^-$ from the $K_S$ decay.
The response is then examined as a function of the distance between the two pions in $\eta-\phi$ after extrapolation to the second layer of the electromagnetic calorimeter.
No isolation criteria are applied to the $\pi^-$.
Because the pions are required to come from a particle with a narrow width, the kinematics of the second pion are constrained, thus removing one possible source of discrepancy between data and MC simulation.

\begin{figure*}[tb]
\begin{center}
\subfloat[]{\label{fig:OpenPEta0Pt0}\includegraphics[width=0.48\textwidth]{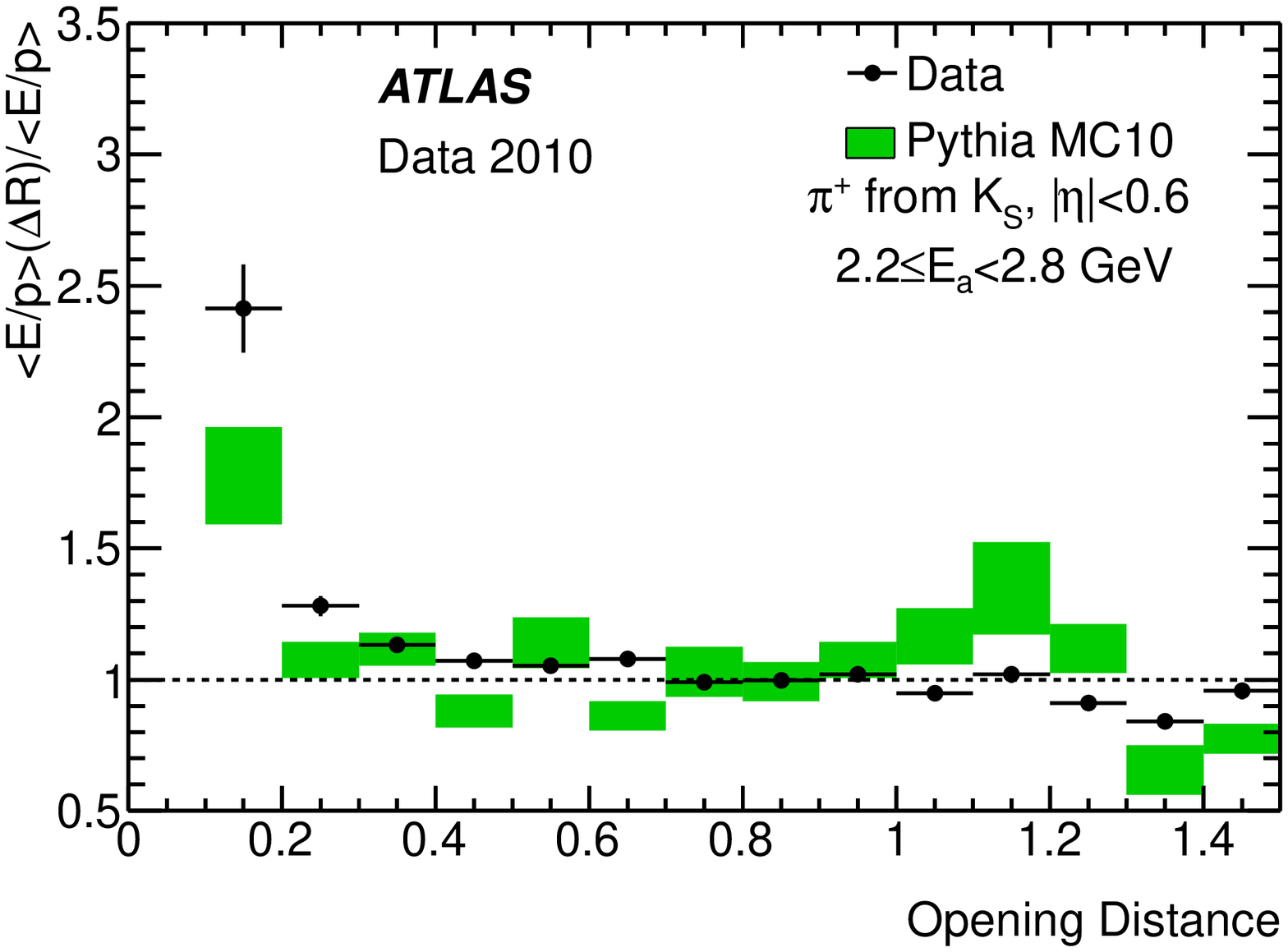}}
\subfloat[]{\label{fig:OpenMEta0Pt0}\includegraphics[width=0.48\textwidth]{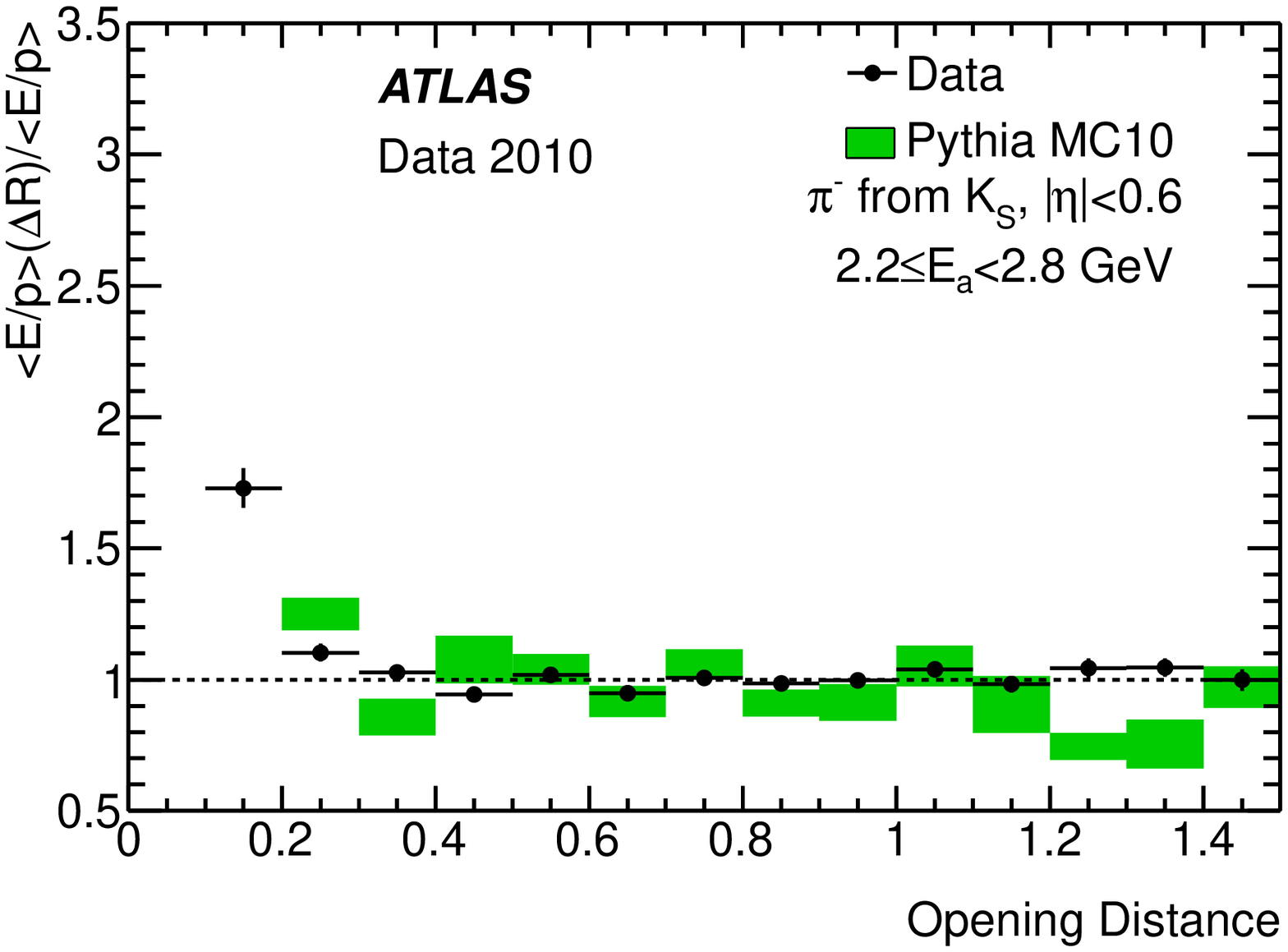}}
\end{center}
\caption{
$\langle E/p \rangle$ for pions as a function of the extrapolated distance between that pion and a pion of the opposite sign.
In all cases the pions are daughters of a reconstructed $K_S$ candidate and are required to be isolated from all tracks in the event except the other daughter of the decay.
The response is shown for low energy ($2.2\leq E_a<2.8$~GeV) \protect\subref{fig:OpenPEta0Pt0} $\pi^{+}$ and \protect\subref{fig:OpenMEta0Pt0} $\pi^{-}$ in the central pseudorapidity bin ($|\eta|<0.6$).
The MC simulation has no tracks passing the selection in the smallest bin (smallest two bins) of opening distance for $\pi^+$ ($\pi^-$).
\label{fig:RespOpen1}}
\end{figure*}

\begin{figure*}[tb]
\begin{center}
\subfloat[]{\label{fig:OpenMEta1Pt0}\includegraphics[width=0.48\textwidth]{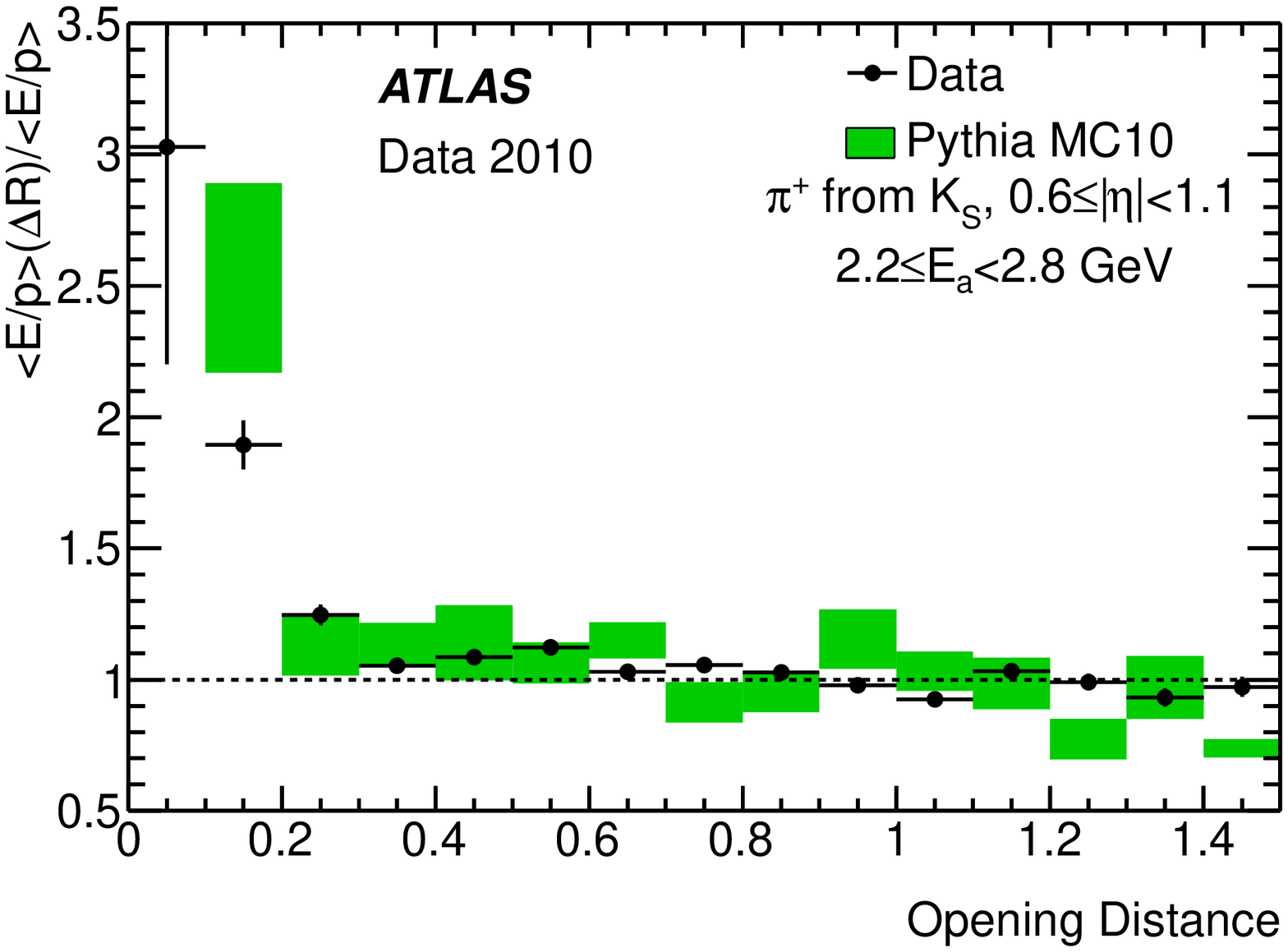}}
\subfloat[]{\label{fig:OpenMEta0Pt1}\includegraphics[width=0.48\textwidth]{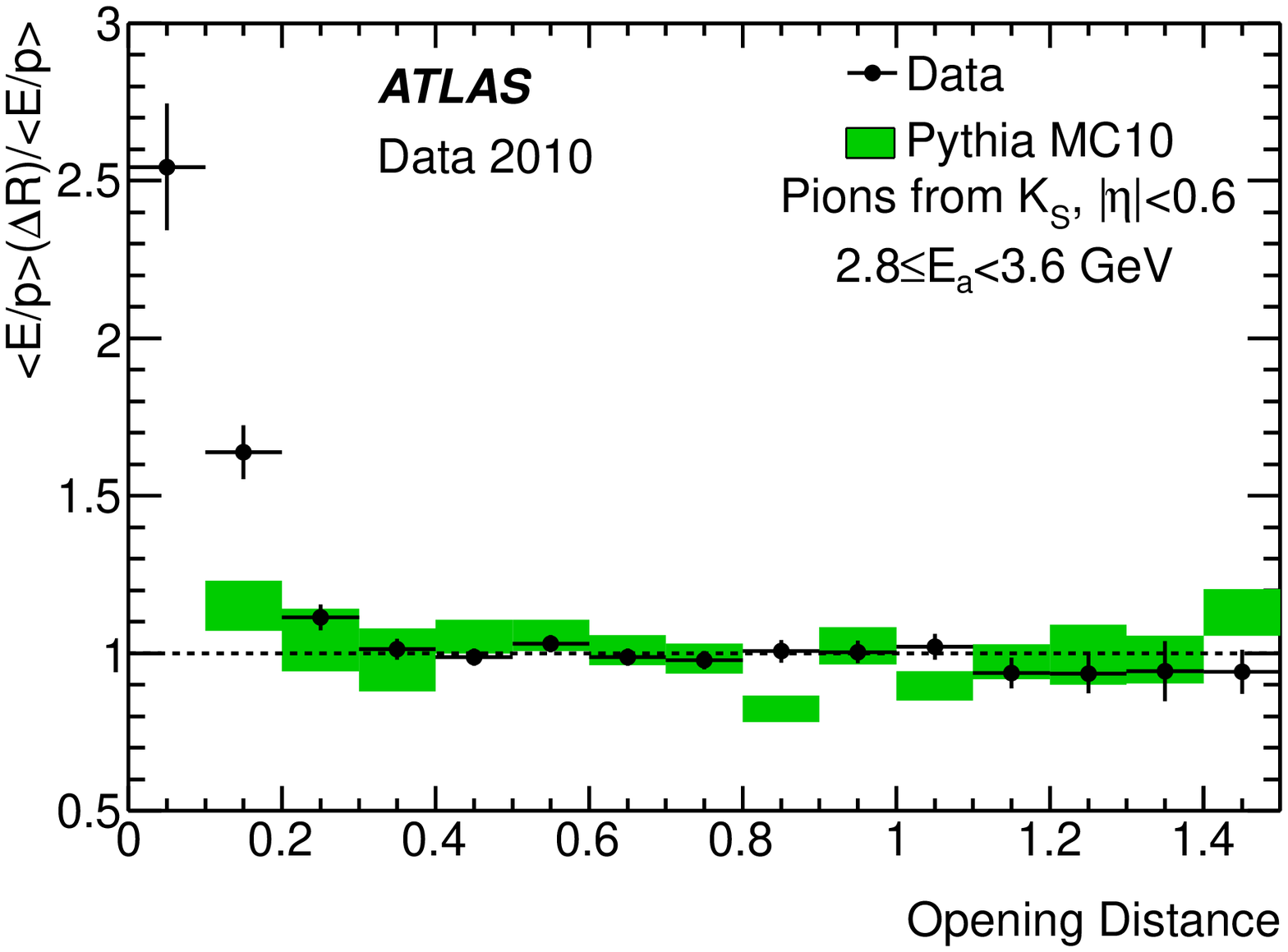}}

\subfloat[]{\label{fig:OpenMEta0Pt2}\includegraphics[width=0.48\textwidth]{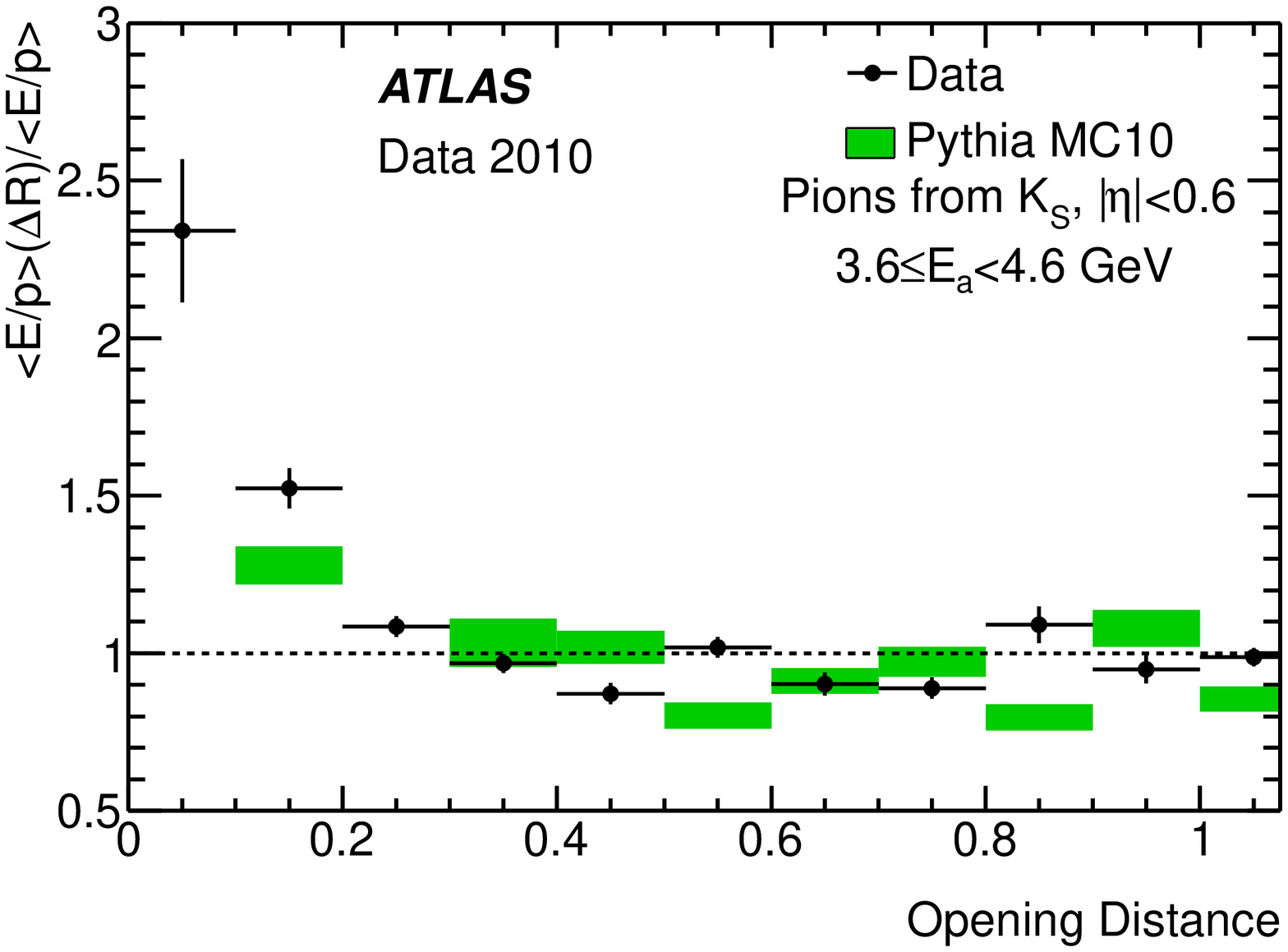}}
\subfloat[]{\label{fig:OpenMEta0Pt3}\includegraphics[width=0.48\textwidth]{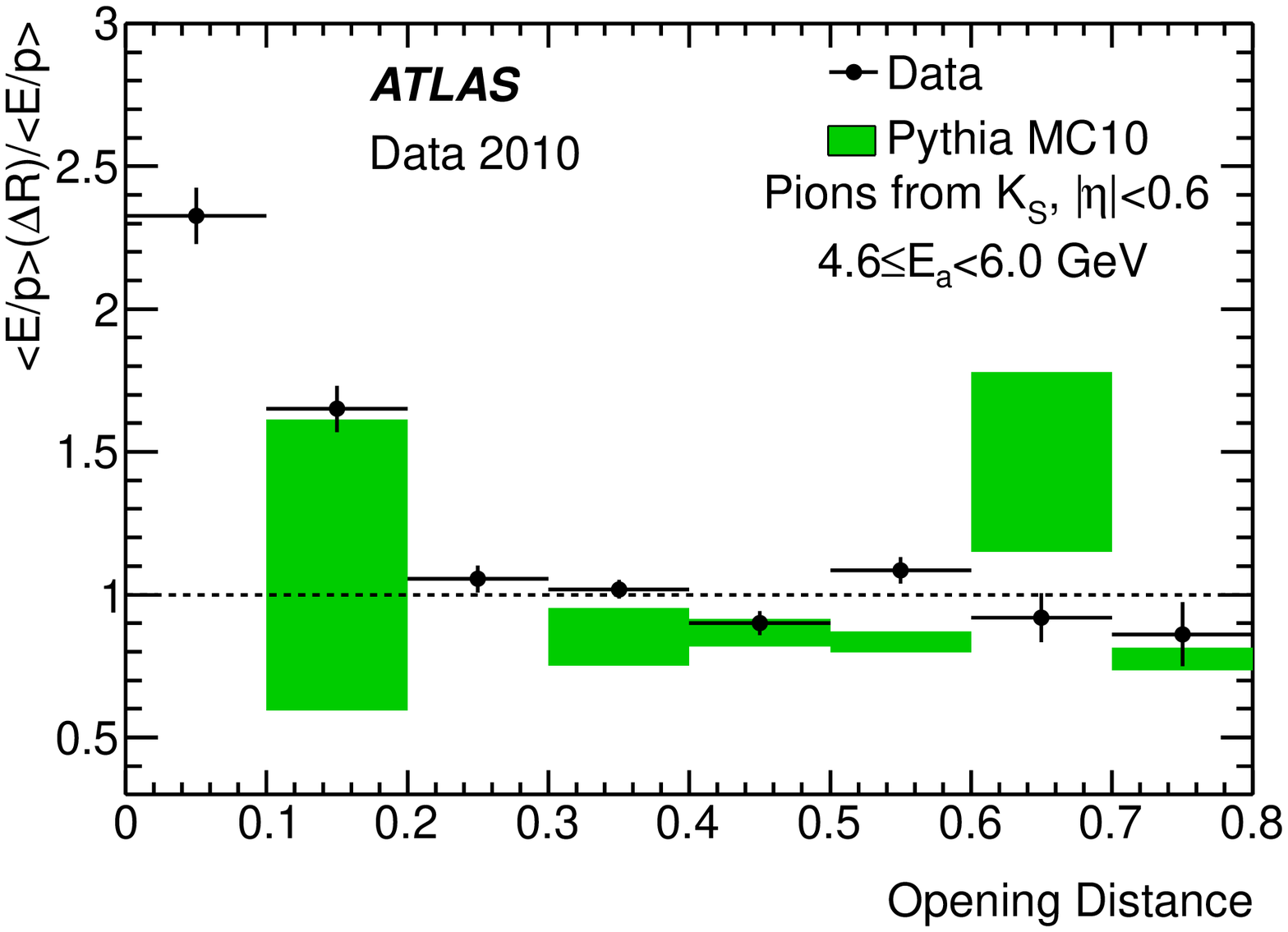}}
\end{center}
\caption{
$\langle E/p \rangle$ for pions as a function of the extrapolated distance between that pion and a pion of the opposite sign.
In all cases the pions are daughters of a reconstructed $K_S$ candidate and are required to be isolated from all tracks in the event except the other daughter of the decay.
The response is shown for \protect\subref{fig:OpenMEta1Pt0} more forward ($0.6\leq|\eta|<1.1$) pions with $2.2\leq E_a<2.8$~GeV and for central ($|\eta|<0.6$) pions in three bins of available energy: \protect\subref{fig:OpenMEta0Pt1} $2.8\leq E_a<3.6$~GeV, \protect\subref{fig:OpenMEta0Pt2} $3.6\leq E_a<4.6$~GeV and \protect\subref{fig:OpenMEta0Pt3} $4.6\leq E_a<6.0$~GeV.
The MC simulation has no tracks passing the selection in the smallest opening distance bin.
\label{fig:RespOpen3}}
\end{figure*}

The ratio $E/p$ as calculated here for a single pion should increase as another particle approaches it and more of the second particle's energy is included in the calorimeter energy, $E$.
As shown in \figurename~\ref{fig:RespOpen1} and~\ref{fig:RespOpen3}, the response does not vary with separation for large extrapolated distances, where the single pion is isolated.
The distributions are normalised to the average $E/p$ in the same $\eta$ and $p$ bin in order to remove any differences from the average response.
In \figurename~\ref{fig:RespOpen3}, pions of both charges are considered in order to increase the available statistics.
The response rises below $\Delta R\approx0.3$, which confirms that the isolation criterion used in Sections~\ref{sec:event_selection} and \ref{sec:ResonantSinglePart} does not lead to a bias.
It increases significantly for $\Delta R<0.2$, when the second pion is inside the cone in which the energy is counted when constructing $E/p$.
The height of the peak at low $\Delta R$ is related to the kinematics of the $K_S$ decay.
In particular, some of the low $\Delta R$ and low available energy bins contain mostly asymmetric decays.

The height of the response distribution for close-by pions and small opening angles is connected to the threshold effects, since it is a measurement of the out-of-cluster energy, to first order.
The slope of the response curve is closely related to the single-hadron shower width.
The agreement between data and MC is good over most of the range of opening angles, but the peak at low opening angles is wider in data, indicating a somewhat broader shower.
This discrepancy is qualitatively in agreement with the discrepancies observed in lateral hadronic shower shape,  seen in test-beam studies and elsewhere~\cite{geant4Physics,CTB,MCCTB1,MCCTB2}.

\section{Calorimeter jet energy scale uncertainty}
\label{sec:method}

The jet energy scale calibration (JES) corrects the measured jet energy for several effects, including calorimeter non-compensation and energy loss in dead material.
The calibration itself is derived from MC simulation.
The calorimeter uncertainty on the JES is calculated from the uncertainty on the energy response of all particles contributing to a jet. Within the MC simulation, the energy contribution of each individual particle to a given jet can be separated. The convolution of the uncertainty on the single particle energy response with the MC jet particle composition is then used to calculate the calorimeter uncertainty on the jet energy scale. 

The calorimeter JES uncertainty is derived for the well-understood central region of the calorimeter. The main reasons for the restriction to the central calorimeter region are the smaller amount of material in front of the calorimeter and the existence of combined test beam measurements with a setup very similar to the final ATLAS configuration. Therefore, the calorimeter JES uncertainty is only evaluated using the single particle response for $|\eta| < 0.8$. The total JES uncertainty, including all calorimeter regions and all effects not related to the calorimeter energy response, is discussed in Ref.~\cite{JES_Summary_r16}.

The JES uncertainty is determined for jets reconstructed from topological clusters~\cite{topoclusters} with the \antiKT jet algorithm~\cite{Cacciari:2008gp} implemented in the FASTJET package \cite{Cacciari:2005hq} for $R=0.4$ and $R=0.6$ jet sizes. 

The analysis is performed with inclusive di-jet Monte Carlo events simulated with {\sc Pythia}. Jets are selected requiring a separation of $\Delta R > 2.0$ to any other jet with EM scale\footnote{The jet energy at the EM scale is the measured energy of a jet before applying any corrections for upstream energy losses and the non-compensating nature of the calorimeter.} $\pt (\mathrm{EM})>7$~GeV. The jet transverse momentum is calibrated from the EM scale to the hadronic scale using a MC-based JES \cite{JES_Summary_r16}.

The numerical evaluation of the uncertainty on the jet energy scale is performed with Monte Carlo pseudo-experiments. In each pseudo-experiment, the jet energy scale is calculated after randomly changing the Monte Carlo single particle energy response within the appropriate uncertainty range given by the measured data/MC ratio. The final uncertainty on the jet energy scale is then given by the spread of the distribution of the jet energy scale over all pseudo-experiments.
Within each pseudo-experiment, all randomly changed parameters are kept fixed, such that the energy response correlations are properly taken into account.

The uncertainties on the particle energy response functions entering the calculation are taken from the $E/p$ measurements described in Sections~\ref{sec:E/P} and \ref{sec:resonances}, ATLAS combined test beam (CTB) measurements~\cite{CTB} and {\sc Geant4} Monte Carlo predictions. The details are described in the following subsections.

\subsection{Additional $E/p$ uncertainties contributing to the jet energy scale}
\label{sec:other_uncertainties_ep}
\begin{figure*}[htb]
\begin{center}
\subfloat[]{\label{fig:eopRatio_etaBin7}\includegraphics[width=0.48\textwidth]{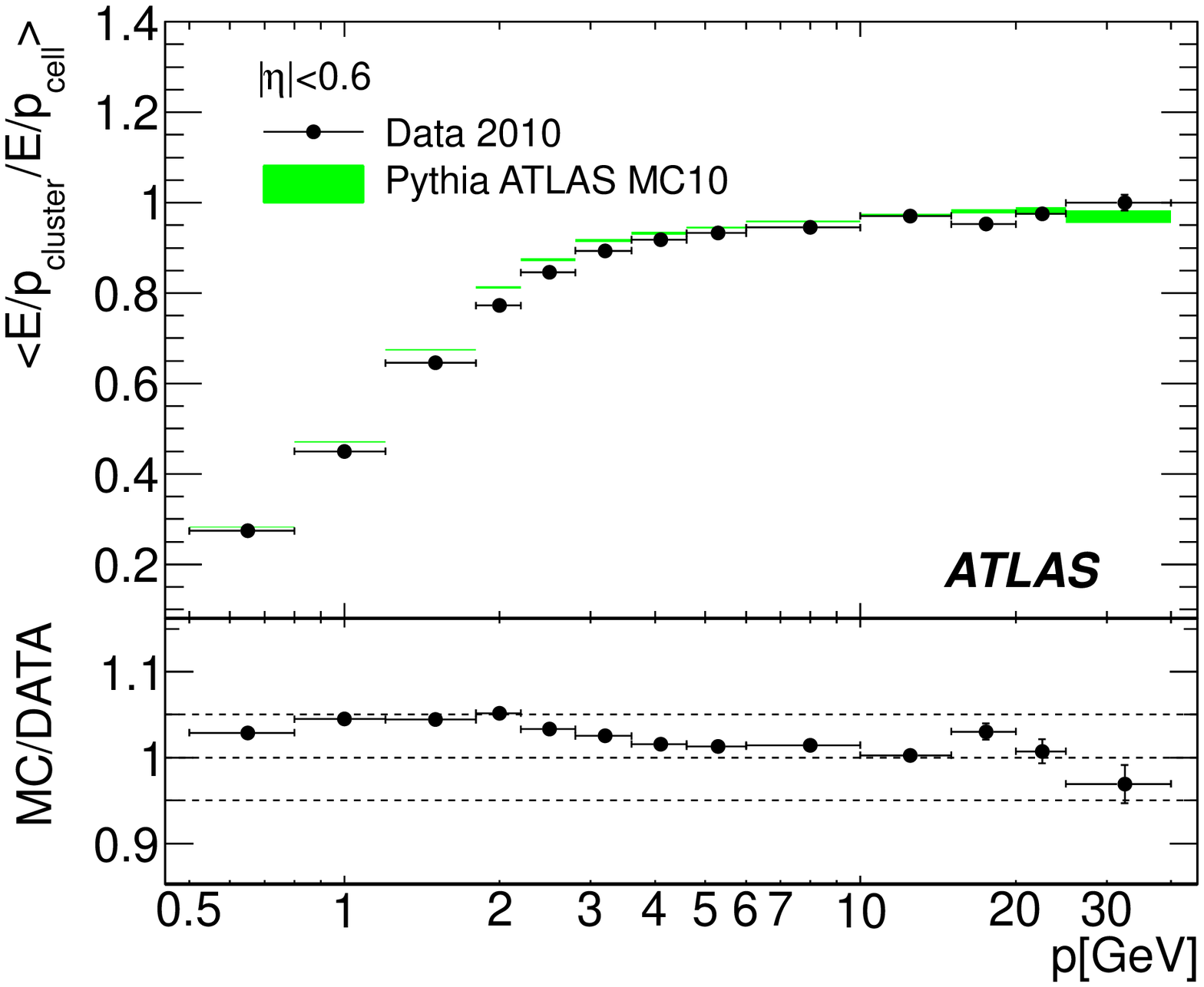}}
\subfloat[]{\label{fig:eopRatio_etaBin6}\includegraphics[width=0.48\textwidth]{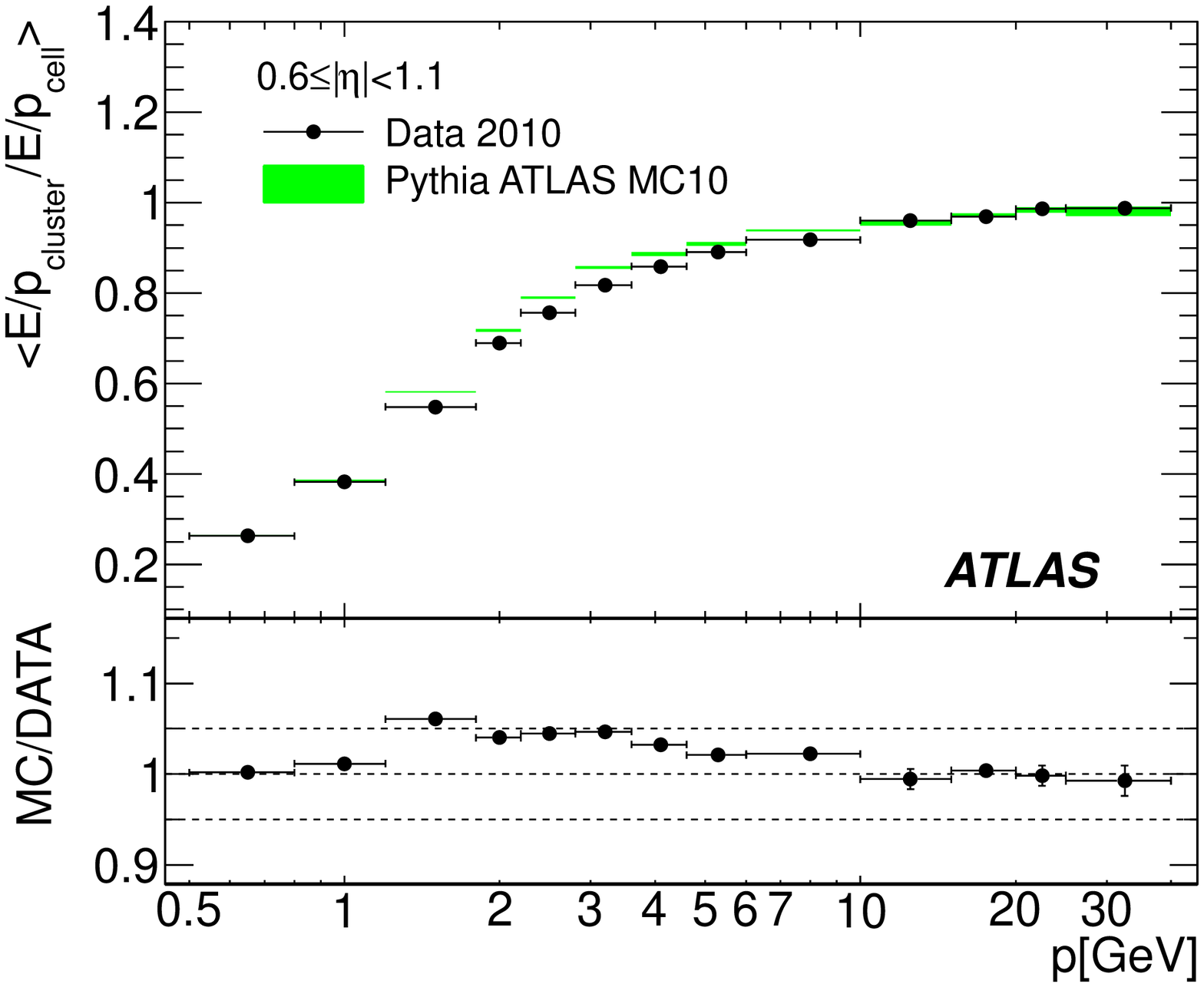}}
\end{center}
\caption{Ratio of the $\langle E/p \rangle$ measurement obtained with topological clusters to that obtained using all calorimeter cells in \protect\subref{fig:eopRatio_etaBin7} the central ($|\eta| < 0.6$) and \protect\subref{fig:eopRatio_etaBin6} forward ($0.6 \leq |\eta| < 1.1$) regions. The inset shows the ratio of the  MC prediction to data. 
\label{fig:cellCluster}}
\end{figure*}

When the single charged hadron response from $E/p$ is used to assess the uncertainty on the jet energy scale, further systematic uncertainties that might affect the propagation of the response to the jet have to be taken into account:

\begin{itemize}
  \item $E/p$ acceptance: it was shown in Section~\ref{sec:definitions} that the probability to find $E/p=0$ is strongly correlated to the amount of upstream material and hence a good measure of the $E/p$ acceptance. A fully correlated (in $p$ and $\eta$) 28\% uncertainty is derived from the maximal observed difference between data and MC simulation in this probability. 

  \item Topological clustering effect: given an amount of energy released in the calorimeter, the energy collected in the topological clusters may differ from the energy released in the calorimeter, depending on how isolated the energy deposit is. This can introduce a bias in the particle response in a jet with respect to the response measured for isolated hadrons. A conservative systematic uncertainty has been computed by comparing the results of the $\langle E/p \rangle$ measurement obtained using clusters as described in Section~\ref{sec:E/P} to those obtained by repeating the measurement using all calorimeter cells in a cone of size $\Delta R < 0.2$. The double ratio of data/MC cluster response to data/MC cell response is used to estimate the uncertainty. The relevant result for the central calorimeter region is presented in \figurename~\ref{fig:cellCluster}, showing discrepancies of $\sim 5\%$ at low $p$ that disappear within the statistical uncertainties for $p \gtrsim 10$~\GeV. 

  \item Out-of-cone $\Delta R > 0.2$ energy deposits: statistical consistent results were found for the $E/p$ data to Monte Carlo ratio for calorimeter energy measured in $\Delta R < 0.3$. Hence no additional uncertainty is assumed.
\end{itemize}

\subsection{Additional uncertainty in  the jet energy scale}
\label{sec:other_uncertainties}

The $E/p$ measurements only cover the response of charged hadronic particles with momenta less than $\sim 20$~GeV. However, depending on the jet momentum, on average between 35\% and 90\% of the energy in jets is carried by particles that are not measured in situ using the isolated track analysis
(mostly photons from $\pi^0$ decays, neutral hadrons and high momentum charged hadrons). Hence, the uncertainty on the energy response to these particles is needed in order to obtain the total calorimeter uncertainty on the jet energy scale. 

\subsubsection{High momentum charged particles}
\label{sec:CTB}
In 2004, an ATLAS Combined Test Beam (CTB) program was carried out at CERN. A ``slice'' of the ATLAS detector composed of the final versions of all sub-detectors in the barrel region was exposed to Super Proton Synchrotron (SPS) test beams. The layout of the sub-detectors was designed to be as close to that of ATLAS as possible.
The setup was used to measure the combined calorimeter response to single charged pions of energies between 20 and 350~GeV for pseudorapidity values of 0.20, 0.25, 0.35, 0.45, 0.55 and 0.65 \cite{CTB,Cojocaru:2004jk,Dowler:2002zp}.

From the measurements in Ref.~\cite{CTB}, the ratio of data to MC simulation predictions is used to supplement the $E/p$ measurements in Section~\ref{sec:E/P} with a larger energy range. However, since these measurements are not made with the same detector, additional systematic uncertainties from the test beam have to be taken into account~\cite{CTB}:
\begin{itemize}
  \item a fully correlated (in $p$ and $\eta$) 0.7\% total energy scale uncertainty for the energies in the LAr calorimeters;
  \item a fully correlated 0.5\% total energy scale uncertainty for the energies in the TileCal;
  \item an additional 0.4\% uncertainty on the LAr uniformity for all measured energies at the same $\eta$ points;
  \item an additional 1.5\% uncertainty on the TileCal uniformity for all measured energies at the same $\eta$ points; and
  \item a 1\% uncertainty from the material in front of the calorimeter.
\end{itemize}

For single particle momenta above 400~GeV no direct measurements in a test beam or in situ exist. Therefore an additional uncertainty of 10\% is added in quadrature to that of the 350~GeV measurement uncertainty in order to cover possible effects from calorimeter non-linearities at high energy densities and longitudinal leakage~\cite{Adragna:2010zz}.

\subsubsection{Absolute calorimeter energy scale}
\label{sec:escale}

The absolute electromagnetic energy scale in ATLAS has been established using $Z \to ee$ decays for the electromagnetic LAr calorimeters and using the energy loss of minimum ionising muons in the TileCal.
For the bulk of the electromagnetic LAr barrel calorimeter, the uncertainty on the cell energy measurement is 1.5\%, and for the LAr presampler the uncertainty is 5\% \cite{egamma_paper_2010}. For the TileCal, the scale uncertainty is 3\% \cite{2010arXiv1007.5423}.  This uncertainty does not affect charged particles with $E/p$ measured in situ, but needs to be considered for all other particles contributing to jets.

\subsubsection{Baryons and neutral hadronic particles}
\label{sec:neutral_hadrons}
Test beam measurements of protons \cite{CTB,Cojocaru:2004jk,Dowler:2002zp,Adragna:2010zz} and $E/p$ measurements for identified pions and protons (Section~\ref{sec:resonances}) have shown that the agreement between data and MC simulation for protons is similar to the data to MC agreement for charged pions. The most significant data to MC difference in the energy response is 25\% for low energy anti-protons. However, on average, these low energy anti-protons contribute no more than 0.5\% to the jet energy. Hence no additional uncertainty for charged baryons is assumed. 

In addition, the charged particle composition of the tracks (mainly $\pi^\pm$, $p^\pm$, $K^\pm$) used for the inclusive $E/p$ measurements in minimum bias events and the charged particle composition in jets is found to be sufficiently similar to cause no additional systematic uncertainty. Thus, even if some identified particle measurements show differences (as for low energy anti-protons), the inclusive $E/p$ measures, on average, the charged particle response that is needed for an application to jets. The related uncertainty is proportional to the MC simulation modelling of the difference in the charged track composition between minimum bias events and jets and found to be negligible.

No test beam measurements for neutral hadronic particles have been carried out. Moreover, the {\sc Geant4} models have large uncertainties. On average 10--12\% of the jet energy is carried by neutral hadrons, mostly $K_S$, $K_L$ and neutrons.
Most of the $K_S$ decay to pions before they reach the calorimeter. Hence the $E/p$ and CTB measurements can be used for $K_S$. 

\begin{figure*}
\begin{center}
    \subfloat[]{
      \includegraphics[width=0.48\textwidth]{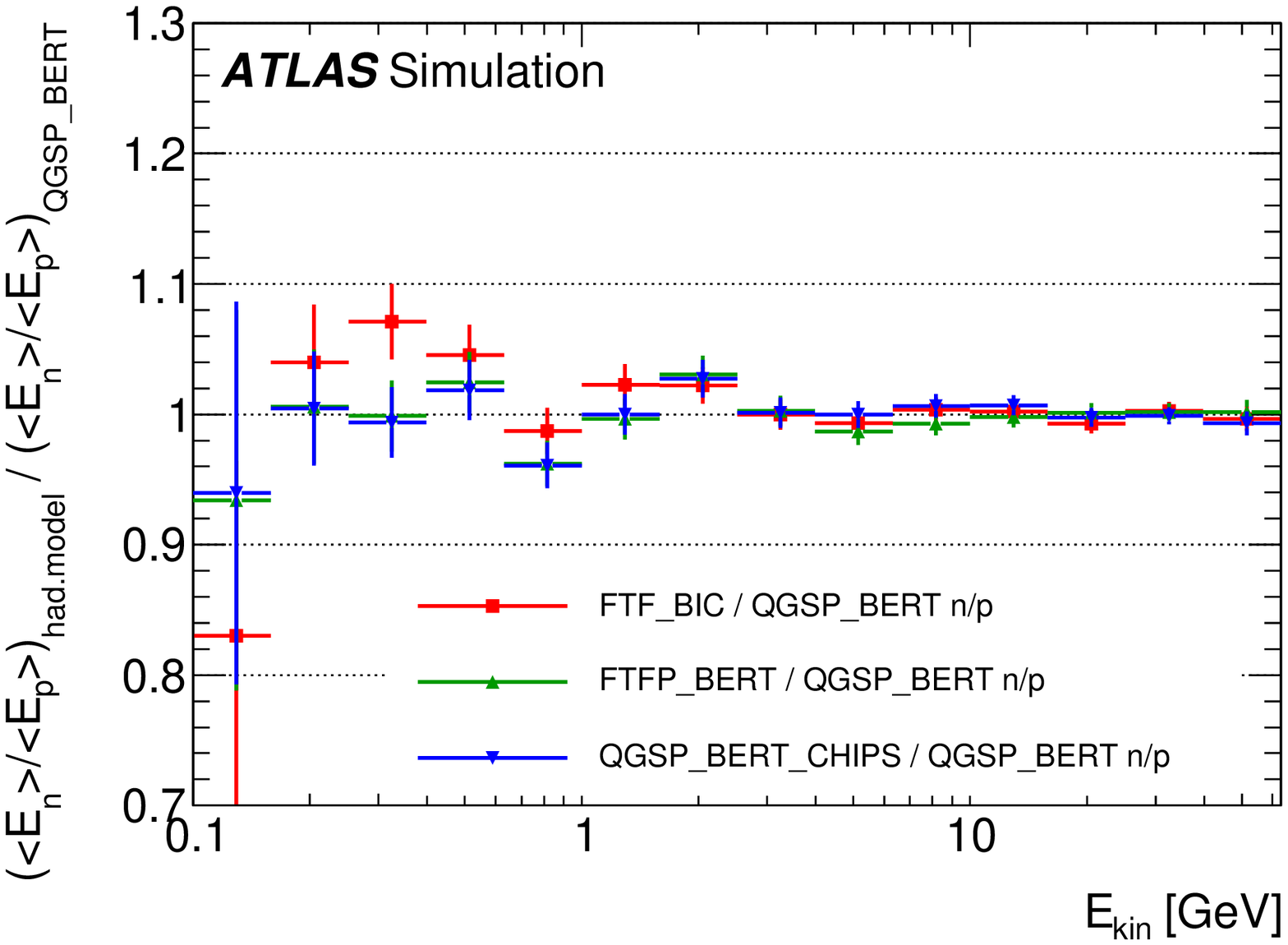}
      \label{fig:G4_neutron_proton_ratio_n_p}      
    }
    \hfill
    \subfloat[]{
      \includegraphics[width=0.48\textwidth]{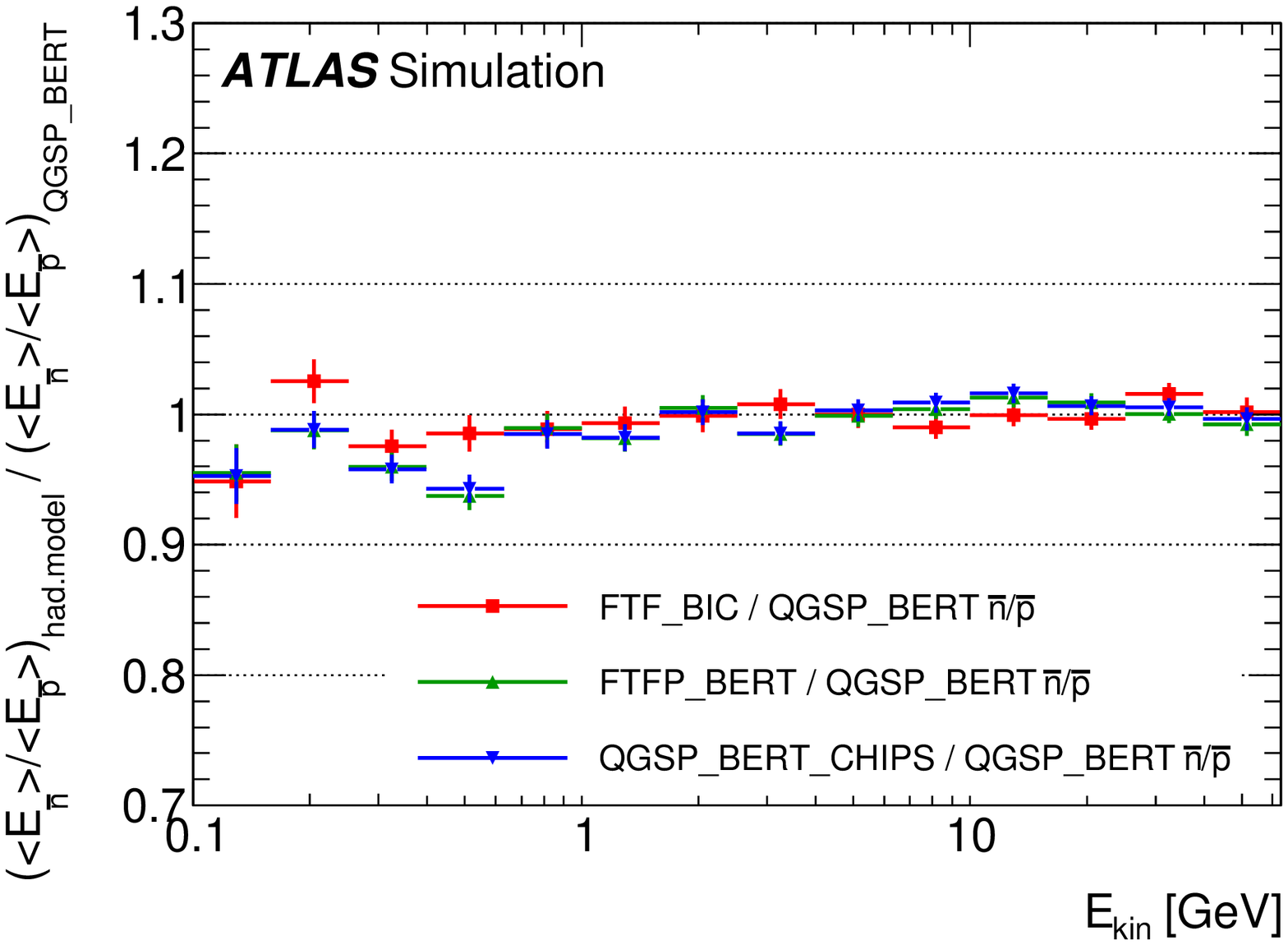}
      \label{fig:G4_neutron_proton_ratio_an_ap}      
    }
    \hfill
    \subfloat[]{
      \includegraphics[width=0.48\textwidth]{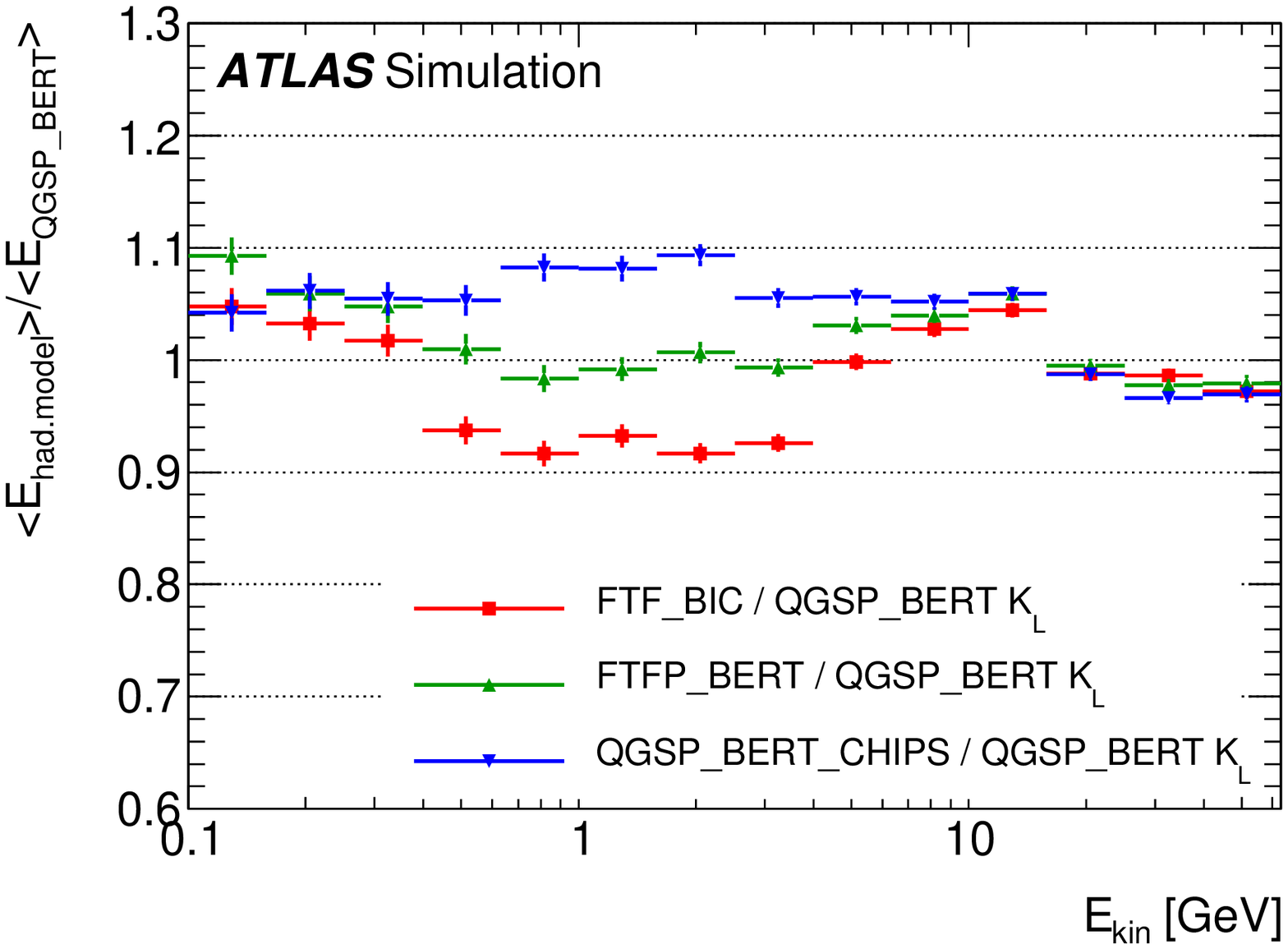}
      \label{fig:G4_KL_ratio}      
    }
    \hfill
\end{center}
\caption{Ratio of the average calorimeter response for alternative {\sc Geant4} hadronic physics models FTF\_BIC, FTFP\_BERT and QGSP\_BERT\_CHIPS~\protect\cite{geant4Physics} to the ATLAS default hadronic physics model QGSP\_BERT as function of the particle kinetic energy $E_{\rm{kin}}$ in the range between 100~\MeV{} and 50~\GeV{} for \protect\subref{fig:G4_neutron_proton_ratio_n_p} the ratio of neutrons $n$ to protons $p$, \protect\subref{fig:G4_neutron_proton_ratio_an_ap} the ratio of anti-neutrons $\bar{n}$ to anti-protons $\bar{p}$ and \protect\subref{fig:G4_KL_ratio} neutral kaons $K_L$.
\label{fig:G4_ratios}}
\end{figure*}

For neutrons and anti-neutrons {\sc Geant4} studies (see \figurename~\ref{fig:G4_neutron_proton_ratio_n_p} and \ref{fig:G4_neutron_proton_ratio_an_ap}) comparing alternative {\sc Geant4} hadronic physics models to the ATLAS-default hadronic physics model QGSP\_BERT~\cite{geant4Physics} show that the (anti-)neutron to (anti-)proton response ratio is determined at the 10\% level for particle momenta below 3~GeV and at the 5\% level for higher momenta. Hence the (anti-)neutron response can be related to the sufficiently well simulated (anti-)proton response with these additional uncertainties.

Few measurements are available for kaon interaction cross sections in materials. While the response of charged kaons is covered by the inclusive $E/p$ measurements, the $K_L$ response has to rely on Monte Carlo simulation predictions. {\sc Geant4} studies comparing alternative {\sc Geant4} hadronic physics models show uncertainties of $\sim$10\% with respect to the ATLAS-default hadronic physics model QGSP\_BERT (see \figurename~\ref{fig:G4_KL_ratio}). However, because of the limited availability of measurements, a more conservative uncertainty of 20\% on the $K_L$ calorimeter response is added in quadrature to the uncertainty for charged particles~\cite{geant4_KL_discussion}.

\subsection{Jet energy scale uncertainty estimation}
\label{sec:combined}
This section combines the single particle uncertainties discussed in Sections~\ref{sec:E/P}, \ref{sec:other_uncertainties_ep} and \ref{sec:other_uncertainties} into an expected shift and uncertainty on the calorimeter jet energy response with respect to the MC simulation prediction. Because of the CTB measurements used for high momentum particles, the estimation is limited to the pseudorapidity region $|\eta| < 0.8$. 

\begin{figure*}[p] 
  \begin{center}
    \subfloat[]{
    \includegraphics[width=.31\textwidth]
      {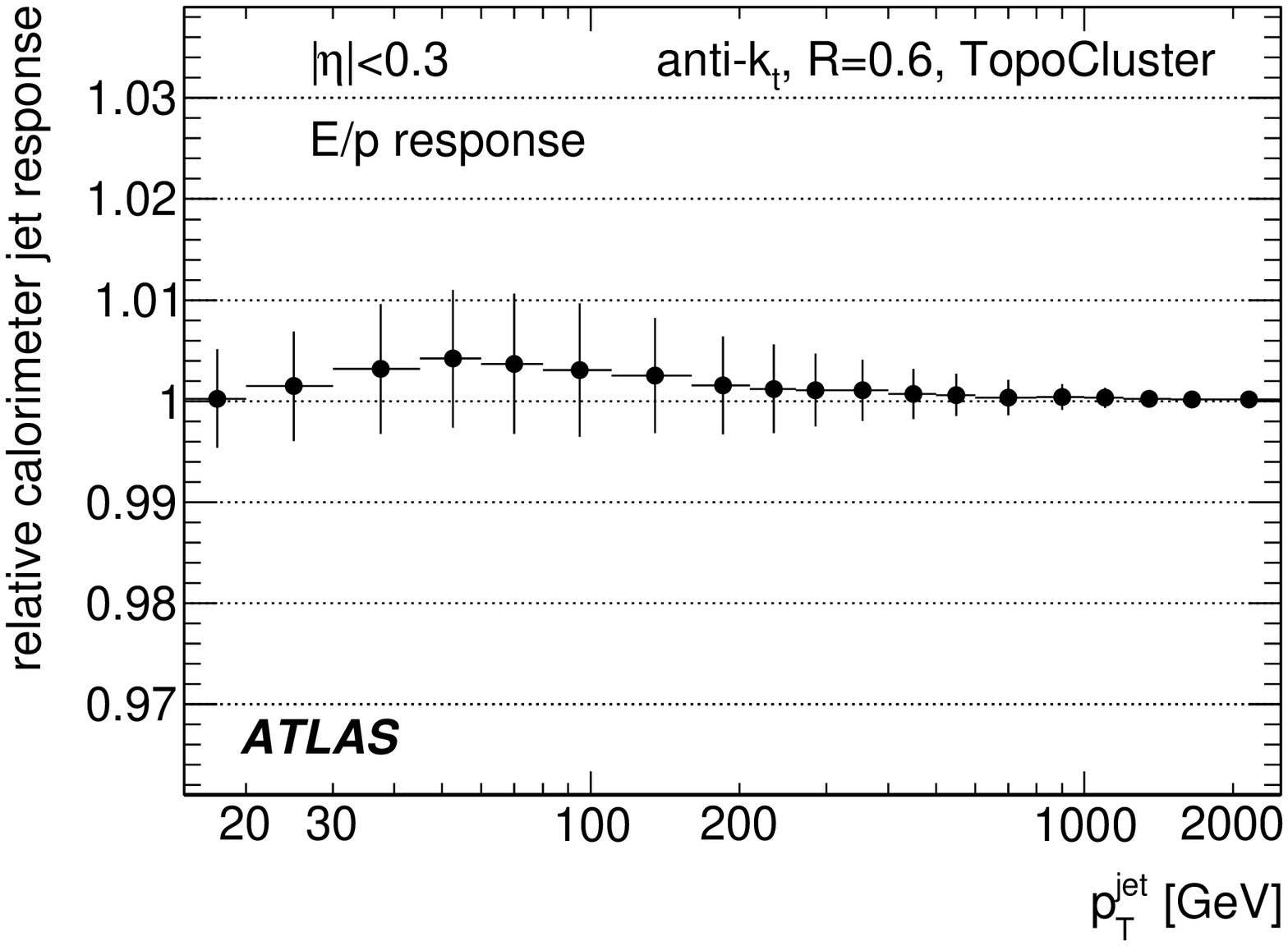}
    \label{fig:JESEoP}
    }
    \hfill
    \subfloat[]{
    \includegraphics[width=.31\textwidth]
      {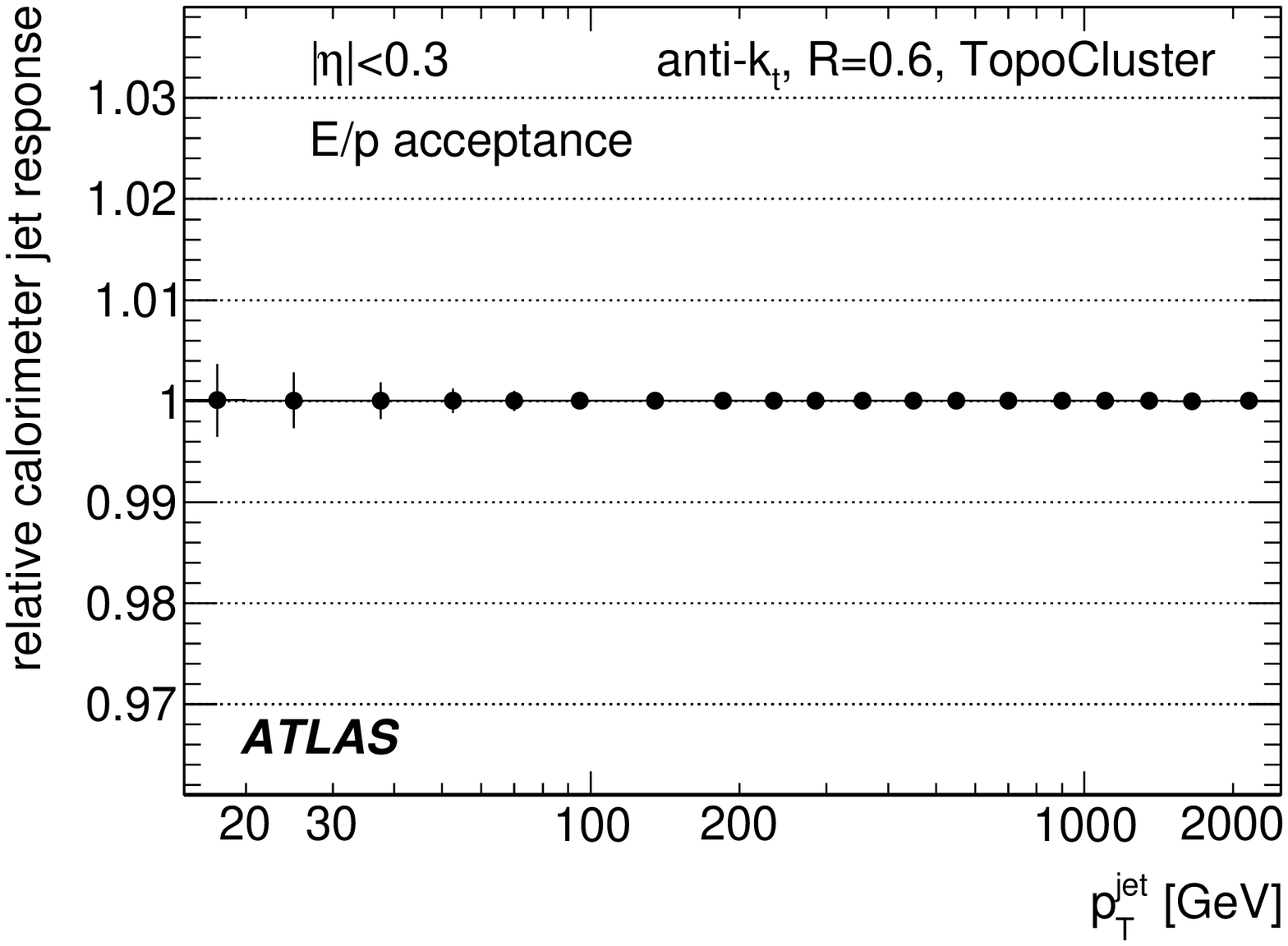}
    \label{fig:JESEoPacc}
    }
    \hfill
    \subfloat[]{
    \includegraphics[width=.31\textwidth]
      {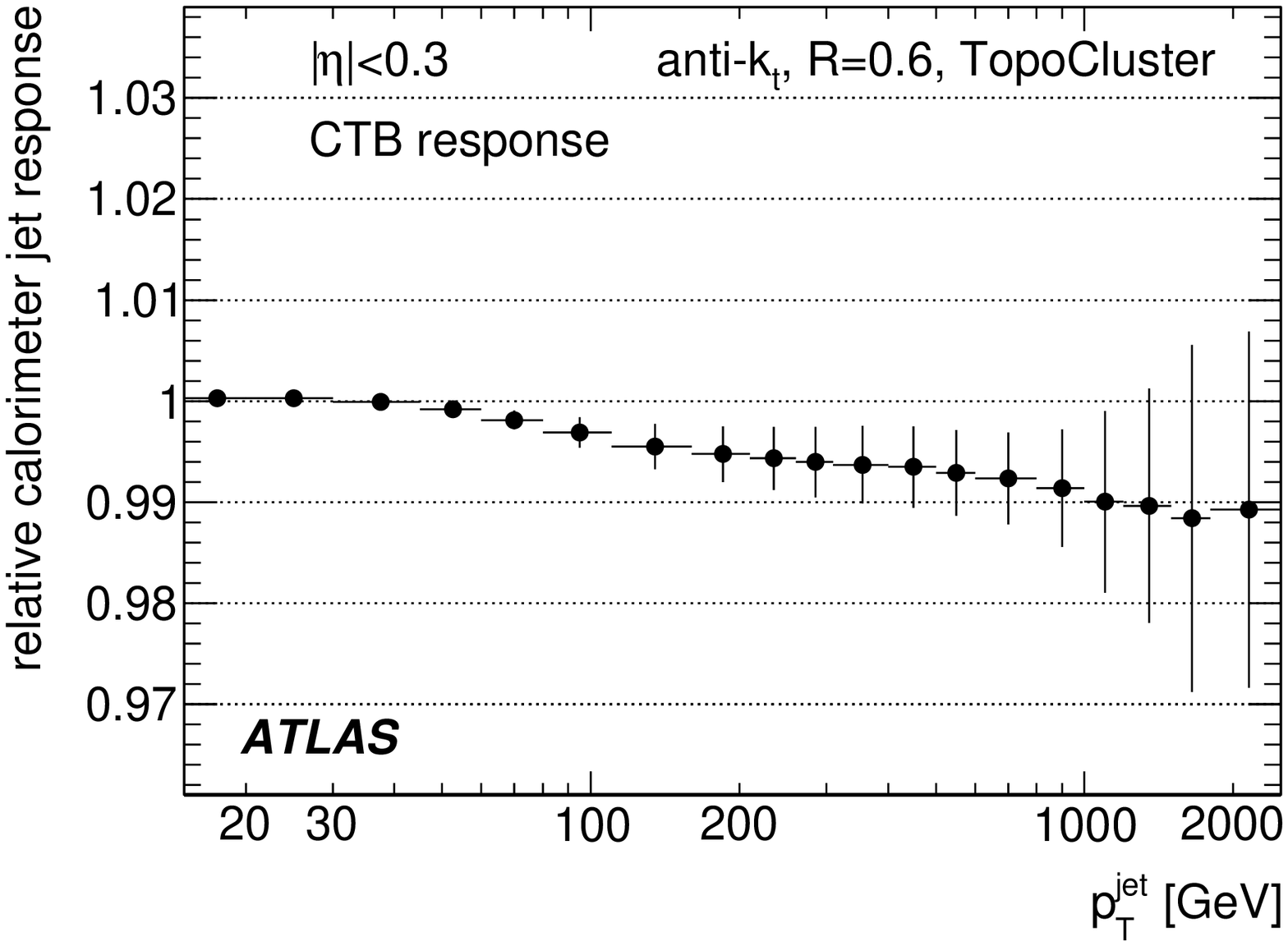}
    \label{fig:JESCTB} 
    }

    \subfloat[]{
    \includegraphics[width=.31\textwidth]
      {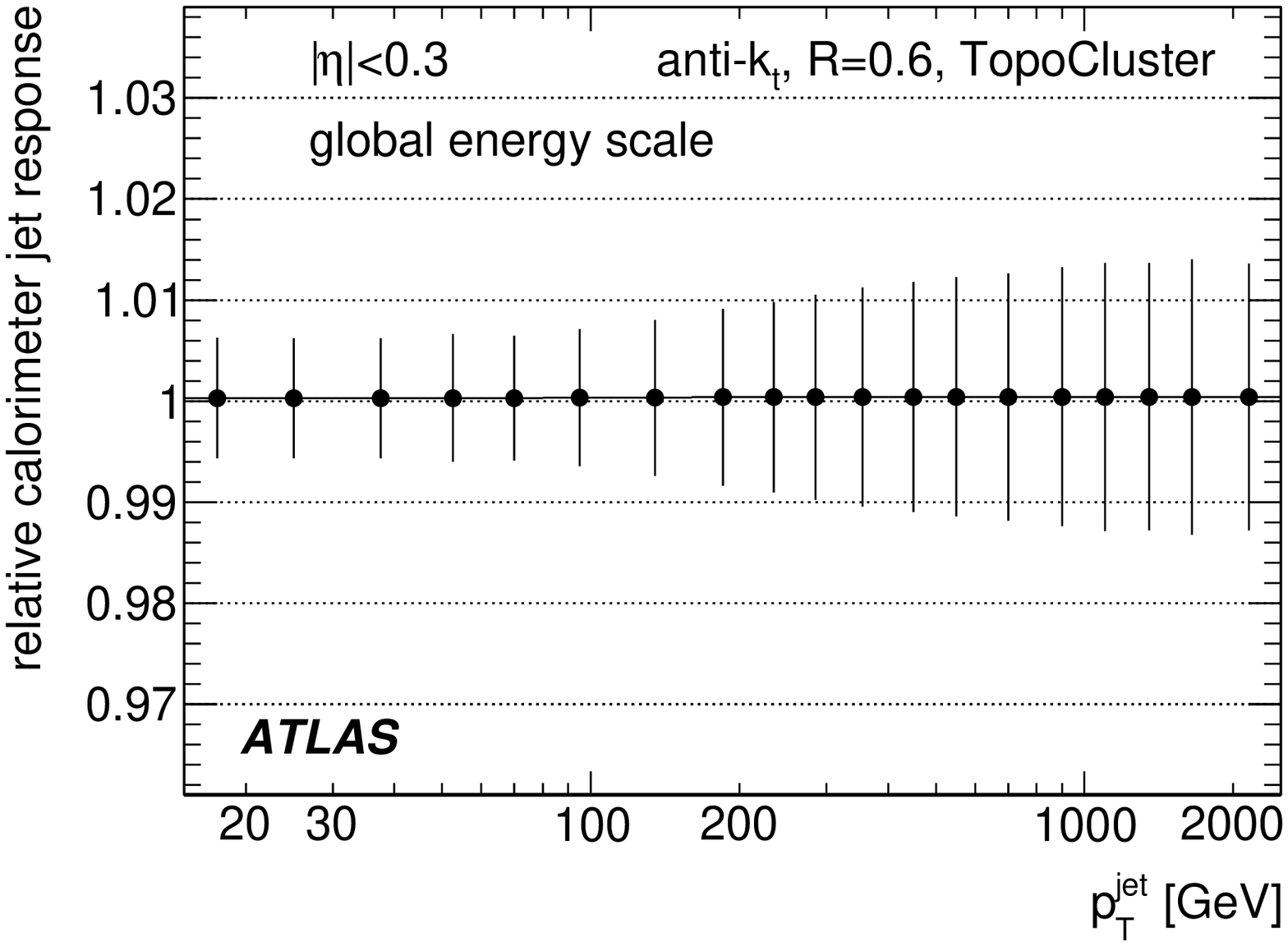}
    \label{fig:escale} 
    }
    \hfill
    \subfloat[]{
    \includegraphics[width=.31\textwidth]
      {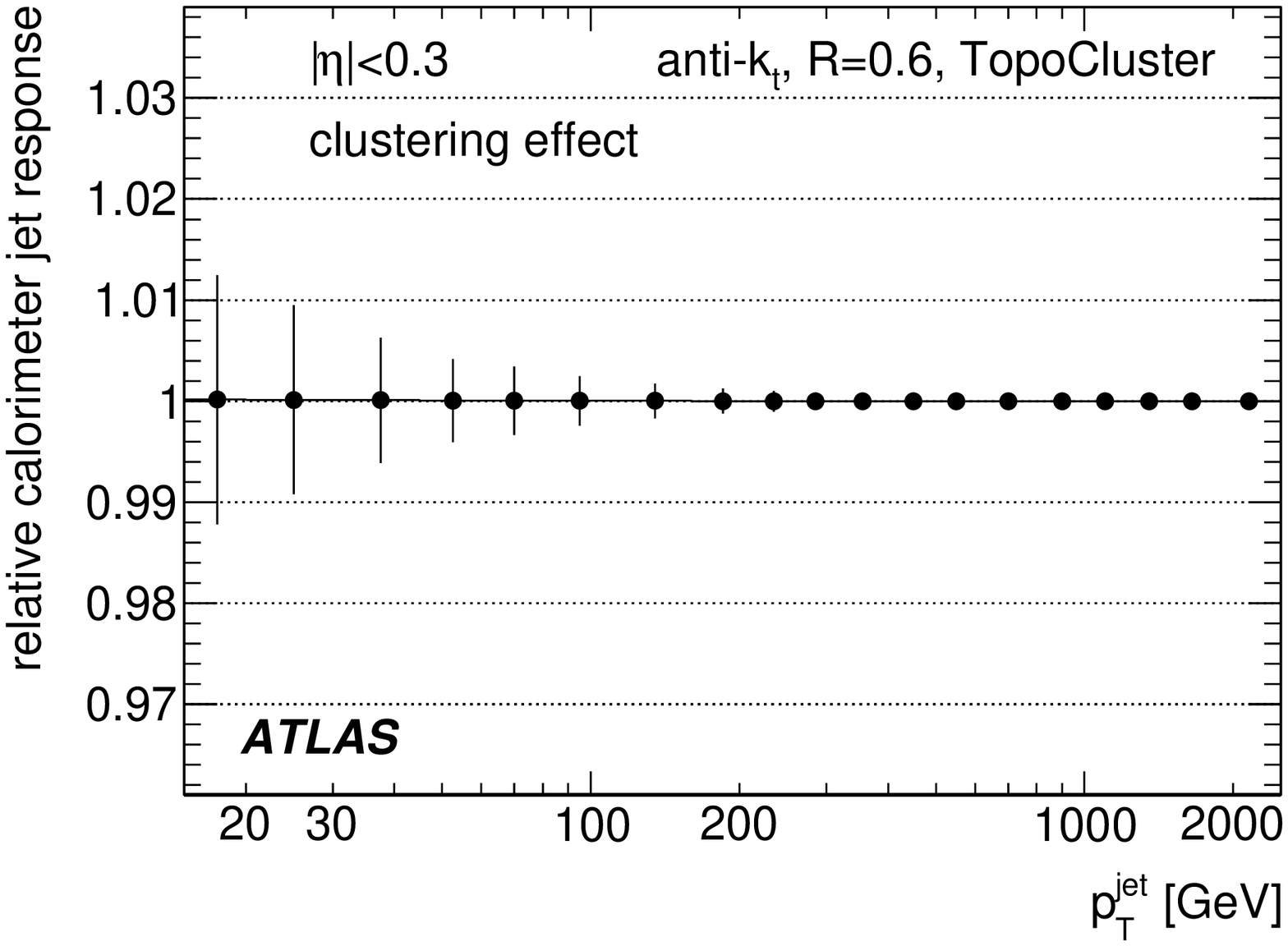}
    \label{fig:JEScell_to_cluster} 
    }
    \hfill
    \subfloat[]{
    \includegraphics[width=.31\textwidth]
      {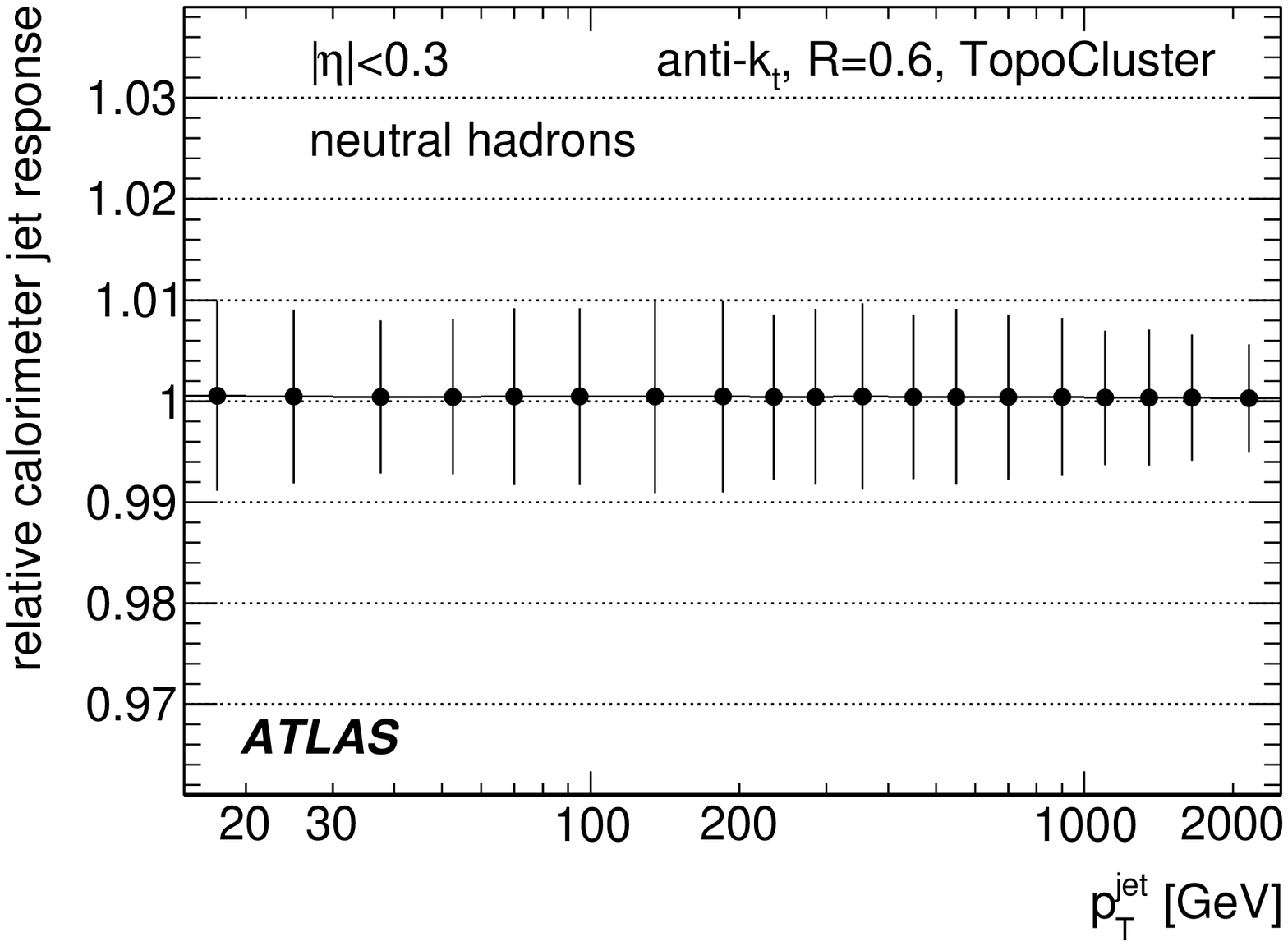}
    \label{fig:JESneutrals} 
    }

    \caption{Expected shift (black markers) and uncertainty (error bars) on the relative calorimeter jet response with respect to the MC simulation  for jets reconstructed with the \antiKT jet algorithm ($R=0.6$) in the range $|\eta|<0.3$ as function of the jet transverse momentum from the uncertainty on \protect\subref{fig:JESEoP} the $E/p$ response, \protect\subref{fig:JESEoPacc} the uncertainty on the $E/p$ acceptance, \protect\subref{fig:JESCTB} the uncertainty from the CTB response, \protect\subref{fig:escale} the uncertainty from the absolute energy scale, \protect\subref{fig:JEScell_to_cluster} the uncertainty from the clustering effect, and \protect\subref{fig:JESneutrals} the uncertainty from neutral hadrons.
    The $x$-axis is the jet transverse momentum calibrated from the EM scale to the hadronic scale using an MC-based JES calibration factor \protect\cite{JES_Summary_r16}.} 
    \label{fig:JEScontributions} 

    \includegraphics[width=.66\textwidth]
      {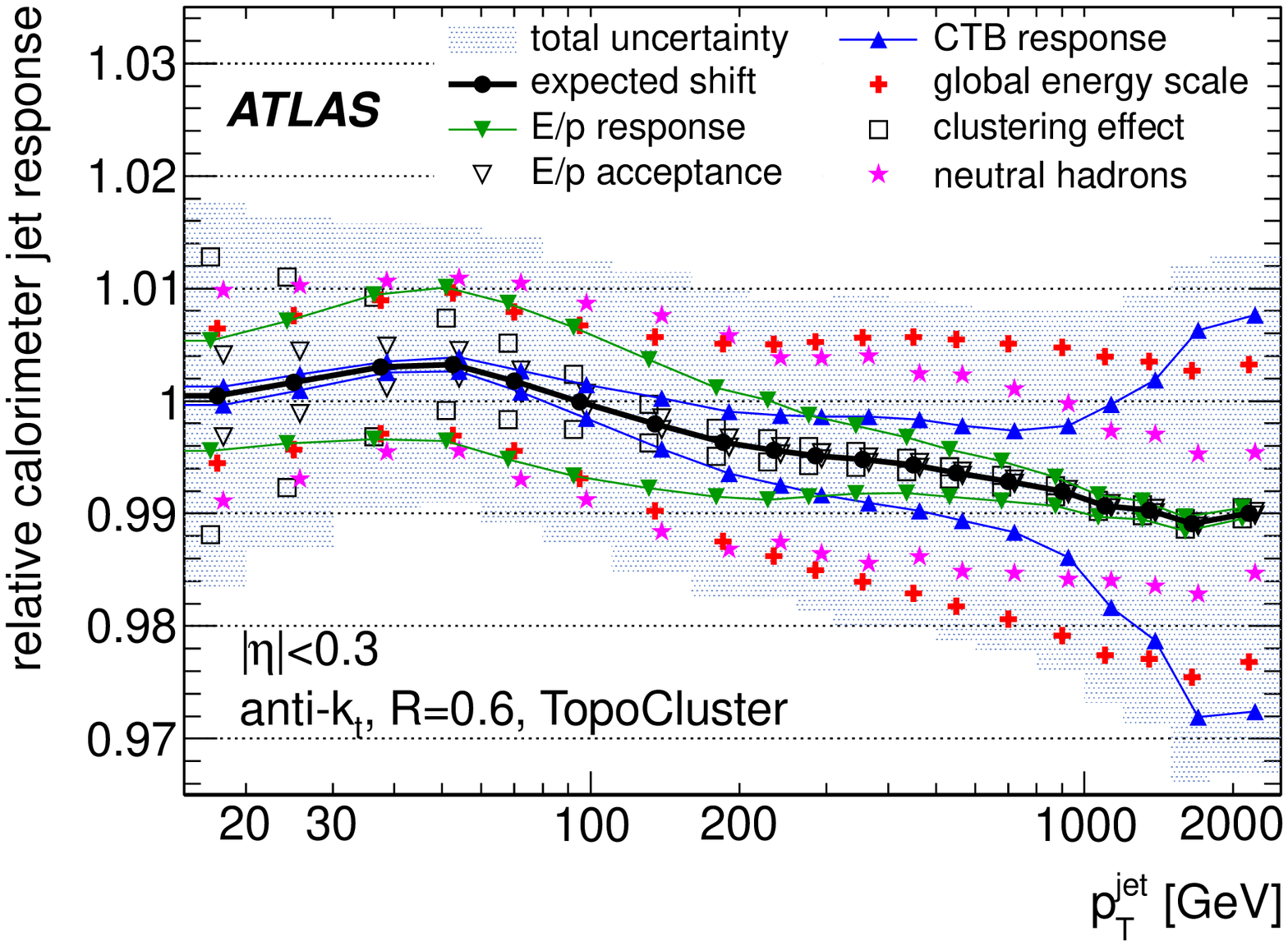}
    \caption{Expected total shift (black dots) and uncertainty (light blue band) on the relative calorimeter jet response with respect to the MC simulation for jets reconstructed with the \antiKT jet algorithm ($R=0.6$) in the range $|\eta|<0.3$ as function of the jet transverse momentum. The uncertainty components from \figurename~\ref{fig:JEScontributions} are shown superimposed on the expected total shift.
    The $x$-axis is the jet transverse momentum calibrated from the EM scale to the hadronic scale using an MC-based JES calibration factor \protect\cite{JES_Summary_r16}.} 
    \label{fig:JESuncert_summary} 
  \end{center}
\end{figure*}

Figure~\ref{fig:JEScontributions} shows the individual contributions to the uncertainty on the jet energy scale for $|\eta| < 0.3$:
\begin{center}
\begin{tabular}{l|l|l}
Uncertainty on & Figure & Details \\\hline
$E/p$ response & \ref{fig:JESEoP} & Section~\ref{sec:eoverp_results} \\
$E/p$ acceptance & \ref{fig:JESEoPacc} & Section~\ref{sec:other_uncertainties_ep}\\
CTB response & \ref{fig:JESCTB} & Section~\ref{sec:CTB} \\
Global energy scale & \ref{fig:escale} & Section~\ref{sec:escale} \\
Clustering effect & \ref{fig:JEScell_to_cluster} & Section~\ref{sec:other_uncertainties_ep}\\
Neutral hadrons & \ref{fig:JESneutrals} & Section~\ref{sec:neutral_hadrons}
\end{tabular}
\end{center}

The $E/p$ response and CTB single particle measurements show a shift of less than $1\%$ between data and Monte Carlo simulation, which propagates into an expected shift of $\sim$~0.5--1\% in the JES (\figurename~\ref{fig:JESEoP} and \ref{fig:JESCTB}). Due to the high precision of the measurements, the uncertainty is small. The visible drop in the response due to CTB measurements is caused by a systematic difference between the CTB data and Monte Carlo simulation for high momenta.

The effects of the $E/p$ acceptance and of the clustering are small (\figurename~\ref{fig:JESEoPacc} and \figurename~\ref{fig:JEScell_to_cluster}). The influence of the absolute electromagnetic energy scale uncertainty is small for low $\pt$ jets, where most of the energy is carried by particles  with $E/p$ measured in situ. However, high $\pt$ jets have only a small fraction of energy in the particle momentum range of the $E/p$ analysis and are hence fully affected by this uncertainty (\figurename~\ref{fig:escale}).

The uncertainty on the calorimeter response to neutral hadrons was estimated in Section~\ref{sec:neutral_hadrons}, resulting in a $\sim$~1\% contribution to the total JES uncertainty (\figurename~\ref{fig:JESneutrals}).

These individual contributions to the uncertainty are summarised in \figurename~\ref{fig:JESuncert_summary} together with the total expected shift and uncertainty. No single component is dominant and depending on the jet momentum several components contribute at approximately the same level to the total uncertainty.

\begin{figure*} 
  \begin{center}
    \subfloat[]{
      \includegraphics[width=.485\textwidth]
      {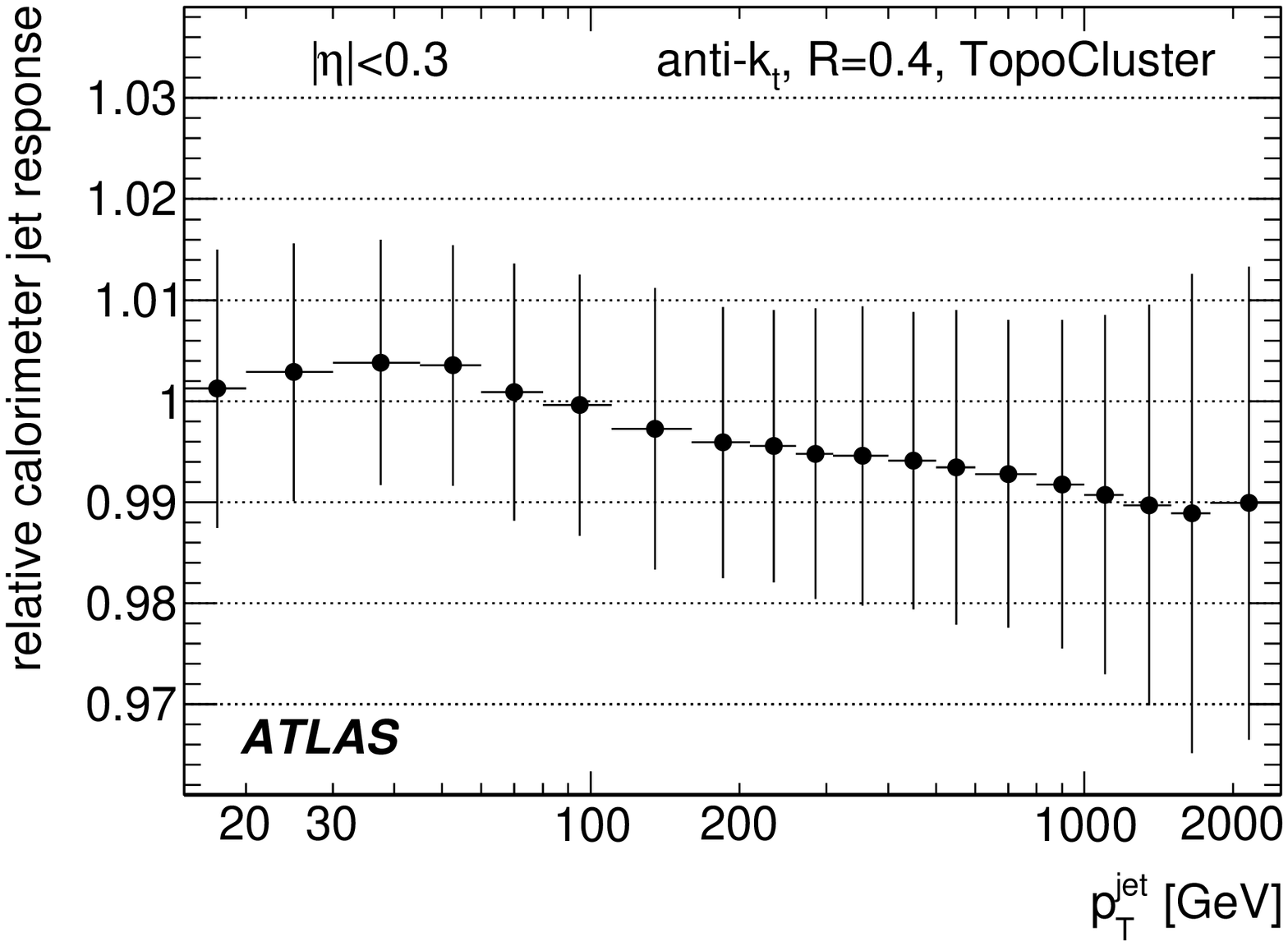}
      \label{fig:JES_total_04_00-03} 
    }
    \hfill
    \subfloat[]{
      \includegraphics[width=.485\textwidth]
      {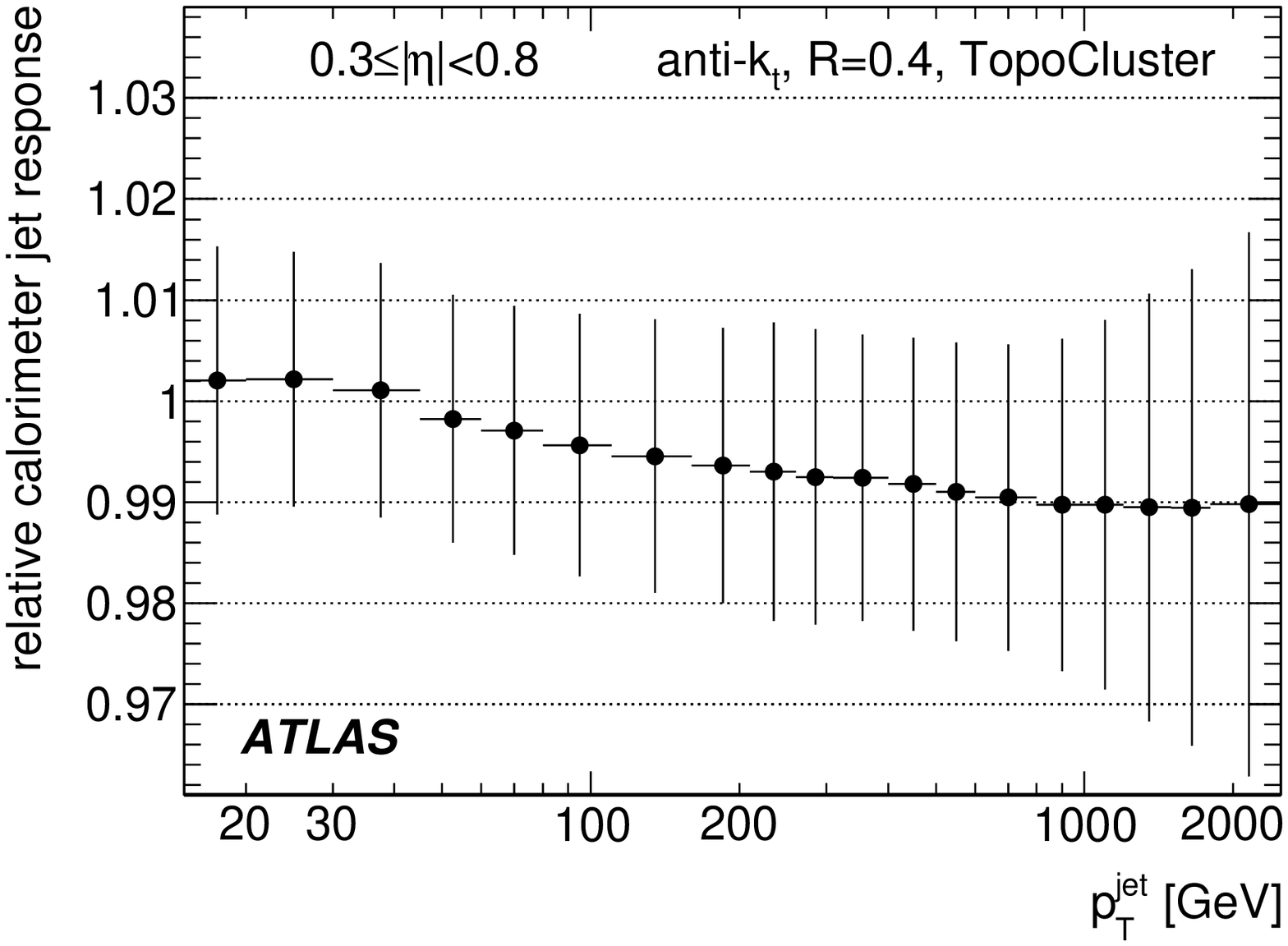}
      \label{fig:JES_total_04_03-08} 
    }

    \subfloat[]{
      \includegraphics[width=.485\textwidth]
      {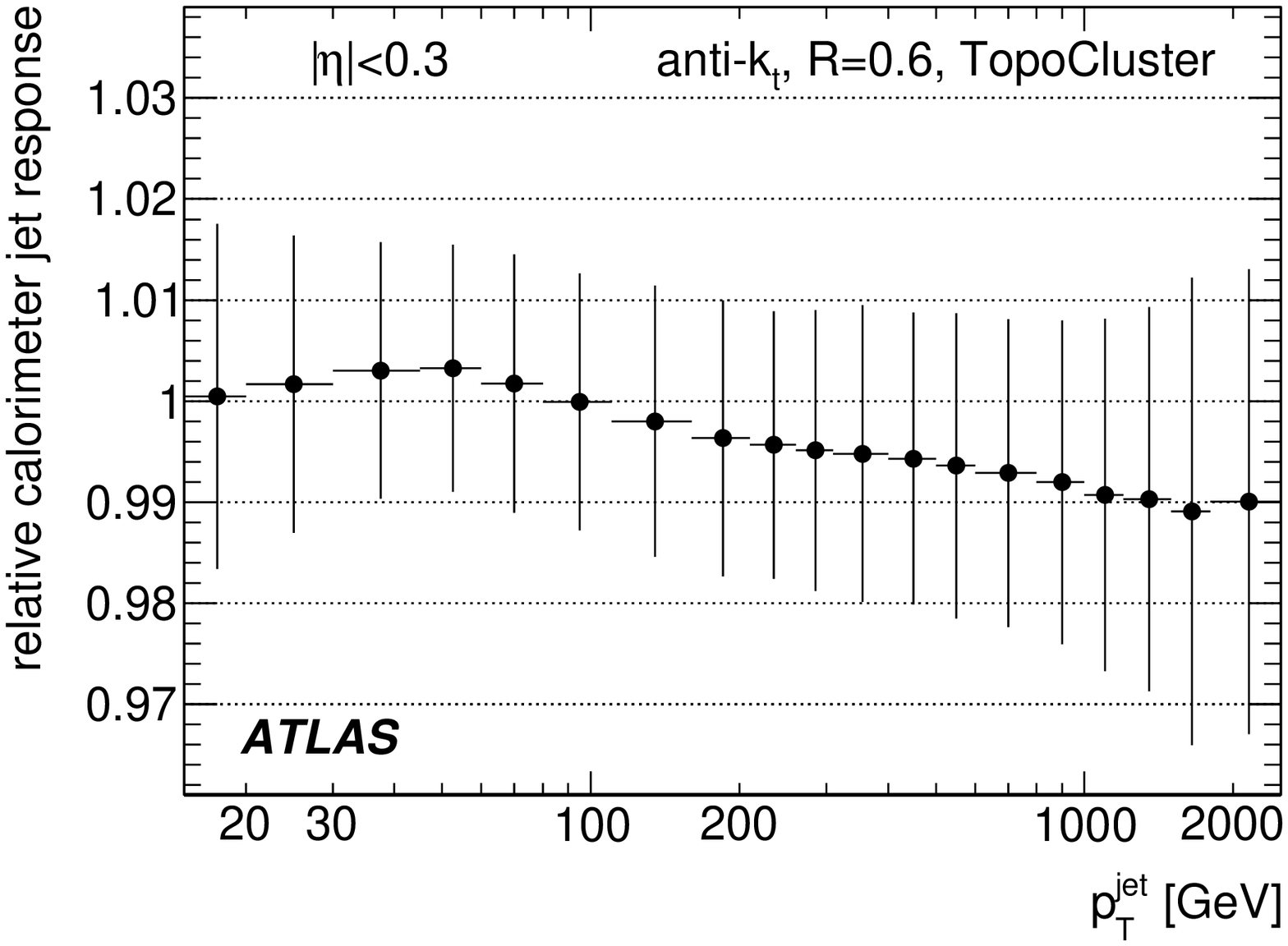}
      \label{fig:JES_total_06_00-03} 
    }
    \hfill
    \subfloat[]{
      \includegraphics[width=.485\textwidth]
      {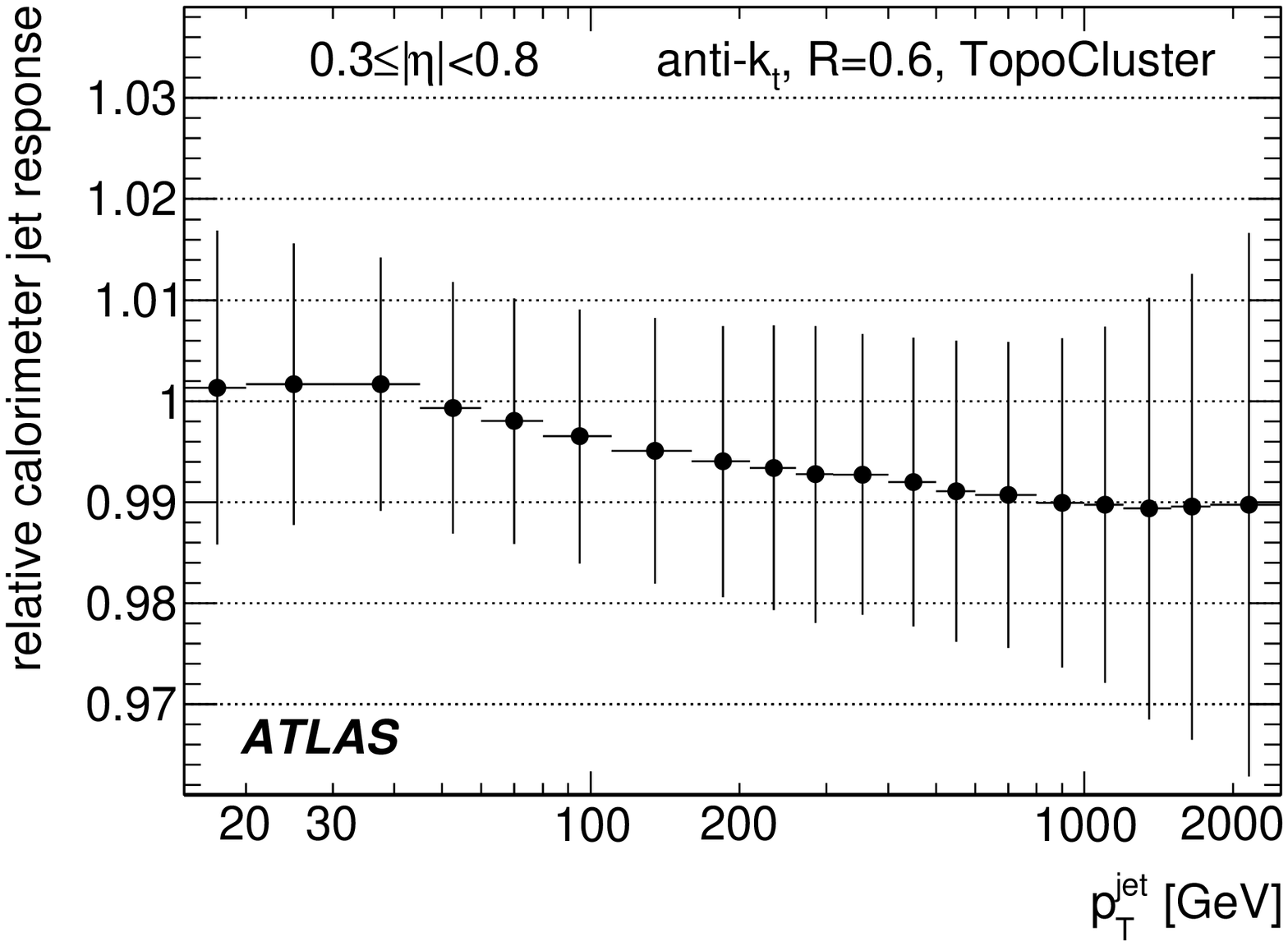}
      \label{fig:JES_total_06_03-08} 
    }
    \caption{Expected shift and total uncertainty on the relative calorimeter jet response with respect to the MC simulation for jets reconstructed with the \antiKT jet algorithm ($R=0.4$ and $R=0.6$) in the range $|\eta|<0.3$ and $0.3 \leq |\eta|<0.8$ as function of the jet transverse momentum.
    The $x$-axis is the jet transverse momentum calibrated from the EM scale to the hadronic scale using an MC-based JES calibration factor \cite{JES_Summary_r16}.} 
    \label{fig:JES_total} 
  \end{center}
\end{figure*}

Finally, the total calorimeter uncertainty on the jet energy scale is shown in \figurename~\ref{fig:JES_total} for \antiKT jets with $R=0.4$ and $R=0.6$ in the pseudorapidity range $|\eta|<0.8$. For both jet sizes the maximum expected shift in the jet energy scale is $\sim$~1\% with an uncertainty of 1--3\%. The envelope of the shift and uncertainty on the calorimeter JES is taken as the contribution to the total JES uncertainty discussed in Ref.~\cite{JES_Summary_r16}.

\subsection{Jet energy scale uncertainty correlations}
\label{sec:correlation}
\begin{figure*} 
  \begin{center}
    \includegraphics[width=.97\textwidth]
    {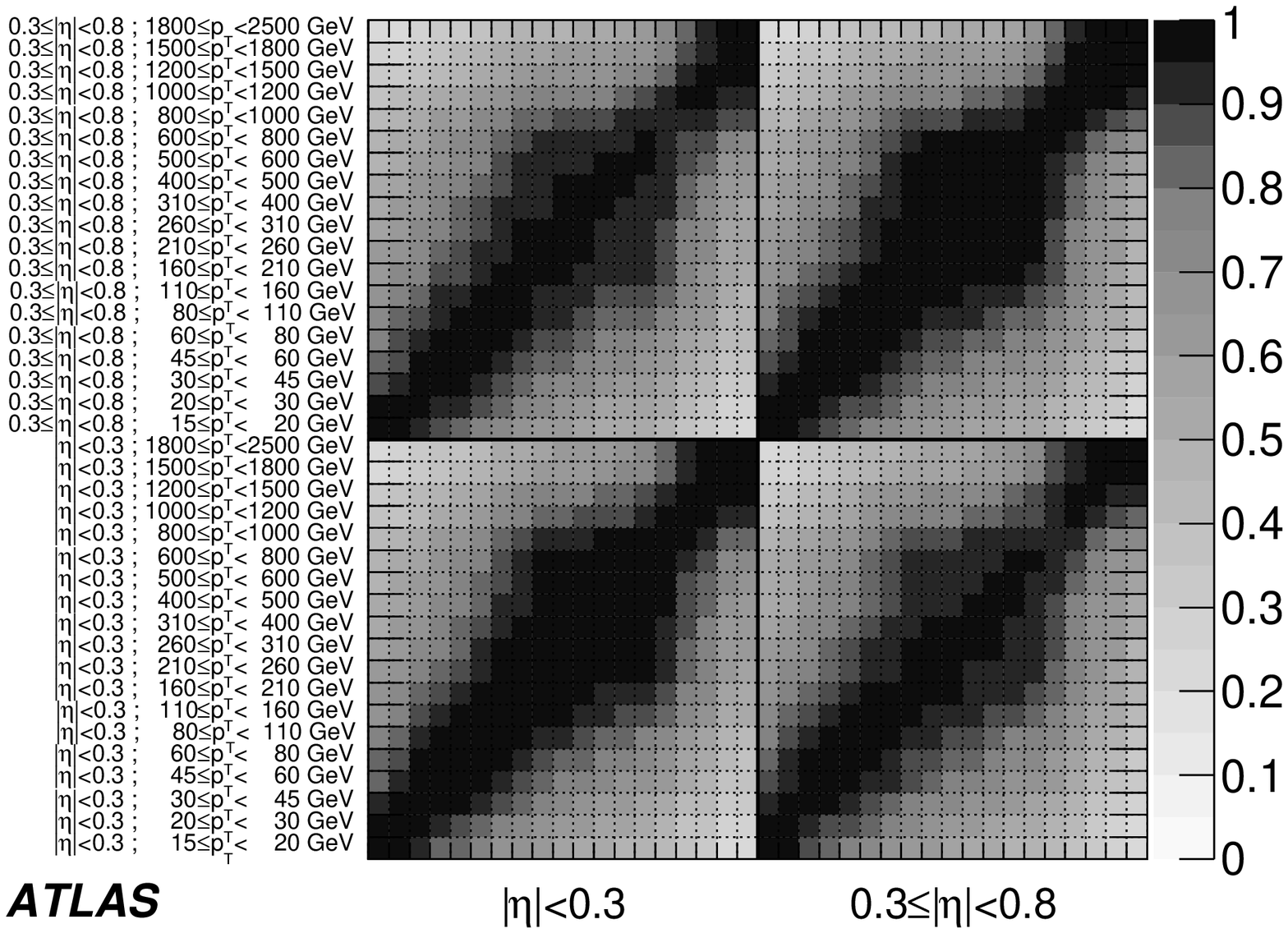}
    \caption{Correlation coefficient of the total uncertainty on the calorimeter jet response for jets reconstructed with the \antiKT jet algorithm ($R=0.6$) in the range $|\eta|<0.8$ and jet transverse momenta between 15~GeV$\leq p_{T}<$2.5~TeV. The bins on the $x$-axis are identical to the bins on the $y$-axis. A dark area indicates highly correlated uncertainties, a light area almost uncorrelated uncertainties.} 
    \label{fig:JES_corr_total06} 
  \end{center}
\end{figure*}

The use of pseudo-experiments for the determination of the JES uncertainty allows a direct extraction of the correlation of uncertainties between different jet momenta, pseudorapidities or algorithms by correlating fluctuations of different quantities within each pseudo-experiment. Figure~\ref{fig:JES_corr_total06} shows the correlation of the JES uncertainty between different $\eta$ and $p$ bins. As expected, the correlation between neighboring bins in $|\eta|$ and $\pt$ is almost 100\%, while widely separated bins show only a $\sim$~30\% correlation. This remaining $\sim$~30\% correlation is mostly caused by the calorimeter energy scale and neutral hadron uncertainty, which contribute identically to all jets.

One specific use for the correlation of the uncertainty between jets is to compare the relative calorimeter response between jets selected from two different categories (different reconstruction algorithm, different originating parton, etc.). For each jet category, the procedure of Section~\ref{sec:combined} is repeated and the expected shift and uncertainty on the JES with respect to the MC simulation is derived\footnote{Because the shifts and uncertainties are evaluated with respect to the MC simulation, any response shift already visible at the MC level will not propagate into either the JES uncertainty or into the relative calorimeter response between jets from different categories.}. For the determination of the uncertainty on the relative calorimeter response, the ratio of two jet categories is taken, cancelling out all common shifts and uncertainties. 

\begin{figure*} 
  \begin{center}
    \subfloat[]{
      \includegraphics[width=.485\textwidth]
      {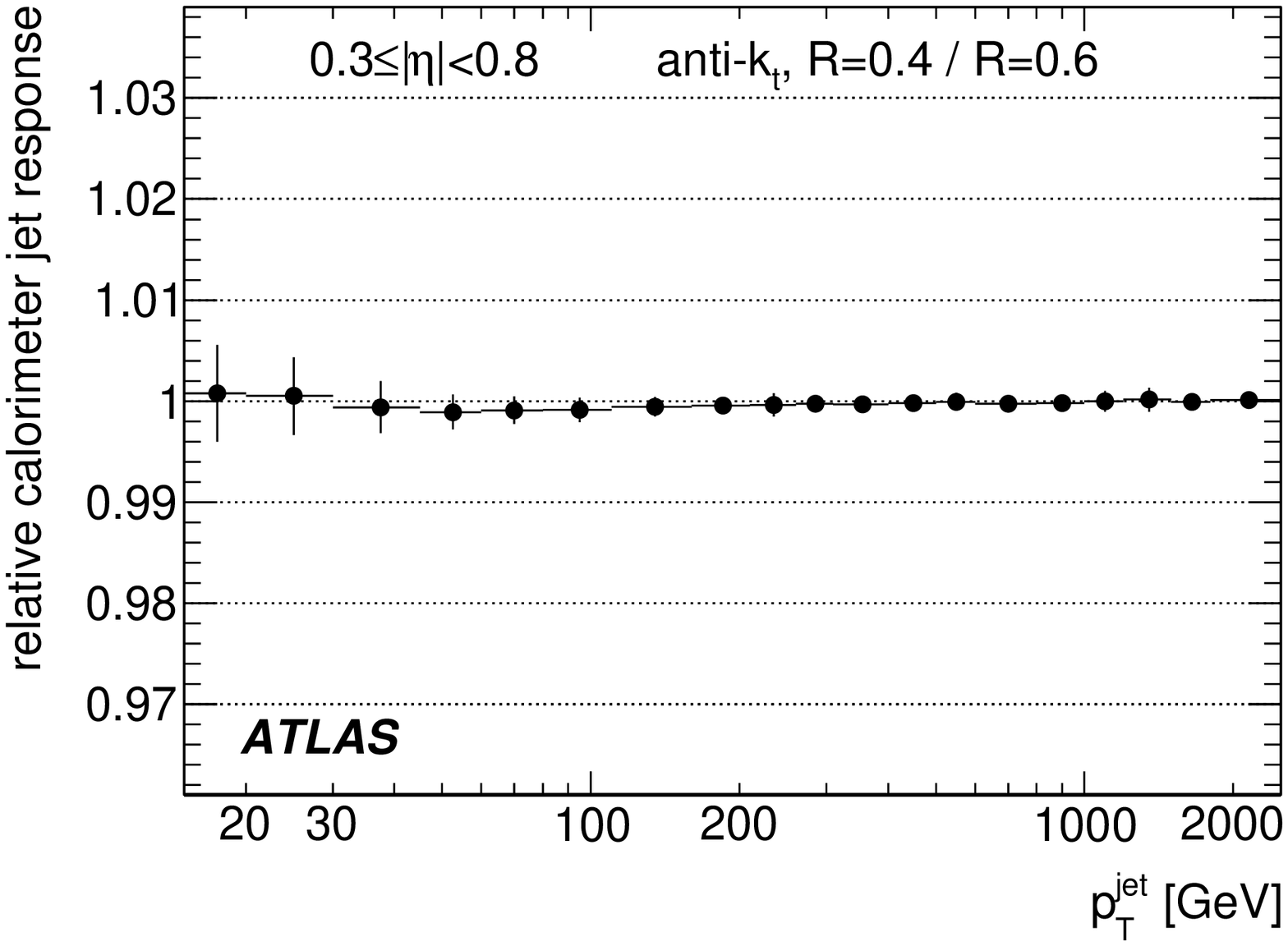}
      \label{fig:JES_corr_ratio_04_06} 
    }
    \hfill
    \subfloat[]{
      \includegraphics[width=.485\textwidth]
      {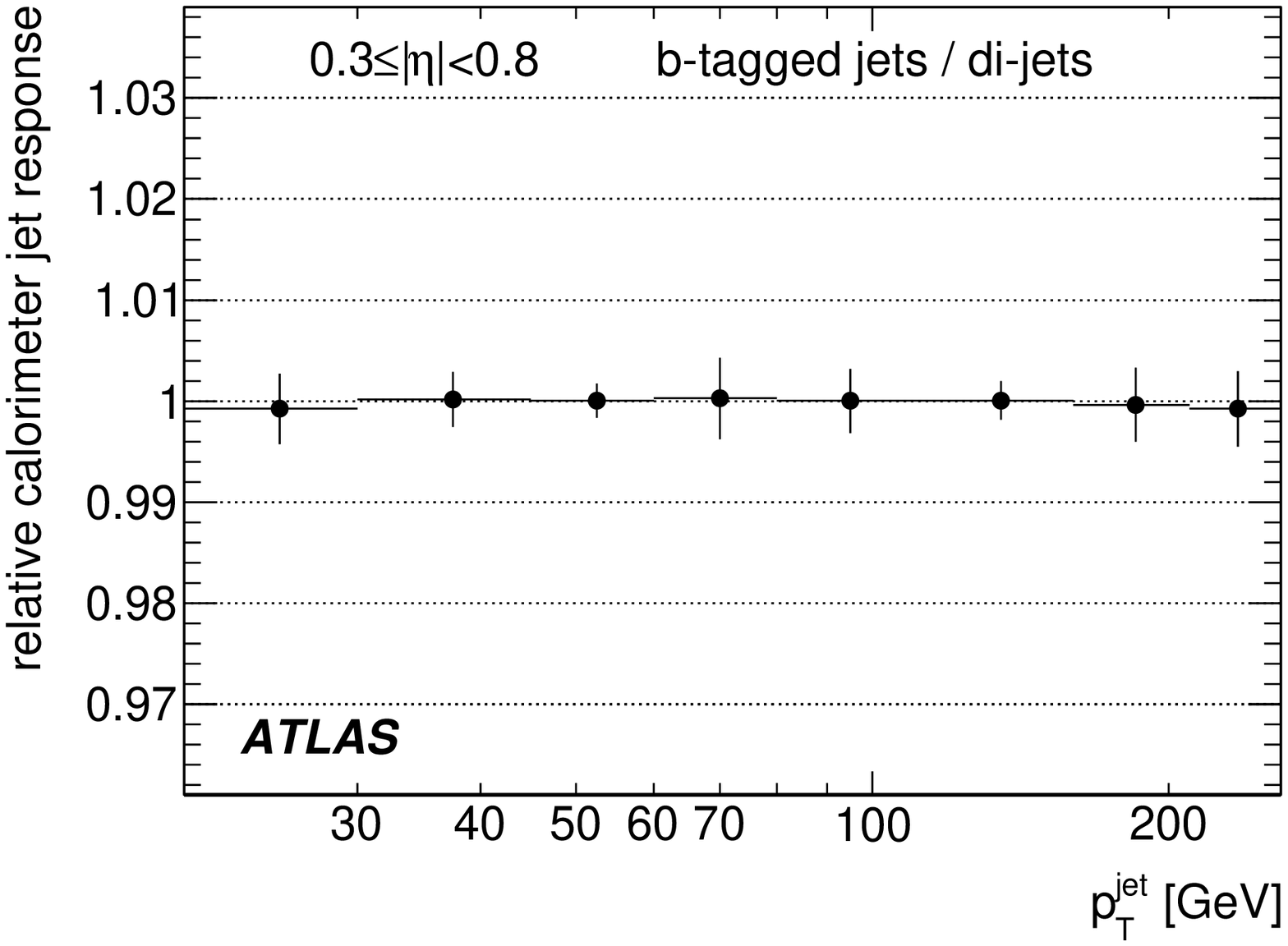}
      \label{fig:JES_corr_ratio_bjet} 
    }

    \subfloat[]{
      \includegraphics[width=.485\textwidth]
      {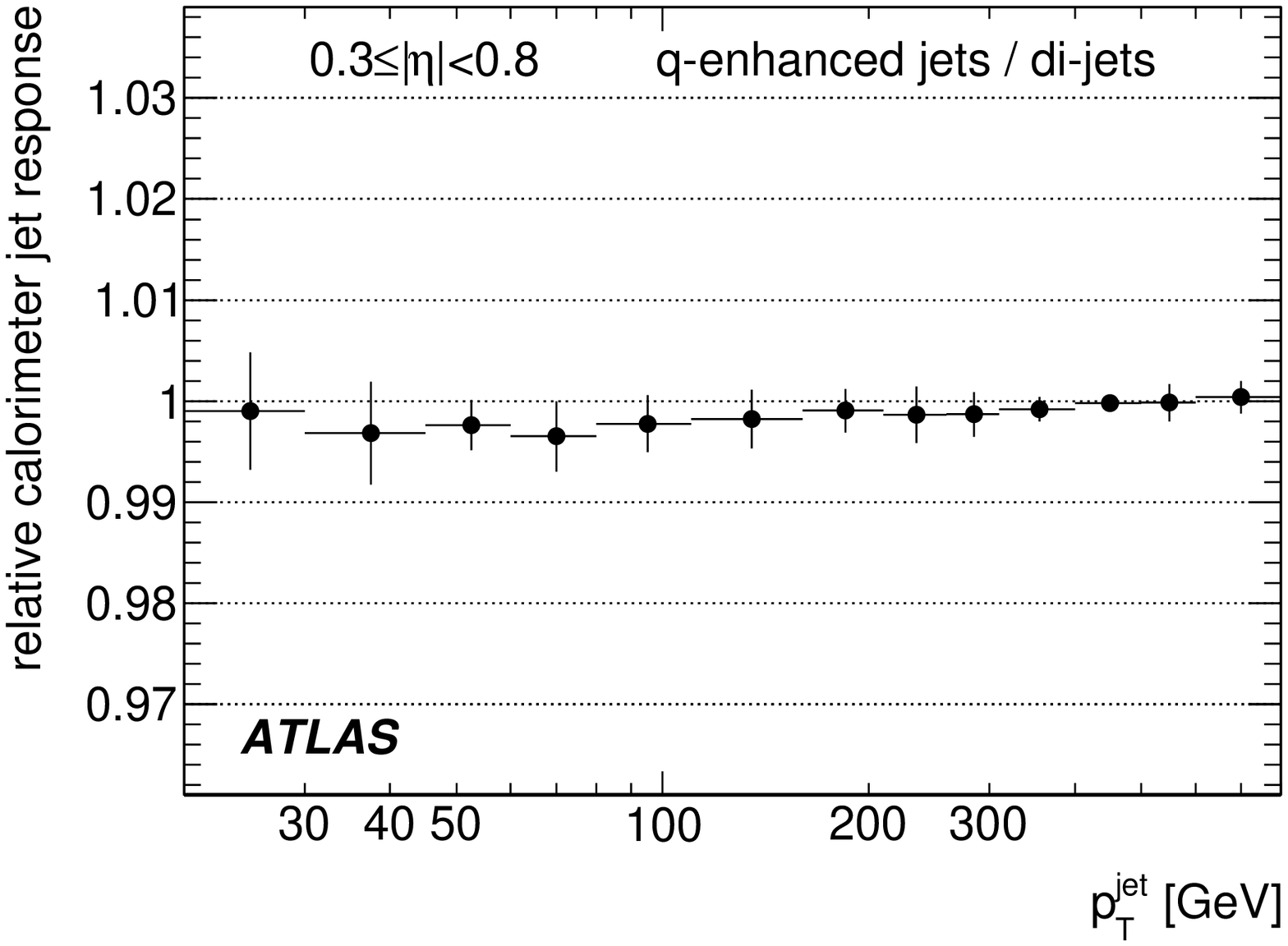}
      \label{fig:JES_corr_ratio_qjet} 
    }

    \caption{Expected shift (black dots) and uncertainty (error bars) on the relative calorimeter jet response as function of the jet transverse momentum for jets in the range $0.3 \leq |\eta|<0.8$: \protect\subref{fig:JES_corr_ratio_04_06} between jets reconstructed with the \antiKT jet algorithm with $R=0.4$ and jets reconstructed with $R=0.6$, \protect\subref{fig:JES_corr_ratio_bjet} between $b$-tagged jets in $t\bar{t}$ events and jets in inclusive di-jet events, and \protect\subref{fig:JES_corr_ratio_qjet} between mostly quark-initiated jets in $\gamma$+jet events and jets in inclusive di-jet events.
    The $x$-axis is the jet transverse momentum calibrated from the EM scale to the hadronic scale using an MC-based JES calibration factor \cite{JES_Summary_r16}.} 
    \label{fig:JES_corr_ratio} 
  \end{center}
\end{figure*}

In \figurename~\ref{fig:JES_corr_ratio_04_06} the relative calorimeter uncertainty on the JES between \antiKT jets reconstructed with $R=0.4$ and $R=0.6$ is shown. 

Figure~\ref{fig:JES_corr_ratio_bjet} shows the relative calorimeter uncertainty for $b$-tagged jets in $t\bar{t}$ events with respect to jets in inclusive di-jet events. Identified $b$-jets are associated with a displaced secondary vertex reconstructed by the SV0 algorithm \cite{ATLAS-CONF-2010-099}, with a weight greater than 5.72. This weight gives an average $b$-tag efficiency of 50\% for $b$-jets from $t\bar{t}$ events.

Finally, \figurename~\ref{fig:JES_corr_ratio_qjet} shows the relative uncertainty for jets initiated mostly by quarks with respect to jets in inclusive di-jet events. Jets initiated mostly by quarks are selected in $\gamma$+jet event simulations using the selection criteria applied in the method of the direct $\pt$ balance between $\gamma$ and jets \cite{JES_Summary_r16}.

Because of the small differences in particle composition between these different jet samples, the relative uncertainty on the pure calorimeter response is found to be below 0.5\% for low $\pt$ and negligible for high $\pt$. 

\section{Conclusions}
\label{sec:Conclusions}

The average calorimeter response to isolated hadrons with respect to the track momentum $\langle E/p \rangle$ has been measured in minimum bias events at $\sqrt{s} = 900$~GeV and $\sqrt{s} = 7$~TeV. A background from neutral hadrons of 4--8\% is subtracted from data and Monte Carlo with a systematic uncertainty of below 1\%.  After the background has been removed, the agreement between data and Monte Carlo simulation is within 2\% for particles with momenta up to 10~GeV and is around 5\% for momenta in the 10--30~GeV range, where the statistical uncertainty dominates the total uncertainty.

The calorimeter response of identified single charged hadrons has been measured using short-lived particle decays. A good agreement between data measured at $\sqrt{s} = 7$~TeV and the Monte Carlo simulation is found for charged pions and protons. However, a disagreement of up to 10\% is found between the Monte Carlo simulation and data for the difference of responses to low momentum anti-pions and anti-protons ($\langle E/p \rangle^{\pi^-}-\langle E/p \rangle^{\bar{p}}$). This difference is attributed to the poor modelling of the anti-proton response in the Monte Carlo simulation.

The ATLAS calorimeter jet energy scale uncertainty has been determined for the well understood central detector region by propagating the energy response uncertainty of all particles contributing to a jet. For charged hadron momenta below 20~GeV, the single charged hadron response has been used, while for higher momenta, the response measured in the ATLAS combined test beam has been included. For the response to neutral pions ($\pi^0\to\gamma\gamma$), the uncertainty on the electromagnetic calorimeter energy scale is dominant, while for all other neutral hadrons an additional uncertainty due to the limited knowledge of the calorimeter response has been incorporated.
 
An uncertainty of 1--3\% on the response to jets in the calorimeter is determined for jets in the central region of the detector with $|\eta| < 0.8$ and a transverse momentum between 15 GeV and 2.5 TeV. An analysis of the correlation in the jet energy scale uncertainty yields an almost complete correlation for jets close in $|\eta|$ or $\pt$, and a correlation of $\sim$~30\% for jets with a large 
transverse momentum separation. The relative jet energy scale uncertainty between \antiKT jets of size $R=0.4$ and $R=0.6$, between $b$-tagged jets and inclusive jets in di-jet events and between mostly quark-initiated jets and inclusive jets in the same event sample was found to be small.

\appendix
\section{Validation of the background subtraction}
\label{app:backval}

The background subtraction procedure relies on the assumption that the energy density deposited by other particles in the cone around a late showering track is constant as a function of the distance with respect to the track impact point on the calorimeter. Although this assumption is valid for low track momenta, the validity of the assumption for high track momenta is questionable, as jet-like structures are expected to emerge.

In order to address this issue, a different background subtraction procedure has been developed. The background is still measured in events where the energy in the EM calorimeter associated to the track is compatible with that of a late showering hadron (MIP-tagged track, in the following). In this case, however, the energy density of the background in small annuli between a cone of radius 0.1 and 0.2 is used to linearly extrapolate the background inside the cone of size 0.1. The differential energy density $\rho_i = \mathrm{d}^2 E/\mathrm{d}\eta \mathrm{d}\phi$ for an annulus of inner radius $R_i = i \times 0.025$  and outer radius $R_{i+1} = (i+1) \times 0.025$ is defined as 

\begin{equation}
\rho_i = \frac{E_{i+1} - E_{i}}{A_{i+1} - A_{i}} ,
\end{equation}

\noindent where $A_i = \pi R_i^2$.

\begin{figure*}[tb]
\centering
\subfloat[]{\label{fig:comparison_true_eta_1500_1800}\includegraphics[width=.45\textwidth]{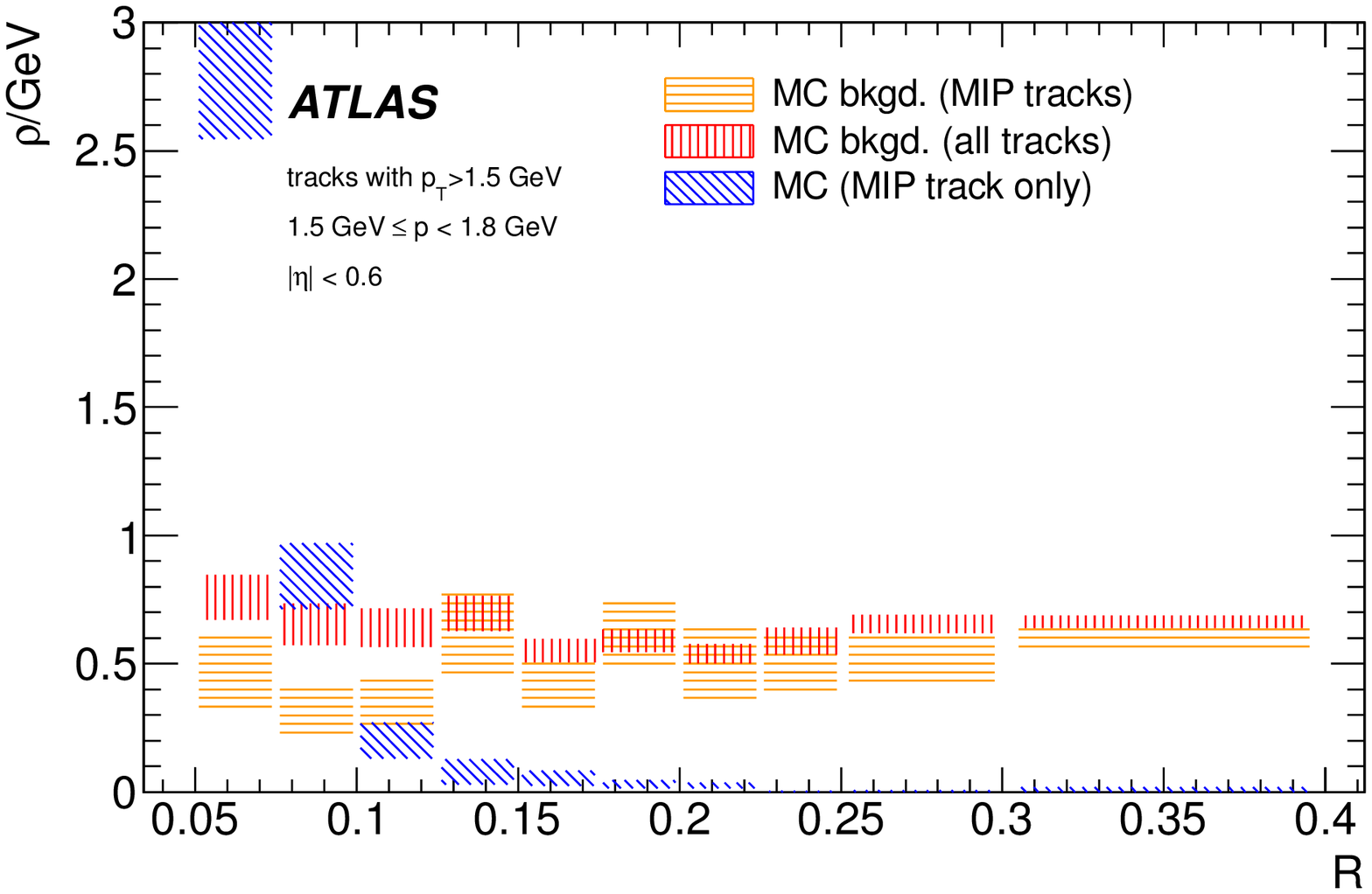}}
\subfloat[]{\label{fig:comparison_true_eta_3600_4600}\includegraphics[width=.45\textwidth]{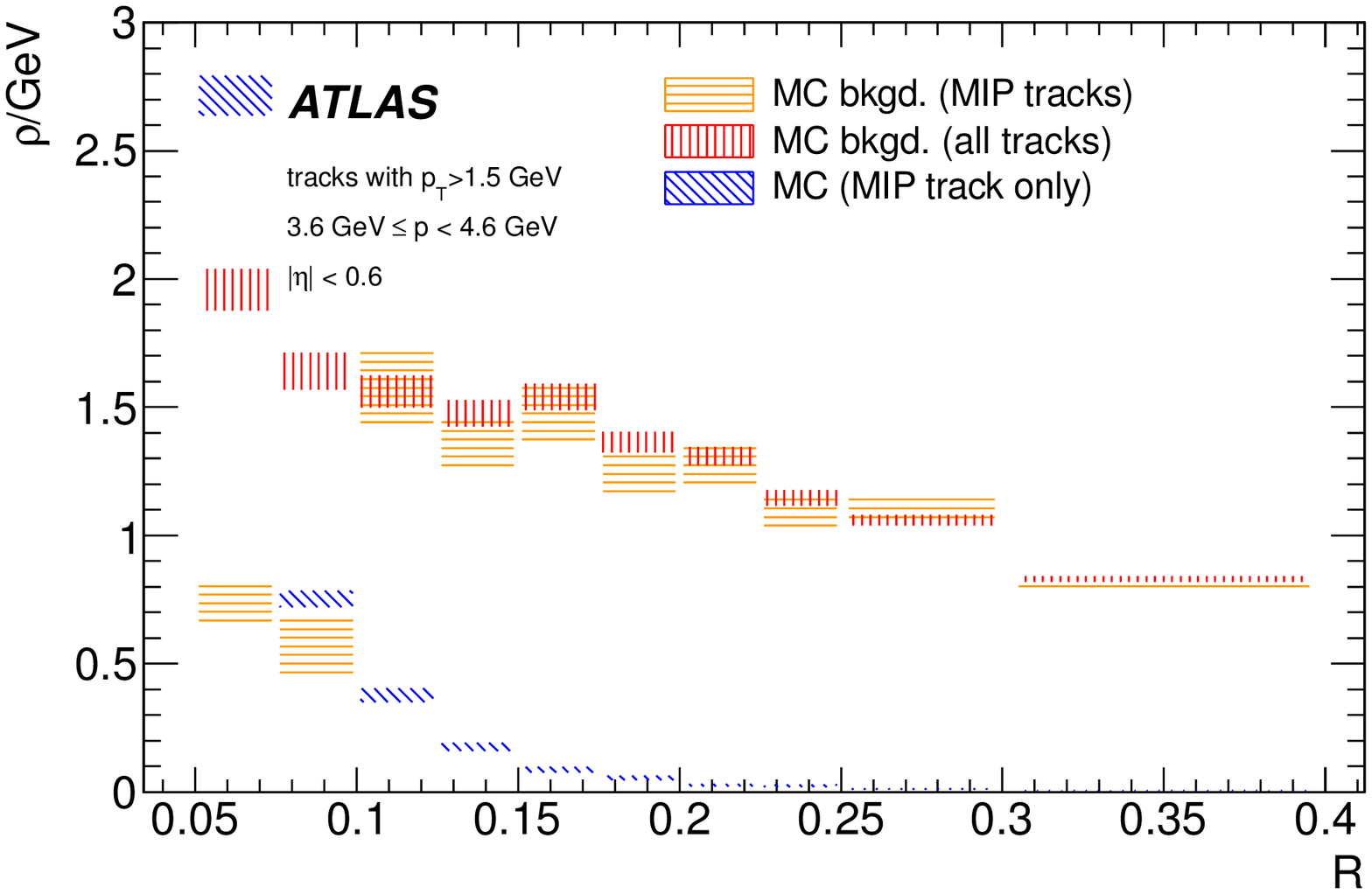}}
\caption{Energy density, $\rho$, as a function of the radial distance $R$ in  $\eta \times \phi$ space from the track impact point for tracks with \protect\subref{fig:comparison_true_eta_1500_1800} $1.5 \leq p < 1.8$~GeV and \protect\subref{fig:comparison_true_eta_3600_4600} $3.6 \leq p < 4.6$~GeV. The blue rectangles  with diagonal hatching show the true energy density associated with MIP tracks. The yellow rectangles with horizontal hatching (red rectangles with vertical hatching) show the true energy density in the EM calorimeter associated to background particles for MIP-tagged (all) tracks. 
}
\label{fig:AltBackSubTrue}
\end{figure*}

The dependence of $\rho_i$ on the distance from the track impact point on the calorimeter has been studied by making use of the true energy deposits as predicted by the {\sc Geant4} simulation. The true energy deposited by the track and by the neutral background for late showering tracks, and by the neutral background in all selected events is shown in \figurename~\ref{fig:AltBackSubTrue} for two different track momentum bins.  The energy density of the MIP-tagged tracks is very narrow in both momentum bins, as expected. The background energy density is constant and independent of the distance to the track impact point at low track momenta. This is consistent with the assumption of constant energy density made to evaluate the background with the baseline method. As the track momentum increases, the energy density close to the track impact point is higher, perhaps indicating a jet-like structure around the track. In both track momentum bins, the background energy density in the annulus $0.1 \leq R < 0.2$ is the same for MIP-tagged and all tracks. 

\begin{figure*}[tb]
\centering
\subfloat[]{\label{fig:comparison_reco_1500_1800}\includegraphics[width=.45\textwidth]{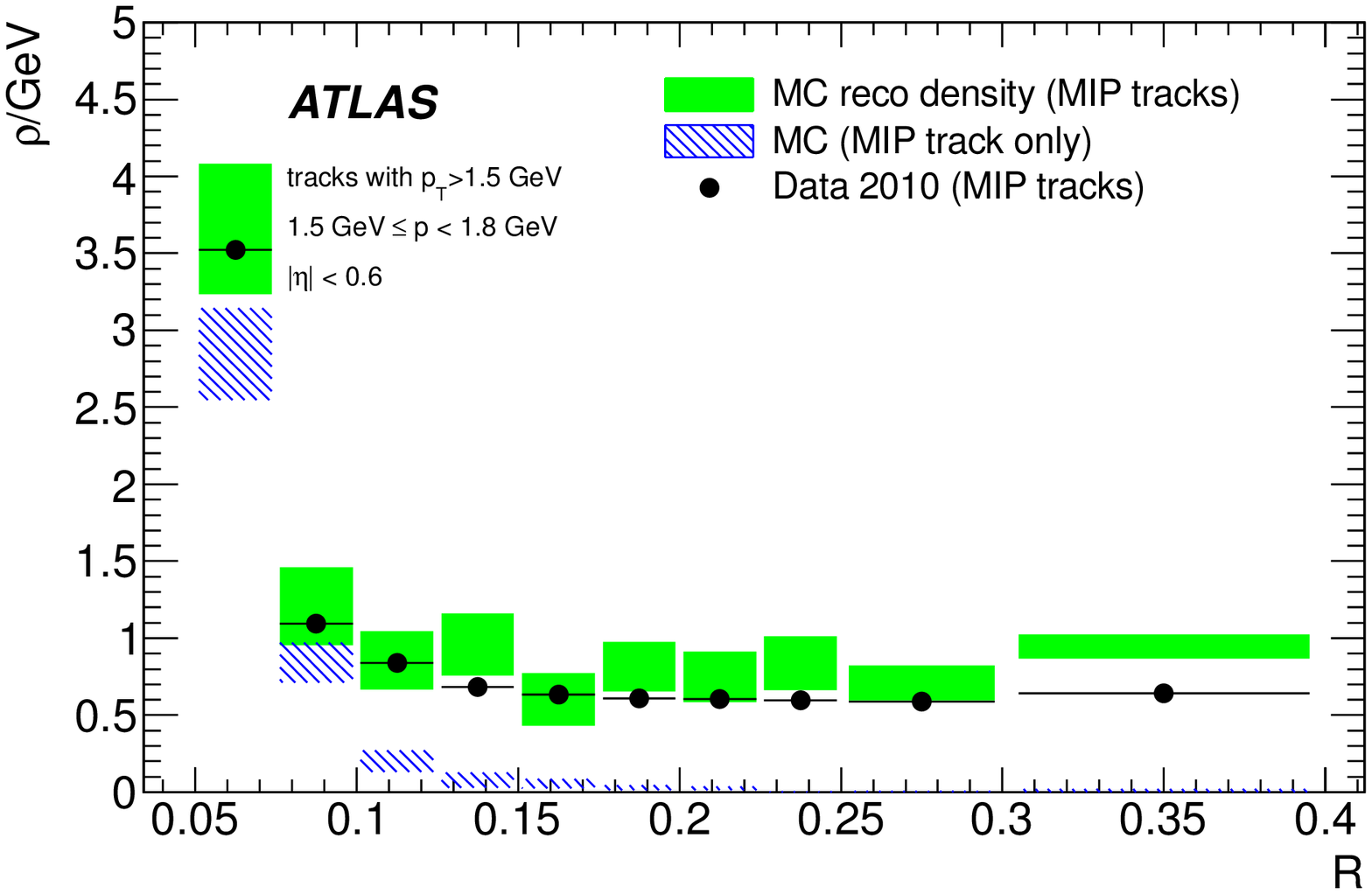}}
\subfloat[]{\label{fig:comparison_reco_3600_4600}\includegraphics[width=.45\textwidth]{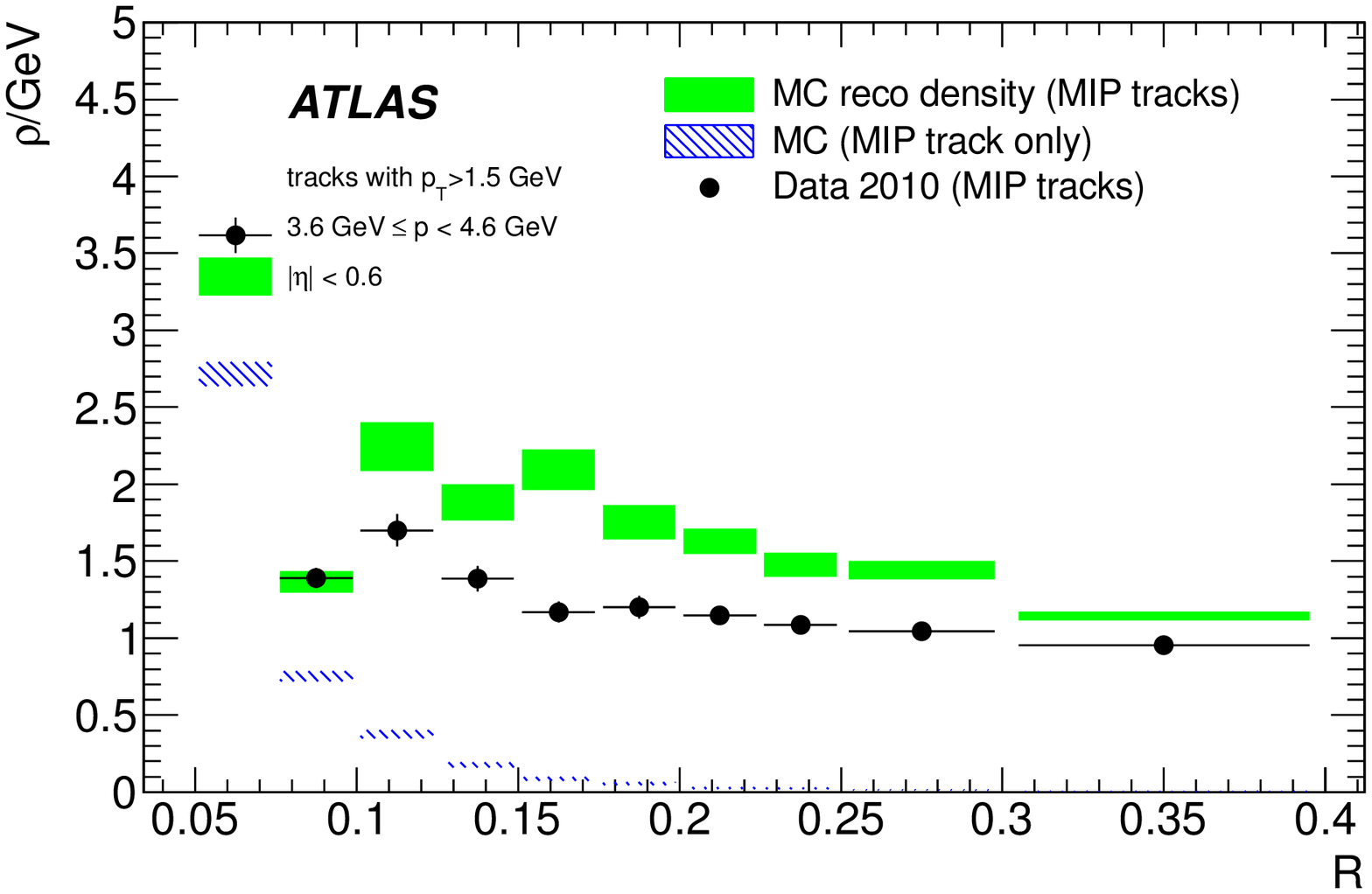}}
\caption{Energy density, $\rho$, as a function of the radial distance $R$ in  $\eta \times \phi$ space from the track impact point for tracks with \protect\subref{fig:comparison_reco_1500_1800} $1.5\leq p < 1.8$~GeV and \protect\subref{fig:comparison_reco_3600_4600} $3.6 \leq p < 4.6$~GeV. The black markers (green rectangles) show the reconstructed energy density associated with MIP-tagged tracks for data (MC). The blue hatched rectangles show the true energy density associated with MIP tracks. 
}
\label{fig:AltBackSubReco}
\end{figure*}

The same features can be observed in \figurename~\ref{fig:AltBackSubReco}. This time, the reconstructed energy density is shown for data and MC simulation (the points corresponding to the true energy density associated to the MIP-tagged track are kept for reference). A comparison of  Fig.~\ref{fig:AltBackSubTrue} and Fig.~\ref{fig:AltBackSubReco} reveals  that the main features observed for the true energy densities still hold for reconstructed energy densities: the density in the halo of the cone is a good measurement of the true background density, while the core of the cone mainly contains energy from the MIP track.

The alternative estimate of the background is obtained by fitting the measured energy density with a linear function between $R_i = 0.1$ and $R_i = 0.25$ and integrating the obtained function between $R_i = 0$ and $R_i = 0.25$:

\begin{equation}
f(R_i) = a + b R_i
\end{equation}

The obtained background estimation gives a total of 10\% more background at high track momenta with respect to the baseline background subtraction procedure, consistently for data and MC simulation. The reason for the small difference is that the two background estimates differ only in the core of the cone, which is a region of small area.

Since the background itself is a 10\% correction to the $\langle E/p \rangle$ measurement, the total effect introduced by the linear parameterisation of the energy density is of the order of 1\% on $\langle E/p \rangle$. Since data and MC points are affected in a similar way, the effect on the agreement between MC and data is negligible. 
  
%


\section*{Acknowledgements}

We thank CERN for the very successful operation of the LHC, as well as the
support staff from our institutions without whom ATLAS could not be
operated efficiently.

We acknowledge the support of ANPCyT, Argentina; YerPhI, Armenia; ARC,
Australia; BMWF, Austria; ANAS, Azerbaijan; SSTC, Belarus; CNPq and FAPESP,
Brazil; NSERC, NRC and CFI, Canada; CERN; CONICYT, Chile; CAS, MOST and NSFC,
China; COLCIENCIAS, Colombia; MSMT CR, MPO CR and VSC CR, Czech Republic;
DNRF, DNSRC and Lundbeck Foundation, Denmark; EPLANET and ERC, European Union;
IN2P3-CNRS, CEA-DSM/IRFU, France; GNAS, Georgia; BMBF, DFG, HGF, MPG and AvH
Foundation, Germany; GSRT, Greece; ISF, MINERVA, GIF, DIP and Benoziyo Center,
Israel; INFN, Italy; MEXT and JSPS, Japan; CNRST, Morocco; FOM and NWO,
Netherlands; RCN, Norway; MNiSW, Poland; GRICES and FCT, Portugal; MERYS
(MECTS), Romania; MES of Russia and ROSATOM, Russian Federation; JINR; MSTD,
Serbia; MSSR, Slovakia; ARRS and MVZT, Slovenia; DST/NRF, South Africa;
MICINN, Spain; SRC and Wallenberg Foundation, Sweden; SER, SNSF and Cantons of
Bern and Geneva, Switzerland; NSC, Taiwan; TAEK, Turkey; STFC, the Royal
Society and Leverhulme Trust, United Kingdom; DOE and NSF, United States of
America.

The crucial computing support from all WLCG partners is acknowledged
gratefully, in particular from CERN and the ATLAS Tier-1 facilities at
TRIUMF (Canada), NDGF (Denmark, Norway, Sweden), CC-IN2P3 (France),
KIT/GridKA (Germany), INFN-CNAF (Italy), NL-T1 (Netherlands), PIC (Spain),
ASGC (Taiwan), RAL (UK) and BNL (USA) and in the Tier-2 facilities
worldwide.


\bibliographystyle{bjets}
\bibliography{atlaspaper}
\clearpage
\onecolumn
\vspace*{0.2cm}
\begin{flushleft}
{\Large The ATLAS Collaboration}

\bigskip

G.~Aad$^{\rm 48}$,
B.~Abbott$^{\rm 111}$,
J.~Abdallah$^{\rm 11}$,
A.A.~Abdelalim$^{\rm 49}$,
A.~Abdesselam$^{\rm 118}$,
O.~Abdinov$^{\rm 10}$,
B.~Abi$^{\rm 112}$,
M.~Abolins$^{\rm 88}$,
O.S.~AbouZeid$^{\rm 158}$,
H.~Abramowicz$^{\rm 153}$
% ...
\bigskip

% ...
$^{10}$ Institute of Physics, Azerbaijan Academy of Sciences, Baku, Azerbaijan\\
$^{11}$ Institut de F\'isica d'Altes Energies and Departament de F\'isica de la Universitat Aut\`onoma  de Barcelona and ICREA, Barcelona, Spain\\
% ...
$^{48}$ Fakult\"{a}t f\"{u}r Mathematik und Physik, Albert-Ludwigs-Universit\"{a}t, Freiburg i.Br., Germany\\
$^{49}$ Section de Physique, Universit\'e de Gen\`eve, Geneva, Switzerland\\
% ...
$^{88}$ Department of Physics and Astronomy, Michigan State University, East Lansing MI, United States of America\\
% ...
$^{111}$ Homer L. Dodge Department of Physics and Astronomy, University of Oklahoma, Norman OK, United States of America\\
$^{112}$ Department of Physics, Oklahoma State University, Stillwater OK, United States of America\\
% ...
$^{118}$ Department of Physics, Oxford University, Oxford, United Kingdom\\
% ...
$^{153}$ Raymond and Beverly Sackler School of Physics and Astronomy, Tel Aviv University, Tel Aviv, Israel\\
% ...
$^{158}$ Department of Physics, University of Toronto, Toronto ON, Canada\\
% ...
\end{flushleft}

\end{document}